%% file: Main.tex
\documentclass[
  journal=mamobx,
  manuscript=article,
  layout=traditional]{achemso}
\setkeys{acs}{articletitle=true}
\usepackage[cp1251]{inputenc}
\usepackage[english]{babel}
\usepackage{amsmath,amssymb,bm}
\usepackage{calc,cancel,caption}
\usepackage[usenames,dvipsnames]{color}
\usepackage{courier,epsfig,epstopdf}
\usepackage{geometry,graphics,graphicx}
\usepackage{helvet}
\usepackage{makeidx,mathptmx,mciteplus}
\usepackage{multirow}
\usepackage{setspace,sistyle,subcaption}
\usepackage{tablefootnote}
\usepackage{tabularx}
\usepackage{verbatim}
\usepackage{xkeyval,xcolor,xr}
\usepackage{wrapfig}

\DeclareMathOperator{\Tr}{Tr}

\author{Igor A. Gula}
\email{gula@sdu.dk}
\affiliation[University of Southern Denmark]{University of Southern Denmark, Campusvej 55, 5230 Odense M, Denmark}

\author{Hossein Ali Karimi-Varzaneh}
\email{ali.karimi@conti.de}
\affiliation[Continental Reifen Deutschland GmbH]{Continental Reifen Deutschland GmbH, J{\"a}dekamp 30, D-30419 Hannover, Germany}

\author{Carsten Svaneborg}
\email{zqex@science.dk}
\affiliation[University of Southern Denmark]{University of Southern Denmark, Campusvej 55, 5230 Odense M, Denmark}

\title{Computational study of the cross-link and the entanglement contributions to the elastic properties of model PDMS networks}

\SectionNumbersOn

\begin{document}

\tableofcontents

\input "Abstract"

\input "1.Intro"

\input "2.0.Theory"

\input "2.1.Kuhn"

\input "2.2.MR"

\input "2.3.Micro"

\input "3.0.Methods"

\input "3.1.KG"

\input "3.2.Mapping"

\input "3.3.Characterization"

\input "3.4.SetUp"

\input "3.5.PPA"

\input "3.6.3PA"

\input "4.0.Results"

\input "4.1.MR"

\input "4.2.PP"

\input "4.3.0.Micro"

\input "4.3.1.Universality"

\input "4.3.2.FitLinearData"

\input "4.3.3.FitFullData"

\input "5.Conclusion"

\begin{acknowledgement}
The simulations were carried out using the LAMMPS molecular dynamics software.\cite{Plimpton1995LAMMPS} Computation/simulation for the work described in this paper was supported by the DeiC National HPC Center, University of Southern Denmark, Denmark. We gratefully acknowledge discussions of these results with R. Everaers and F. Fleck. As this is the revised version of the manuscript, the great thanks go to the reviewers for the time they dedicated for careful reading the manuscript, reviewing it and for the useful comments and suggestions for improvements.
\end{acknowledgement}

\bibliography{biblio}

\end{document}

%% file: Abstract.tex
\section*{Abstract}
\label{sct:Abstract}

We built randomly cross-linked model PDMS networks and used Molecular Dynamics Methods to obtain stress-strain curves. Mooney-Rivlin~(MR) analysis was used to estimate the shear moduli. We applied Primitive Path analysis~(PPA) and its variation, Phantom Primitive Path analysis~(3PA), to estimate the entanglement and the cross-link moduli, respectively.
The MR moduli estimates are in good agreement with the sum of the entanglement and the cross-link moduli, and we observe that the stress-strain data collapse to a universal form when reduced with the PPA and 3PA moduli.
We studied how the MR parameters $\mathrm{C}_1$, $\mathrm{C}_2$ vary from cross-link to entanglement dominated networks. For the latter, we observed a $40\%$, $60\%$ contribution of $2\,\mathrm{C}_1$, $2\,\mathrm{C}_2$ to the shear modulus, respectively.
Finally, we fitted several models to the data. While all fits are good, the estimates for the entanglement and the cross-link moduli vary significantly when compared to our PPA and 3PA benchmarks.

%% file: 1.Intro.tex
\section{Introduction}
\label{sct:Intro}

Polymers are long chain-like molecules consisting of repeating chemical units. Cross-linking a polymer melt leads to the formation of an amorphous solid, an elastomer. Elastomers are unique due to their ability to reversibly sustain large deformations that can be up to several times their undeformed size. Comprehensive knowledge of the mechanical properties of elastomers and deep understanding of their microscopic origin are of great fundamental and practical importance.~\cite{Strobl1997PhysPolym, Mark2007PhysPropertPolymHandbook}

Elastomers represent, in essence, a single huge molecule, a polymer network, which consists of topologically entangled polymer chains~\cite{Edwards1967StatMechTopologyI} chemically connected by cross-links.~\cite{Edwards1967StatMechPolymMat, DeamEdwards1976ThrRubElast} The large number of conformational degrees of freedom of the polymer chains gives rise to the enormous reversible deformability of the elastomers and explains why their stress response is dominated by entropic effects. Cross-links and topological entanglements localize thermal fluctuations. These two effects are qualitatively different. Once two or more monomers are joined by a cross-link, their relative motion is constrained. Polymer chains can not move through each other, the result is topological entanglements. In melts, entanglements can slide freely along the chains, while in networks the mobility of entanglements is restricted by the cross-links.~\cite{KhokhlovNechaev1990PolymChainArrObstac}

Historically, the affine network model~\cite{
Kuhn1936,
KuhnGrun1942,
Kuhn1946,
Wall1942StatTDRubI,
Wall1942StatTDRubII,
Wall1943StatTDRubIII,
FloryRehner1943StatMechXlnkdPolymNtwkI,
Treloar1943ElasticityOfNtwkI}
and the phantom network model~\cite{
JamesGuth1943ThrElstPropRub,
James1947StatPropNetwFlexChains,
JamesGuth1947,
JamesGuth1949,
Flory1976}
were the first attempt to describe rubbery materials with a microscopic statistical mechanical model. However, numerous experimental results showed significant deviations from the predictions of these models.~\cite{Treloar1975PhysRubElasticity} These deviations were due to additional topological constraints, entanglements, not captured by these early models.~\cite{Edwards1967StatMechTopologyI} In polymer melts, entanglements are solely responsible for the transient elastic properties characterized by the melt plateau modulus. In networks, the entanglements are permanently ''trapped'' by cross-links and hence preserve their contribution to elastic properties, i.e. the tube topology is frozen by cross-links. Numerous microscopic models for elasticity of entangled polymer networks have been proposed, e.g., the non-affine tube model~\cite{RubinPanyuk1997NFFDfrmElstctPlmNtwk} and the slip tube model~\cite{RubinPanyuk2002ElastPolymNetw} of Rubinstein and Panyukov, the extended tube model~\cite{KaliskeHeinrich1999ExtendedTubeModel} of Kaliske and Heinrich, the double tube model~\cite{MergellEveraers2001TubeModels} of Mergell and Everaers, the non-affine network model of Davidson and Goulbourne~\cite{DavidsonGoulbourne2013NffNtwkModel}, and the general constitutive model of Xiang et al.~\cite{XiangZhongWangMaoYuQu2018GnrlModelSoftElast}. These models have been compared to experiment, see, e.g., Refs.~\cite{
GottliebGaylord1983I1D,
GottliebGaylord1984IISwelling,
GottliebGaylord1987III2D,
HiggsGaylord1990,
KawamuraUrayamaKohjiya2001MltXDfrmOfELPolymerNtwrkII}

The most well-known macroscopic phenomenological approach for the description of the mechanical behaviour of rubbery materials is the Mooney-Rivlin~(MR) model.~\cite{
Mooney1940Origin,
Rivlin1948LrgLstDfrmI,
Rivlin1948LrgLstDfrmII,
Rivlin1948LrgLstDfrmIII,
Rivlin1948LrgLstDfrmIV,
RivlinSaunders1951LrgLstDfrmVII}
The MR model is based on an empirical expression, relating the strain energy density and deformation tensor. It was shown that the predictive power of the model was limited, and the model was only able to describe uniaxial stretching.~\cite{Rivlin1948LrgLstDfrmIII} Moreover, the microscopic interpretation of the MR model parameters is not clear~\cite{Mark1975MRConstants, MarkSullivan1977ModelNetworksOfELPDMSI, SharafMark1994Interpretation} and has been a topic for discussion in the literature for more than $50$ years.~\cite{SchloglTrutschelChasseRiessSaalwachter2014}

Cross-linking is the process, by which a precursor melt is converted into a random network. The result is rubber materials with varying elastic properties. However, their microscopic structure remains largely unknown and uncontrolled. Hence, systematic investigations of structure-property relations are faced with difficulties, since experimentally it is nearly impossible to accurately and independently characterize the network structure of the studied samples. Furthermore, different cross-linking chemistries give rise to different types of network defects such as chain scission and dangling ends that acerbate the complexity of characterizing the network structure.~\cite{SharafMark1994Interpretation}

Computer simulations and in particular Molecular Dynamics~(MD) methods offer a useful alternative to experiment for systematic investigations of structure-property relations. The greatest advantages of computer modelling compared to experiments are its ability to ''look inside'' the material under study, to control the cross-linking process, and to comprehensively characterize the resultant network structure as well as its topological state, for instance, by analyzing its strand length and cross-link functionality distributions. Furthermore, ideal defect free models materials can be made.

Reproducibility of simulation results hinges on the ability to produce well equilibrated precursor melts. A polymer chain has the structure of an extended random walk, and many molecules pervade the volume spanned by a single chain. Hard interactions between chains are required in computational polymer models to prevent chain from moving through each other. The aim of equilibration is to produce well equilibrated model melts, where the statistics of each chain is consistent with the desired polymer chemistry, and density fluctuations are absent due to the melts incompressibility. Brute force relaxation of precursor melts of practical interest would require simulations exceeding what is possible with current hardware. However, recent advances in melts equilibration techniques, see Refs.~\cite{SvaneborgKarimiHojdisFleckEveraers2016MultiscaleApproach, ZhangMoreiraStuehnDaoulasKremer2014EquilHierarchStrat, AuhlEveraersGrestKremerPlimpton2003EQLongChainPlmMltInCmpSim}, make it possible to generate well equilibrated, huge, highly entangled model precursor melts for computational studies.

We generated well-equilibrated KG model precursor melts, having $500$ chains with $N_b=10\,000$ beads in each, following the approach described in Ref.~\cite{SvaneborgKarimiHojdisFleckEveraers2016MultiscaleApproach}. The Kuhn number of the KG polymer model was chosen to match PDMS.~\cite{EveraersKarimiFleckHojdisSvaneborg2020KGMap} The ends of the precursor chains were initially bonded to neighbouring beads to form an end-linked $3$-functional network. Subsequent bonds were introduced between random bead pairs to produce a predominantly $4$-functional model polymer materials with various cross-link densities. The resulting model networks do not contain dangling ends, since the long precursor chains are initially cross-linked. However, we observe that $13-17$\% of the strands form loops, which appear as the result of intramolecular cross-linking.~\cite{LangGoritzKreitmeier2005IntraReactions} This effect would also occur in real materials.

The resulting model materials were uniaxially stretched. We estimate the elastic moduli of model PDMS networks by two approaches: analysis of the simulation stress-strain curves within the scope of the MR empirical model and static structural Primitive Path methods. Elastic moduli can be estimated from simulation stress-strain data, however, such simulations are expensive due to the very long relaxation times of polymer materials. We invested in excess of $200$ core years of computer time in estimation of the equlibrium stresses. The resulting stress-strain curves were analyzed within the scope of the MR model to estimate the parameters $\mathrm{C}_1$, $\mathrm{C}_2$ and, consequently, to obtain the shear modulus $\mathrm{G}$. Static analysis methods such as Primitive Path Analysis~(PPA)~\cite{EveraersSukuGrestSvaneborgSivaKremer2004RheolTopol} and Phantom Primitive Path Analysis~(3PA)~\cite{SvaneborgEveraersGrestCurro2008StressContributions} allows us to independently estimate the entanglement modulus $\mathrm{G}_E$ of the precursor network and the cross-links moduli $\mathrm{G}_X$ of the networks, respectively.

To provide a microscopic interpretation of the empirical MR coefficients, we made Langley plots of $\mathrm{C}_1$, $\mathrm{C}_2$ and also studied the cross-link and entanglement modulus contributions to the shear modulus. Using the entanglement and the cross-link shear moduli, we were also able to reduce our stress-strain data to a single universal curve. We fitted the non-affine tube model~\cite{RubinPanyuk1997NFFDfrmElstctPlmNtwk}, the slip tube model~\cite{RubinPanyuk2002ElastPolymNetw}, the extended tube model~\cite{KaliskeHeinrich1999ExtendedTubeModel}, the double tube model~\cite{MergellEveraers2001TubeModels}, the non-affine network model of Davidson and Goulbourne~\cite{DavidsonGoulbourne2013NffNtwkModel} and the general constitutive model of Xiang~\cite{XiangZhongWangMaoYuQu2018GnrlModelSoftElast} to the simulation data, and compared obtained the cross-link and the entanglement moduli to those independently obtained by the Primitive Path methods. This provides a computational calibration standard of these parameters, which is useful when interpreting fits to experimental results where the cross-link and the entanglement moduli are not independently available.

The paper is organized as follows. Sect.~\ref{sct:Theory} summarizes theoretical background used for the interpretation and the analysis of our KG model simulation results. In Sect.~\ref{sct:Methods}, we explain how we build the model networks, characterize them, set up the MD simulations, and describe the PPA techniques. Results are summarized and discussed in Sect.~\ref{sct:Results}. In Sect.~\ref{sct:Conclusion}, we present our conclusions.

%% file: 2.0.Theory.tex
\section{Theoretical models}
\label{sct:Theory}

The present section contains the necessary theoretical background. We begin with the introduction of the unified Kuhn notation for description of polymers~[Sect.~\ref{sbsct:Kuhn}]. Next we present the macroscopic Mooney-Rivlin model of incompressible hyperelastic solids~[Sect.~\ref{sbsct:MR}] and continue with microscopic models for polymer elasticity~[Sect.~\ref{sbsct:Micro}].

%% file: 2.1.Kuhn.tex
\subsection{Kuhn model}
\label{sbsct:Kuhn}

In this section, we summarize the Kuhn approach for the description of polymers.~\cite{Kuhn1934} Kuhn's seminal insight was the idea to map a polymer chain to an equivalent freely jointed chain model~(FJC), matching the contour length and end-to-end distance of the polymer chain.~\cite{Flory1953PrincipPolymerChem}

The static configuration of a single polymer molecule can be characterized by its contour length $L$ and mean-square end-to-end distance $\left<R^2\right>$. The equivalent Kuhn model chain consists of $N_K$ Kuhn segments of length $l_K$, which is denoted as the Kuhn length. The values of $N_K$ and $l_K$ are chosen to match the mean-square end-to-end distance and the contour length of the molecule, hence, $L=l_K\,N_K$, $\left<R^2\right>=l_K^2\,N_K$, therefore, one obtains $l_K=\left<R^2\right>/L$ for the Kuhn length and $N_K=L^2/\left<R^2\right>$ for the number of Kuhn segments. We assume chains are long enough, so that finite-chain length effects can be neglected. The molar mass of a Kuhn segment is $M_K=M_c/N_K$ where $M_c$ denotes the molar mass of the whole polymer molecule.

To describe the static properties of a system of many interpenetrating long polymer molecules, we introduce the number density of Kuhn segments $\rho _K = \rho _c\,N_K$, where $\rho _c$ is the number density of chains. The volume spanned by a single chain is $V=\left<R^2\right>^{3/2}=l_K^3\,N_K^{3/2}$. The degree of chain interpenetration is described by the Flory number, $n_F = \rho _c\,V$, which estimates the number of neighbors a single chain can interact with.~\cite{Flory1949ConfigRealPolyChains} The Flory number can be expressed in Kuhn parameters as $n_F = \rho _K\,l_K^3\,N_K^{1/2}$. While the chain length $N_K$ varies from melt to melt, the prefactor $n_K=\rho _K l_K^3$ called the Kuhn number plays the role of dimensionless density and depends only on the specific polymer chemistry.

The characteristic time scales for the dynamics of a single polymer molecule are obtained from the Rouse model.~\cite{Rouse1953Origin} The dynamics of a Kuhn segment can be characterized by a friction $\zeta _K$ or, equivalently, by the Kuhn time $\tau _K \sim l_K^2/D_K \sim l_K^2\,\zeta _K/\left(k_B\,T\right)$, which is the time it takes a Kuhn segment to diffuse a Kuhn length $l_K$, where $D_K=k_B\,T/\zeta _K$ is the Kuhn segment diffusion constant, $k_B$ is the Boltzmann constant, and $T$ is temperature. The dynamics of a whole chain can be characterized by the Rouse time $\tau _R \sim \left<R^2\right>/D_c \sim l_K^2\,N_K^2\,\zeta _K/\left(k_B\,T\right)$, which is the time required for a chain to diffuse its own size, where $D_c=k_B\,T/(N_K\,\zeta _K)$ is the chain diffusion coefficient. Consequently, the Rouse and Kuhn times are related as $\tau _R = \tau _K\,N_K^2$.

The major advantage of the Kuhn description of polymer physics is that the universal properties (dominated by conformational entropy) become apparent. For a melt, the emergent macroscopic properties such as the relaxation time scales, chain size, time dependent shear modulus and the viscosity depend only on two dimensionless parameters, the Kuhn number $n_K$ and the number of Kuhn segments per chain $N_K$, i.e. chain length of the precursor melt, in addition to the dimensional parameters of the Kuhn length $l_K$, Kuhn time $\tau_K$ and the energy $k_B\,T$.~\cite{EveraersKarimiFleckHojdisSvaneborg2020KGMap} For model networks, the number of Kuhn segments between cross-links $N_{XK}$, which denotes the average strand length in the rest of the paper, replaces melt chain length as a parameter. This is due to the fact that the chain length of the precursor melt~($N_b = 10\,000$ beads) is so large that it is irrelevant for the determination of the strand length distribution of the network. If the networks were formed by shorter end-linking chains or from a bimodal precursor melt, then additional dimensionless parameters could be required to characterize the strand length distribution.

%% file: 2.2.MR.tex
\subsection{Mooney-Rivlin material model}
\label{sbsct:MR}

The empirical Mooney-Rivlin model~(MR)~\cite{Mooney1940Origin, Rivlin1948LrgLstDfrmI, Rivlin1948LrgLstDfrmII, Rivlin1948LrgLstDfrmIII, Rivlin1948LrgLstDfrmIV, RivlinSaunders1951LrgLstDfrmVII} for incompressible hyperelastic material is often used for the analysis of the experimental stress-strain data and elastic moduli estimation of rubbery polymers. The model relates the strain energy density function $\mathrm{W}$ stored by an incompressible solid and its deformation as:
\begin{equation}
\label{eqn:MRModel}
  \mathrm{W}(\bm{\Phi}) = \mathrm{C}_1\,\left(\mathrm{I}_1(\bm{\Phi})-3\right) + \mathrm{C}_2\,\left(\mathrm{I}_2(\bm{\Phi})-3\right)\,,
\end{equation}
where $\mathrm{C}_1$, $\mathrm{C}_2$ are the material constants, $\mathrm{I}_1(\bm{\Phi}) = \Tr\bm{\Phi}$, $\mathrm{I}_2(\bm{\Phi}) = \left(\left(\Tr\,\bm{\Phi}\right)^2 - \Tr\left(\bm{\Phi}^2\right)\right)/2$ are the $1^{st}$ and the $2^{nd}$ invariants of the Finger tensor $\bm{\Phi} = \mathbf{E}\cdot\mathbf{E}^T$, respectively, $\mathbf{E}=\left(\stackrel{0}{\bm{\nabla}}\mathbf{R}\right)^T$ is the deformation gradient tensor and $\mathbf{R}$ is a coordinate vector of a material point in the actual configuration. The operator $\stackrel{0}{\bm{\nabla}}$ indicates that differentiation is performed with respect to coordinates of a material point in the reference configuration.

The MR parameters $\mathrm{C}_1$, $\mathrm{C}_2$ are related to the shear modulus $\mathrm{G}$ as:
\begin{equation}
\label{eqn:MR2G}
  \mathrm{G}^{MR} = 2\,\left(\mathrm{C}_1 + \mathrm{C}_2\right)\,.
\end{equation}

Let us consider uniaxial deformation of a solid along $Ox$ axis by a factor $\lambda$, then the normal tension $\sigma _N$ is given by:
\begin{equation}
\label{eqn:MR_nt}
  \sigma _{N} \equiv \sigma _{xx} - \dfrac{\sigma _{yy} + \sigma _{zz}}{2} = 2\,\left(\mathrm{C}_1 + \mathrm{C}_2\,\lambda^{-1}\right)\,\left(\lambda^2 - \lambda^{-1}\right)\,,
\end{equation}
here $\sigma _{xx}$, $\sigma _{yy}$ and $\sigma _{zz}$ are the diagonal components of the Cauchy stress tensor $\bm{\sigma}$:
\begin{equation*}
  \bm{\sigma} = \dfrac{\partial\mathrm{W}}{\partial\mathbf{E}}\cdot\mathbf{E}^T-\mathrm{P}\,\mathbf{I}\,,
\end{equation*}
where $\mathrm{P}$ is pressure, $\mathbf{I}$ is the unit tensor. For analysis and convenient representation of experimental results, the reduced normal tension $\widetilde{\sigma}_N$ is defined as:
\begin{equation}
\label{eqn:MR_rnt}
  \widetilde{\sigma}_N(\lambda) \equiv \dfrac{\sigma _N}{\lambda^2 - \lambda^{-1}} = 2\,\left(\mathrm{C}_1 + \mathrm{C}_2\,\lambda^{-1}\right)\,,
\end{equation}
hence, the MR model postulates a linear dependency of $\widetilde{\sigma}_N$ on the inverse elongation $\lambda^{-1}$.

It was shown that the MR model provided a good description of experiments on uniaxial stretching of rubbery materials, whereas it failed to describe other deformation modes nor did it correctly describe the materials stress response at large deformations.~\cite{Treloar1975PhysRubElasticity,GottliebGaylord1983I1D,KawamuraUrayamaKohjiya2001MltXDfrmOfELPolymerNtwrkI}

%% file: 2.3.Micro.tex
\subsection{Microscopic models of polymer elasticity}
\label{sbsct:Micro}

The MR model is empirical, hence the model parameters $\mathrm{C}_1$, $\mathrm{C}_2$ do not \emph{a priori} have a physical interpretation. To identify them, one would have to elucidate their microscopic origin, e.g. relate them to the cross-link and the entanglement moduli $\mathrm{G}_X$, $\mathrm{G}_E$ of rubber model materials.~\cite{MarkSullivan1977ModelNetworksOfELPDMSI, SharafMark1994Interpretation} Here, we denote $\mathrm{G}_X$ as the phantom modulus of the network and $\mathrm{G}_E$ as the entanglement modulus of the precursor melt.

The first microscopic theories of polymer elasticity incorporated only the contribution from network connectivity. The affine network model~\cite{
Kuhn1936,
KuhnGrun1942,
Kuhn1946,
Wall1942StatTDRubI,
Wall1942StatTDRubII,
Wall1943StatTDRubIII,
FloryRehner1943StatMechXlnkdPolymNtwkI,
Treloar1943ElasticityOfNtwkI}
assumes that all strands are monodisperse and pinned to an affinely deforming background. The model predicts the cross-link modulus as:
\begin{equation}
\label{eqn:FF_ntwrk_GX}
  \mathrm{G}_X^{aff} \equiv \dfrac{\sigma _N}{\lambda^2 - \lambda^{-1}} = \dfrac{\rho_K\,k_B\,T}{N_{XK}}\,,
\end{equation}
where $N_{XK}$ is the number of Kuhn segments per network strand. On the other hand, within the scope of the phantom network model~\cite{
JamesGuth1943ThrElstPropRub,
James1947StatPropNetwFlexChains,
JamesGuth1947,
JamesGuth1949,
Flory1976}
the thermal fluctuations of the network junctions are not constrained rigidly. For the cross-link modulus the following relation was derived:~\cite{
Graessley1975,
Flory1976,
EdwardsViglis1988TubeModelThrRubbElast}
\begin{equation}
\label{eqn:Phnt_ntwrk_GX_org}
  \mathrm{G}_X^{ph} = \left(\rho_S^{good} - \rho_X^{good}\right)\,k_B\,T\,,
\end{equation}
where $\rho_S^{good}$, $\rho_X^{good}$ are the densities of ''good'' strands and cross-links, respectively. During cross-linking, network defects such as dangling ends and loops are created. They do not carry stress upon macroscopic deformation, hence, do not contribute to the elastic properties of the cross-linked material. ''Bad'' strands are only connected to the network by one end, and as such can not carry a load. Loops are formed when two monomers belonging to the same polymer chain are cross-linked. Most such loops are short and do not capture entanglements with other chains.

In the case where all the junctions are assumed to have the same functionality $f$ and the network strands are assumed to be monodisperse with the length $N_{XK}$, the phantom network model Eq.~(\ref{eqn:Phnt_ntwrk_GX_org}) predicts the phantom modulus as:
\begin{equation}
\label{eqn:Phnt_ntwrk_GX_prt}
  \mathrm{G}_X^{ph} = \left(1-\dfrac{2}{f}\right)\,\dfrac{\rho_K\,k_B\,T}{N_{XK}}\,.
\end{equation}

The prefactor $1-2/f$ was interpreted as being due to the fact that the strand ends are not directly pinned to the deforming background as in the affine network model, but rather via an infinite Cayley tree of monodisperse strands and $f$-functional junctions.~\cite{
EdwardsViglis1988TubeModelThrRubbElast,
RubinPanyuk1997NFFDfrmElstctPlmNtwk} Refinements of this approximation, including the effects of loops, have been recently proposed in Refs.~\cite{
ZhongWangKawamotoOlsenJohnson2016RENT,
Lang2018,
Panyukov2019,
LinWangJohnsonOlsen2019ElasticityRealGaussPhtnmNtwrk}.

The effect of topological entanglements can be accounted for in multiple ways, and the tube model, introduced in Refs.~\cite{Edwards1967StatMechTopologyI, Edwards1967StatMechPolymMat}, is the most successful. The idea is that thermal fluctuations of the network chains are restricted in space not only by the cross-links, but also by entanglements. Numerous approaches have been proposed~\cite{
EdwardsViglis1986EffOfEntanglRubbElast,
EdwardsViglis1988TubeModelThrRubbElast,
RubinPanyuk1997NFFDfrmElstctPlmNtwk,
KaliskeHeinrich1999ExtendedTubeModel,
MergellEveraers2001TubeModels,
RubinPanyuk2002ElastPolymNetw,
DavidsonGoulbourne2013NffNtwkModel,
XiangZhongWangMaoYuQu2018GnrlModelSoftElast}
, and below we present several models, which contain the cross-link and the entanglement moduli $\mathrm{G}_X$, $\mathrm{G}_E$ as parameters. In addition, we point out the relations of $\mathrm{G}_X$, $\mathrm{G}_E$ to the MR model parameters $\mathrm{C}_1$, $\mathrm{C}_2$. The expressions below for the reduced normal tension $\widetilde{\sigma}_N$ correspond to the uniaxial deformation.

In the Warner-Edwards tube model, it is assumed that the tube moves affinely with the imposed deformation, while its diameter is strain independent.~\cite{DeamEdwards1976ThrRubElast, WarnerEdwards1978NeutronScatter} The resulting stress-strain relation is the same as in the phantom network model, hence, the Warner-Edwards tube model predicts that both the cross-link and the entanglement moduli $\mathrm{G}_X$, $\mathrm{G}_E$ contribute only to $\mathrm{C}_1$ MR model parameter, whereas $\mathrm{C}_2$ remains equal to zero.~\cite{MergellEveraers2001TubeModels}

The non-affine tube model of Rubinstein and Panyukov~\cite{RubinPanyuk1997NFFDfrmElstctPlmNtwk} represents the confinement potential due to entanglements as an additional set of harmonic virtual chains, acting along the real network chains. These virtual chains connect the strand ends to the nonfluctuating elastic background. The spring constant of the virtual chains, which determines the strength of the confinement potential, is supposed to be deformation dependent in such the way that the fluctuations of the virtual strands change affinely with the network deformation. The main result of the model is that the diameter of the confining tube deformes non-affinely as $d_{\mu} \sim \lambda_{\mu}^{1/2},\,\mu = x,\,y,\,z$, where $d_{\mu}$ is the tube diameter along the coordinate axis $\mu$. For the reduced normal tension, the non-affine tube model predicts:
\begin{equation}
\label{eqn:RNT_Micro2MR_NffT}
  \widetilde{\sigma}_N(\lambda) = \mathrm{G}_X + \dfrac{\mathrm{G}_E}{\lambda-\lambda^{1/2}+1}\,,
\end{equation}
where $\mathrm{G}_X$ is related to the cross-links modulus and $\mathrm{G}_E$ is related to the entanglement modulus. Expanding Eqs.~(\ref{eqn:RNT_Micro2MR_NffT}) and (~\ref{eqn:MR_rnt}) at low strains $\varepsilon$, where $\varepsilon=\lambda-1$, one obtains the following relations between the MR model parameters $\mathrm{C}_1$, $\mathrm{C}_2$ and microscopic moduli $\mathrm{G}_X$, $\mathrm{G}_E$:
\begin{equation}
\label{eqn:Micro2MR_NffT}
  2\,\mathrm{C}_1 = \mathrm{G}_X + 0.5\,\mathrm{G}_E\;,\;
  2\,\mathrm{C}_2 = 0.5\,\mathrm{G}_E\,.
\end{equation}

The slip tube model~\cite{RubinPanyuk2002ElastPolymNetw} refines the non-affine tube model~\cite{RubinPanyuk1997NFFDfrmElstctPlmNtwk} of Rubinstein and Panyukov. The attachment junctions between the virtual chains and the network strands are replaced by the slip links, which are allowed to slide along the network chains, but not to pass through each other. Consequently, the chain can redistribute its contour length along the tube upon the network deformation. Within the scope of the slip tube model, the reduced normal tension is given by:
\begin{equation}
\label{eqn:RNT_Micro2MR_SlT}
  \widetilde{\sigma}_N(\lambda) = \mathrm{G}_X + \dfrac{\mathrm{G}_E}{0.74\,\lambda+0.61\,\lambda^{-1/2}-0.35}\,,
\end{equation}
and the MR model parameters are identified as:
\begin{equation}
\label{eqn:Micro2MR_SlT}
  2\,\mathrm{C}_1 = \mathrm{G}_X + 0.565\,\mathrm{G}_E\;,\;
  2\,\mathrm{C}_2 = 0.435\,\mathrm{G}_E\,.
\end{equation}

The double tube model~\cite{MergellEveraers2001TubeModels} postulates that the ends of the confinement springs of the confinement potentials deform affinely, while the localization strength is assumed to be sub-affine for the entanglement confinement springs and strain-independent for the cross-link confinement springs. The reduced normal tension within the scope of the double tube model can be expressed as:
\begin{equation}
\label{eqn:RNT_Micro2MR_DT}
  \widetilde{\sigma}_N(\lambda) = \dfrac{\lambda^2-1}{\lambda^2-\lambda^{-1}}\,
                                  \dfrac{\mathrm{G}_X^2+2\,\left(\dfrac{\mathrm{G}_E}{\lambda}\right)^2}
                                        {\sqrt{\mathrm{G}_X^2+4\,\left(\dfrac{\mathrm{G}_E}{\lambda}\right)^2}} +
                                  \dfrac{1-\lambda^{-1}}{\lambda^2-\lambda^{-1}}\,
                                  \dfrac{\mathrm{G}_X^2+2\,\mathrm{G}_E^2\,\lambda}
                                        {\sqrt{\mathrm{G}_X^2+4\,\mathrm{G}_E^2\,\lambda}}\,,
\end{equation}
and the MR model parameters are related to $\mathrm{G}_X$, $\mathrm{G}_E$ as:~\cite{MergellEveraers2001TubeModels}
\begin{equation}
\label{eqn:Micro2MR_DT}
  2\,\mathrm{C}_1 = \dfrac{\mathrm{G}_X^4 + 6\,\mathrm{G}_X^2\,\mathrm{G}_E^2 + 4\,\mathrm{G}_E^4}{\left(\mathrm{G}_X^2 + 4\,\mathrm{G}_E^2\right)^{3/2}}\;,\;
  2\,\mathrm{C}_2 = \dfrac{4\,\mathrm{G}_E^4}{\left(\mathrm{G}_X^2 + 4\,\mathrm{G}_E^2\right)^{3/2}}\,,
\end{equation}
and we note that both the cross-link and the entanglement moduli $\mathrm{G}_X$, $\mathrm{G}_E$ contribute to both MR model parameters $\mathrm{C}_1$, $\mathrm{C}_2$ and that the network modulus is not just a simple sum $\mathrm{G}_X+\mathrm{G}_E$.

The models presented above are all based on Gaussian distribution for chain configurations:
\begin{equation}
\label{eqn:GaussDstr}
  P(N_K,\,\mathrm{R}) \propto \exp\left(-\dfrac{3\,\mathrm{R}^2}{2\,N_K\,l_K^2}\right)\,,
\end{equation}
where $\mathrm{R}$ is the absolute value of the chain end-to-end vector, $N_K$ is number of Kuhn segments in the chain, $l_K$ is the Kuhn length defined in Sect.~\ref{sbsct:Kuhn}. At large deformations, when a polymer molecule is stretched up to its contour length~($\mathrm{R} \sim N_K\,l_K$), this approximation is not reliable as it does not capture the finite extensibility of a polymer molecule. The models presented below are based on more realistic approximations, which take finite chain length effects into account.

The extended tube model developed by Kaliske and Heinrich~\cite{KaliskeHeinrich1999ExtendedTubeModel} is essentially a more detailed version of the non-affine tube model~\cite{RubinPanyuk1997NFFDfrmElstctPlmNtwk}. The authors modified the Gaussian distribution by introducing a singularity as proposed in Ref.~\cite{EdwardsViglis1986EffOfEntanglRubbElast}. The singularity is controlled by the finite extensibility parameter $\delta \equiv \left<\left(\partial\mathbf{R}(s)/\partial s\right)^2\right>$, where $\mathbf{R}$ is the end-to-end vector of the network chain, and $s$ is the contour coordinate. By construction, the choice $\delta=0$ neglects finite extensibility effects. The tube diameter is supposed to depend on strain as $d_{\mu} \sim \lambda_{\mu}^{\alpha\,\beta}\;,\;\mu = x\,,y\,,z$, where $\alpha = 1/2$ as in the non-affine tube model, $\beta \in (0;\,1)$ is taken as an empirical fit parameter, which is interpreted as indicating the completeness of the cross-linking reaction. The relaxation of the dangling ends is assumed to affect the effective strain dependency of the tube diameter making it more isotropic~($\beta \rightarrow 0$). The value $\beta \approx 1$ corresponds to a defect free network.~\cite{KaliskeHeinrich1999ExtendedTubeModel} The reduced normal tension was derived as:
\begin{equation}
\label{eqn:RNT_Micro2MR_ExT}
  \widetilde{\sigma}_N(\lambda) = \mathrm{G}_X + \mathrm{G}_E\,\varphi\left(\lambda\right)\,,
\end{equation}
where
\begin{gather*}
  \varphi\left(\lambda\right) = \dfrac{2}{\beta}\,\dfrac{\lambda^{\beta/2}-\lambda^{-\beta}}{\left(\lambda^2-\lambda^{-1}\right)\,g(D\,,\delta)}\,,\\
  g(D\,,\delta) = \dfrac{1-\delta^2}{\left(1-\delta^2\,\left(D-3\right)\right)^2} - \dfrac{\delta^2}{1-\delta^2\,\left(D-3\right)}\,,
\end{gather*}
and $D = \lambda^2+2\,\lambda^{-1}$. Consequently, the MR model parameters are found as:
\begin{equation}
\label{eqn:Micro2MR_ExT}
  2\,\mathrm{C}_1 = \mathrm{G}_X + \mathrm{G}_E\,\dfrac{2-\beta}{4-8\,\delta^2}\;,\;
  2\,\mathrm{C}_2 = \mathrm{G}_E\,\dfrac{2+\beta}{4-8\,\delta^2}\,.
\end{equation}

In particular case, when finite extensibility effects are neglected, $\delta=0$, and $\beta=0$, Eq.~(\ref{eqn:Micro2MR_ExT}) reduce to Eq.~(\ref{eqn:Micro2MR_NffT}). On the other hand, if $\beta=1$, then Eq.~(\ref{eqn:Micro2MR_ExT}) reduces to $2\,\mathrm{C}_1 = \mathrm{G}_X + 0.25\,\mathrm{G}_E$, $2\,\mathrm{C}_2 = 0.75\,\mathrm{G}_E$.

The non-affine network model of Davidson and Goulbourne~\cite{DavidsonGoulbourne2013NffNtwkModel} generalizes the phantom network model and combines it with the non-affine tube model of Rubinstein and Panyukov.~\cite{RubinPanyuk1997NFFDfrmElstctPlmNtwk} The authors repeated the derivation of Eq.~(\ref{eqn:Phnt_ntwrk_GX_prt}) as in Ref.~\cite{RubinPanyuk2002ElastPolymNetw}, but for statistics of the effective strands they used the exact solution of the freely jointed chain model derived in Ref.~\cite{KuhnGrun1942}:
\begin{equation}
\label{eqn:LangDstr}
  P(N,\,\mathrm{R}) \propto \exp\left\{-\dfrac{\mathrm{R}}{l_K}\,\mathcal{L}^{-1}\left(\dfrac{\mathrm{R}}{N_K\,l_K}\right) - N_K\,\ln\left[\dfrac{\mathcal{L}^{-1}\left(\dfrac{\mathrm{R}}{N_K\,l_K}\right)}{\sinh\left(\mathcal{L}^{-1}\left(\dfrac{\mathrm{R}}{N_K\,l_K}\right)\right)}\right]\right\}\,,
\end{equation}
and used a Pad{\'e} approximant~\cite{Cohen1991} for the inverse Langevin function $\mathcal{L}^{-1}\left(x\right) = \coth(x)-1/x$. For the reduced normal tension, they obtained:
\begin{equation}
\label{eqn:RNT_Micro2MR_DG}
  \widetilde{\sigma}_N(\lambda) = \mathrm{G}_X\,\dfrac{\lambda^2+2\,\lambda^{-1}-9\,\lambda_{max}^2}{3\,\left(\lambda^2+2\,\lambda^{-1}-3\,\lambda_{max}^2\right)}
                                + \mathrm{G}_E\,\dfrac{\lambda-\lambda^{-1}-\lambda^{-1/2}+\lambda^{1/2}}{\lambda^2-\lambda^{-1}}\,,
\end{equation}
where $\lambda_{max}$ is the maximum stretch. The MR model parameters are expressed as:
\begin{equation}
\label{eqn:Micro2MR_DG}
  2\,\mathrm{C}_1 = \mathrm{G}_X\,\dfrac{1-3\,\lambda_{max}^2}{3\,\left(1-\lambda_{max}^2\right)} + 0.5\,\mathrm{G}_E\;,\;
  2\,\mathrm{C}_2 = 0.5\,\mathrm{G}_E\,.
\end{equation}

When finite extensibility of the network chains is not taken into account, Eq.~(\ref{eqn:Micro2MR_DG}) reduces to the non-affine tube model Eq.~(\ref{eqn:Micro2MR_NffT}).

Xiang et al.~\cite{XiangZhongWangMaoYuQu2018GnrlModelSoftElast} combined a generalized version of the affine network model Eq.~(\ref{eqn:FF_ntwrk_GX}) with the tube deformation hypotheses of Heinrich and Straube~\cite{HeinrichStraube1984} and the three-chain model~\cite{JamesGuth1943ThrElstPropRub} to derive their generalized constitutive model. Generalization of the affine network model is based on the freely jointed chain distribution~(Eq. \ref{eqn:LangDstr}) with an approximation for the inverse Langevin function due to Kr{\"o}ger.~\cite{Kroeger2015}. Within the scope of Xiang et al. model, the reduced normal tension at uniaxial deformation has the following form:
\begin{equation}
\label{eqn:RNT_Micro2MR_XI}
  \widetilde{\sigma}_N(\lambda) = \mathrm{G}_X\,\left[\left(1-\dfrac{\mathrm{I}_1}{3\,N_{XK}}\right)\,\left(1+\dfrac{1}{2}\,\dfrac{\mathrm{I}_1}{3\,N_{XK}}\right)\right]^{-1}
                                + 2\,\mathrm{G}_E\,\dfrac{\lambda^{1/2}-\lambda^{-1}}{\lambda^2-\lambda^{-1}}\,,
\end{equation}
where $\mathrm{I}_1 = \lambda^2+2\,\lambda^{-1}$ is the $1^{st}$ invariant of the Finger deformation tensor, $N_{XK}$ is the network strand length, and the MR coefficients are derived as:
\begin{equation}
\label{eqn:Micro2MR_XI}
  2\,\mathrm{C}_1 = \mathrm{G}_X\,\dfrac{2\,N_{XK}^2}{2\,N_{XK}^2-N_{XK}-1} + \dfrac{1}{4}\,\mathrm{G}_E\;,\;
  2\,\mathrm{C}_2 = \dfrac{3}{4}\,\mathrm{G}_E\,.
\end{equation}

Here, we redefined the entanglement modulus $\mathrm{G}_E$ by including a prefactor of $2$ compared to the original paper, so that the sum $2\,\mathrm{C}_1+2\,\mathrm{C}_2$ gives the full contribution from entanglement effects $\mathrm{G}_X + \mathrm{G}_E$. When finite extensibility of the network chains is not taken into account, the expression for the reduced normal tension~(\ref{eqn:RNT_Micro2MR_XI}) reduces to the extended tube model Eq.~(\ref{eqn:RNT_Micro2MR_ExT}) with $\beta=1$, $\delta=0$.

Hence, to summarize, the classical affine and phantom models, which neglect entanglements, correspond to the MR model with $2\,\mathrm{C}_1=\mathrm{G}_X$, $2\,\mathrm{C}_2=0$. This also applies to the Warner-Edwards tube model, which has a strain independent tube diameter. The tube models with strain dependent tube diameters put the cross-link modulus entirely into the $\mathrm{C}_1$ parameter and split the entanglement effects between $\mathrm{C}_1$ and $\mathrm{C}_2$. Among all the models, only the slip tube model accounts for the chain contour length redistribution between parallel and perpendicular tube sections upon deformation. The non-affine tube model, the slip tube model, and the double tube model assume Gaussian chain statistics and neglect finite extensibility effects. The extended tube model of Kaliske and Heinrich, the non-affine network model of Davidson and Goulbourne, and the general constitutive model of Xiang et al. take finite extensibility into account using various approaches to describe non-Gaussian chain statistics at the expense of introduction additional model parameters. It is worth noting that the estimates for the shear modulus provided by extended tube model, the non-affine network model and the general constitutive model depend on their finite extensibility parameters. All models except the double tube assume additivity of network connectiviy and the entanglement effects.

%% file: 3.0.Methods.tex
\section{Methods}
\label{sct:Methods}

In this section, we briefly introduce the Kremer-Grest~(KG) polymer model~[Sect.~\ref{sbsct:KG}] and mapping relations between the KG model, the Kuhn representation and PDMS~[Sect.~\ref{sbsct:Mapping}], describe our procedure for generating model polymer networks from an equilibrated precursor melt and characterize their microscopic structure~[Sect.~\ref{sbsct:Characterization}], present our protocol for the model networks deformation and describe the results postprocessing~[Sect.~\ref{sbsct:SetUp}]. In Sect.~\ref{sbsct:PPA}, Sect.~\ref{sbsct:3PA} we introduce the Primitive Path Analysis~(PPA) and the Phantom Primitive Path Analysis~(3PA), respectively.

%% file: 3.1.KG.tex
\subsection{Kremer-Grest model}
\label{sbsct:KG}

To study structure-property relations for PDMS rubbers, we utilize the KG model in combination with an angular potential. Within the scope of the generic Kremer-Grest model~(KG)~\cite{GrestKremer1986MDSimPolymHeatBath, KremerGrest1990DynEntangLinPolyMelt} polymers are represented as linear bead-spring chains. Pair interactions between beads are described by the Weeks-Chandler-Anderson~(WCA) potential:
\begin{equation}
\label{eqn:WCA}
  U_{WCA}(r) = 4\,\epsilon\,\left[\left(\dfrac{\sigma}{r}\right)^{12} - \left(\dfrac{\sigma}{r}\right)^{6} + \dfrac{1}{4}\right]\;,\,r<\sigma\,\sqrt[6]{2}\,,
\end{equation}
and the interactions of bonded beads are described by the FENE potential:
\begin{equation}
\label{eqn:FENE}
  U_{FENE}(r) = -\dfrac{1}{2}\,k\,R_0\,\ln\left[ 1 - \left(\dfrac{r}{R_0}\right)^2 \right]\,,
\end{equation}
where $\epsilon$ is energy scale, $\sigma$ is chosen as the simulation unit of length. The standard choice for the spring constant is $k=30\,\epsilon\sigma^{-2}$, $R_0$ is set to $1.5\,\sigma$, which leads to average bond length $l_b=0.965\,\sigma$. The simulation unit of time is defined as $\tau = \sqrt{m\,\sigma^2/\epsilon}$, where $m$ denotes the mass of a bead which we choose as our simulation mass scale.

The KG model is a generic polymer model. To adapt it to specific chemical polymer species~\cite{EveraersKarimiFleckHojdisSvaneborg2020KGMap, FallerKolbMullerPlathe1999LocalChainOrder}, we introduce an additional bending interaction:
\begin{equation}
\label{eqn:Bend}
  U_{bend}(\theta) = \kappa\varepsilon\,\left( 1 - \cos\theta \right)\,,
\end{equation}
where $\kappa$ is the stiffness constant. The angular potential~(\ref{eqn:Bend}) was introduced in Ref.~\cite{FallerKolbMullerPlathe1999LocalChainOrder}. For a discussion of the choice of the stiffness constant, see Sect.~\ref{sbsct:Mapping}.

The dynamics of the KG model is governed by the Langevin equation:
\begin{equation}
\label{eqn:LangevinDyn}
  m\,\dfrac{\partial\mathbf{R}_n}{\partial t^2} = -\nabla_{\mathbf{R}_n}U - \Gamma\,\dfrac{\partial\mathbf{R}_n}{\partial t} + \bm{\xi}_n\,,
\end{equation}
where $\mathbf{R}_n$ is the position on $n$-th bead, $U$ is interaction potential. The bead friction $\Gamma$ is set to $0.5\,m\tau^{-1}$. The stochastic force terms $\bm{\xi}_n$ obey statstics $\left<\bm{\xi}_n\right>=0$ and $\left<\bm{\xi}_m(t)\bm{\otimes}\bm{\xi}_n(t')\right> = 6\,k_B\,T\,\Gamma/\Delta t\,\delta_{mn}\,\delta(t-t')\,\mathbf{I}$, where $\mathbf{I}$ is the unit tensor and $\Delta t$ is the time step. The Langevin equation was integrated with time step $0.01\,\tau$ using the GJ-F Langevin integrator~\cite{GronJensFarago2013GJF}, which is a Verlet-type algorithm~\cite{Verlet1967, PressTeukolskyVetterlingFlannery1992NumRecInC}, implemented in LAMMPS.~\cite{Plimpton1995LAMMPS}

%% file: 3.2.Mapping.tex
\subsection{Mapping of units for PDMS-KG}
\label{sbsct:Mapping}

By its nature, the KG model does not correspond to any specific of chemical polymer. The introduction of the bending potential Eq.~(\ref{eqn:Bend}) allows one to tune the Kuhn number $n_K(\kappa)$ of a KG model to match the Kuhn number of any desired polymer.~\cite{EveraersKarimiFleckHojdisSvaneborg2020KGMap, Fetters2006KGMap} For instance, to match PDMS, which has $n_K=2.82$, one has to choose the bending constant $\kappa=0.013$,~\cite{EveraersKarimiFleckHojdisSvaneborg2020KGMap} which is nearly identical to the standard KG model. This particular KG model we denote PDMS-KG model. It has a Kuhn length of $l_K=1.853\,\sigma$,~\cite{SvaneborgkEveraers2020KGModel} which is identified with the experimental value $l_K=11.42$~\AA.~\cite{FettersLohseColby2007ChainDimEntaSpac} Hence, $1\,\sigma=6.17$~\AA. The PDMS-KG model also has $C_{\infty}=1.921$ beads per Kuhn segment, and the mass of a Kuhn segment $M_K=C_{\infty}\,m_b$ expressed in KG units is identified with the molar mass $M_K=309.28$~g/mol of a PDMS Kuhn segment. Consequently, $m_b=M_K/C_{\infty}=160.99$~g/mol. The entanglement time of PDMS-KG model $\tau_E=8\,500\,\tau$~\cite{SvaneborgkEveraers2020KGModel} is identified with the experimental entanglement time of PDMS $\tau_E = 0.11\,\mu s$.~\cite{ChavezSaalwachter2011TimeDomainNMRObservZPolymDyn} Therefore, $1\,\tau=0.013$~ns. Finally, the energy scale $\varepsilon$ of the KG model is identifed with $k_B\,T$, where $T$ denotes the temperature at which the experimental reference system is characterized. Therefore, the energy density conversion for PDMS-KG is obtained as $1\,\varepsilon\,\sigma^{-3} = 17.6\,\mathrm{MPa}$,~\cite{EveraersKarimiFleckHojdisSvaneborg2020KGMap} which we use to convert PDMS-KG stress data to SI units. Following these relations, results expressed in simulation units of the KG model with PDMS stiffness can be converted into PDMS specific SI or Kuhn units.~\cite{EveraersKarimiFleckHojdisSvaneborg2020KGMap} For KG models of other polymers and their mapping relations, see Refs.~\cite{EveraersKarimiFleckHojdisSvaneborg2020KGMap, Fetters2006KGMap}.

Note, that it is not necessary to match independently, for instance, the entanglement length of the KG model with the entanglement length of PDMS, as these structural properties emerge automatically as a result of the proper choice of the Kuhn number due to universality.~\cite{EveraersKarimiFleckHojdisSvaneborg2020KGMap} According to the packing argument,~\cite{Lin1987, KavassalisNoolandi1987} $N_{EK}=\alpha^2/n_K^2$, where $\alpha=19.0 \pm 2$~\cite{RosaEveraers2014RingPolymInMelt, FettersLohseRichterWittenZirkel1994Connection} is the number of entanglement strands in an entanglement volume. Moreover, in dimensionless Kuhn form, one can write $\mathrm{G}_E\,l_K^3 / \left(k_B\,T\right)=n_K/N_{EK}=n_K^3/\alpha^2$. Hence, our PDMS-KG model with Kuhn number $n_K=2.82$ automatically reproduces the emergent entanglement related properties of PDMS melts. The model has an entanglement strand length $N_{EK} = 33.1$, which is in excellent agreement with the experimental value of $31.3$,~\cite{EveraersKarimiFleckHojdisSvaneborg2020KGMap}, hence, the model is expected to accurately reproduce the entanglement modulus of PDMS melts.

%% file: 3.3.Characterization.tex
\subsection{Generation and characterization of model networks}
\label{sbsct:Characterization}

\begin{figure}[!h]
  \begin{subfigure}{8.25cm}
    \includegraphics[width=\textwidth]{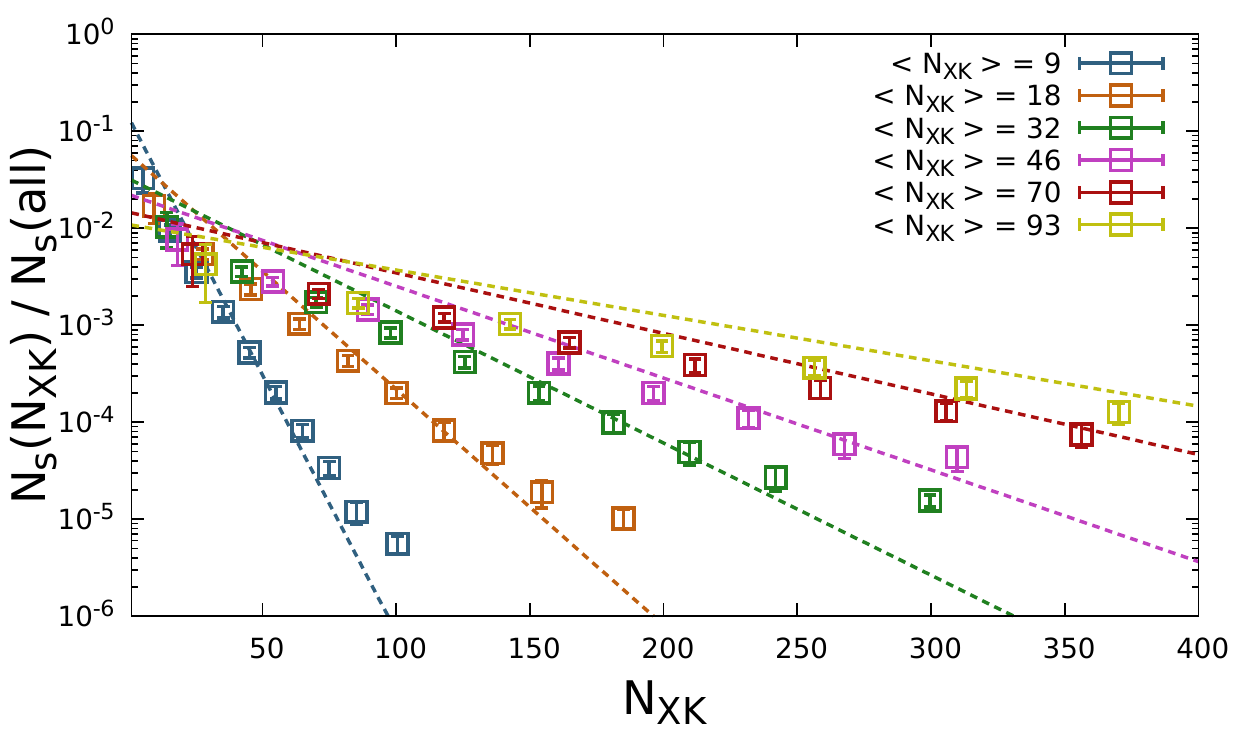}
    \caption{}
    \label{sbfgr:good_strands}
  \end{subfigure}
  \begin{subfigure}{8.25cm}
    \includegraphics[width=\textwidth]{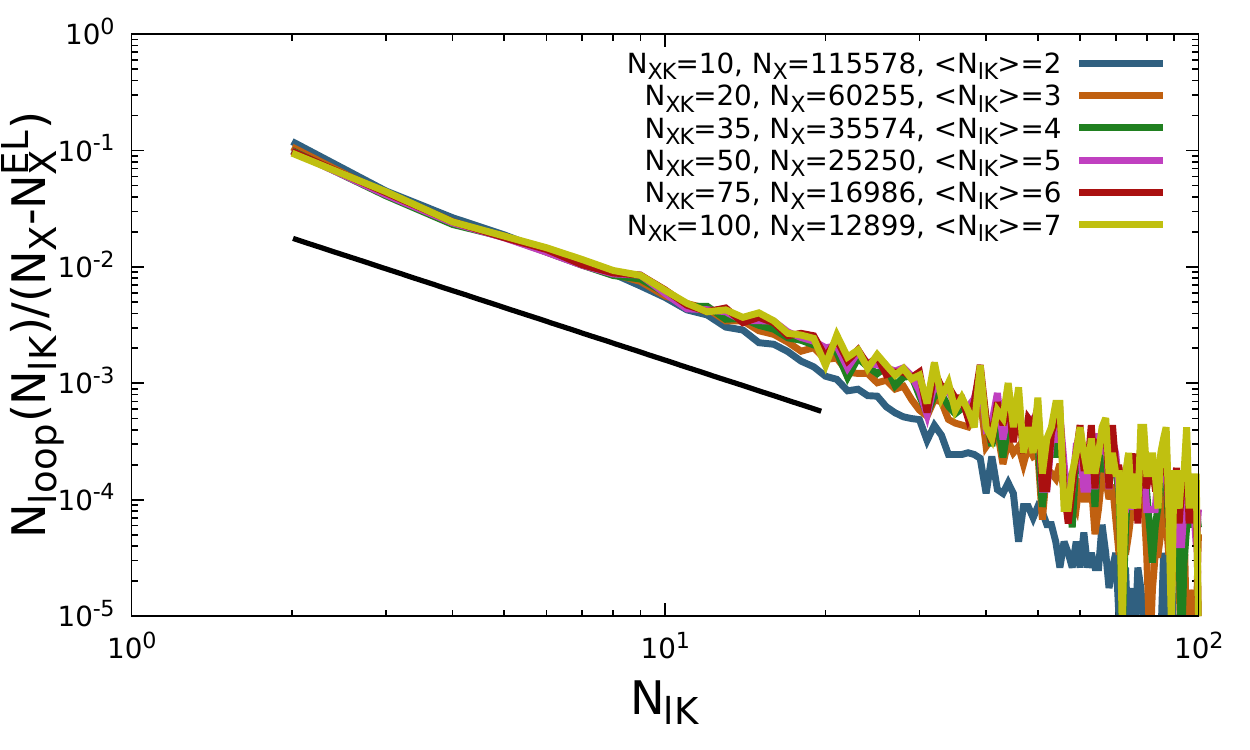}
    \caption{}
    \label{sbfgr:loops}
  \end{subfigure}
  \begin{subfigure}{8.25cm}
    \includegraphics[width=\textwidth]{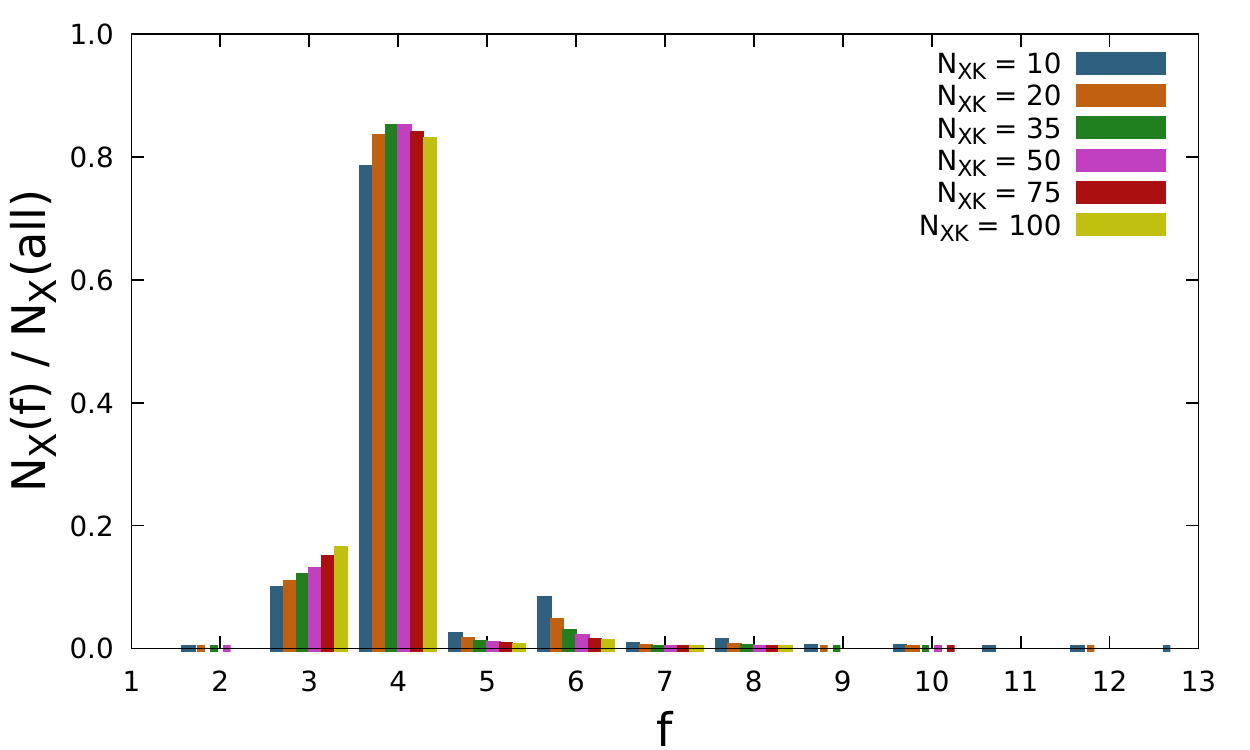}
    \caption{}
    \label{sbfgr:Xs}
  \end{subfigure}
  \caption{Microscopic characterization of model networks. \subref{sbfgr:good_strands}: Distribution of network strand lengths. Dashed lines correspond to empirical relation~(\ref{eqn:ProbabilityNXK}) written for each model network. \subref{sbfgr:loops}: Distribution of loops normalized by the number of cross-links in the model network. The black line indicate the power law $x^{-3/2}$. \subref{sbfgr:Xs}: Distribution of cross-link functionalities.}
  \label{fgr:characterization}
\end{figure}

To build model polymer networks, we start from the melt generation procedure of Svaneborg et al.~\cite{SvaneborgKarimiHojdisFleckEveraers2016MultiscaleApproach} The approach allows one to generate large equilibrated polymer melts in a computationally effective way. Briefly, a polymer melt is generated as a random walk on a cubic lattice. As lattice parameter we choose the tube diameter of the polymer. Multiple chains ($\alpha=19$) are allowed to occupy the same site in the lattice. A Monte-Carlo simulated annealing algorithm is used to minimize large scale density fluctuations. This results in equilibration of melt configuration above the tube scale. Next, the lattice melt conformation is transformed into a bead-spring conformation, and we simulate a Rouse dynamics with a force-capped WCA potential and a stiffness tuned to reproduce the target Kuhn length. This introduces the desired random walk chain structure below the tube scale, while preventing the growth of density fluctuations. Finally, the local bead packing structure is equilibrated via a short simulation with the full KG force field. Following the procedure, we generated an equilibrated KG polymer melt model, comprising of $500$ chains with $10\,000$ beads each, corresponding to about $Z=85$ entanglements per chain and systems with $5$ million beads. For comparison, recently in Ref.~\cite{HsuKremer2019ClusterZPoints}, Hsu et al. generated systems with a size of $2$ million beads.

Earlier, in Sect.~\ref{sbsct:Mapping} we presented the PDMS-KG model and its mapping relations. The present results are generated for KG melts with a slightly larger bending stiffness $\kappa=0.206$~($n_K=3.0$). We started these simulations based on a preliminary version of the analysis in Ref.~\cite{EveraersKarimiFleckHojdisSvaneborg2020KGMap} and continued to use the same stiffness for the sake of consistency. The corresponding error is estimated to be $\approx 8$\% for the Kuhn number, $\approx 4$\% for the Kuhn length, $\approx 15$\% for the entanglement length, and $\approx 12$\% for the energy density. These errors are comparable to the experimental error of the Kuhn number, on which the mapping relations are based. Hence, we continue to use the PDMS-KG mapping relations presented in Sect.~\ref{sbsct:Mapping} for the present results.

Cross-linking was initiated by first connecting chain ends to the closest neighbouring beads in the precursor melt to avoid dangling ends. Afterwards, random bead pairs within a distance $1.3\sigma$ were linked by a FENE bond if they were not already connected. Following the procedure, we generated a set of the PDMS-KG model networks, ranging from weakly to strongly cross-linked, characterized by average network strand length values $\left<N_{XK}\right>$ equal to $100$, $75$, $50$, $35$, $20$, $10$ and number of additional FENE bonds from $1\,000$ to $130\,000$. For the end-linked network, we expect entanglement effects to dominate over the network connectivity as there are $\left<N_{XK}\right>/N_{EK}=56$ entanglements per network strand, where for the present model the entanglement strand length $N_{EK} \approx 33.1$.~\cite{SvaneborgkEveraers2020KGModel}. For the most strongly cross-linked system, $N_{EK} / \left<N_{XK}\right>=3.7$, hence, we have approximately $4$ cross-links per entanglement strand. The value $\left<N_{XK}\right>=35 \approx N_{EK}$ is of special interest as contributions from the network connectivity and the entanglement effect to the shear modulus are expected to be comparable. We repeated the cross-linking process several times from different random initial seeds and verified that the systems were large enough to be self-averaging, hence, we expect our results to be reproducible and report results only for one of the networks.

To analyse the network structure, we coarse-grain the beads and bonds into strands and cross-links. We start by identifying beads based on their functionality: an end bead has a single bond, an internal bead has two bonds, and a cross-link bead has $3$ or more bonds. Then, we identified network strands as connected chains of internal beads, and network cross-links as clusters of interconnected cross-link beads. The functionality of a cross-link is the number of strands that emanate from it.

Afterwards, we proceeded to analyze the network structure in terms of good or defect strands and cross-links. Defect cross-links have $1$ or less good strands connected to them, since two good strands are required for a cross-link to carry a load. A strand is a defect if it is a loop~(starts and terminates at the same cross-link) or if at least one strand end is an end bead or a defect cross-link, since such strands can not carry a load either. We apply this algorithm repeatedly until no new defects were identified.

In Fig.~\ref{fgr:characterization}, we present a characterization of the generated networks. The plot in Fig.~\ref{sbfgr:good_strands} indicates the exponential distribution of network strand lengths $N_{XK}$ expressed as the number of Kuhn units between cross-links with pre-defined average $\left<N_{XK}\right>$:
\begin{equation}
\label{eqn:ProbabilityNXK}
  P(N_{XK}) = \frac{1}{\left<N_{XK}\right>}\,\exp\left(-\frac{N_{XK}}{\left<N_{XK}\right>}\right)\,.
\end{equation}

Relative number of strands $N_s(N_{XK}) / N_s$, having specific length $N_{XK}$ in Kuhn units, where $N_s$ is the total number of strands, is shown as a function of $N_{XK}$. Average values of network strand lenghts $\left<N_{XK}\right>$, shown in the legend, are calculated based on the network analysis. The value $\left<N_{XK}\right> \approx 1850$, obtained for the end-linked melt, differs from the chain length $10,000$ beads. This is due to the effect that chain ends are connected to a random bead in their neighbourhood, which is most likely an internal bead in another chain.

The plot in Fig.~\ref{sbfgr:loops} shows the histogram of the loop sizes. Note that loops here refer to those strands, which start and terminate at the same cross-link. Intramolecular cross-links can pinch off a section of a chain, thus forming a loop.~\cite{LangGoritzKreitmeier2005IntraReactions} The loop distributions were normalized by the number of cross-links added beyond the percolation threshold. We observe an excellent collapse of the distributions from different networks. The distributions are characterized by the $-3/2$ exponent expected from random walk theory.~\cite{JacobsonStockmayer1950} The total number of loops is about $13-17$\% of the number of strands, which is consistent with Refs.~\cite{SvaneborgEveraersGrestCurro2008StressContributions, LangGoritzKreitmeier2005IntraReactions}. Since most of the loops are much smaller than the entanglement length, they do not thread other strands and hence do not contribute additional entanglements.

The plot in Fig.~\ref{sbfgr:Xs} shows the distribution of cross-link functionalities. We observe that the vast majority of cross-links are $4$-functional, we also observe cross-links with higher functionality. These are created, when adjacent beads along a chain are chosen for creating random bonds with neighboring chains, thus effectively forming a cross-link cluster with higher functionality. For the end-linked model network, most of the initial cross-links are three functional since in this case most ends are bonded to an internal bead in a neighboring chain. We emphasize that the model network statistics is observed to be the same for multiple runs of the cross-linking procedure, which is due to the fact that our system sizes are large enough to be effectively self-averaging.

%% file: 3.4.SetUp.tex
\subsection{Network deformation}
\label{sbsct:SetUp}

\begin{figure}[!h]
  \begin{subfigure}{8.25cm}
    \includegraphics[width=\textwidth]{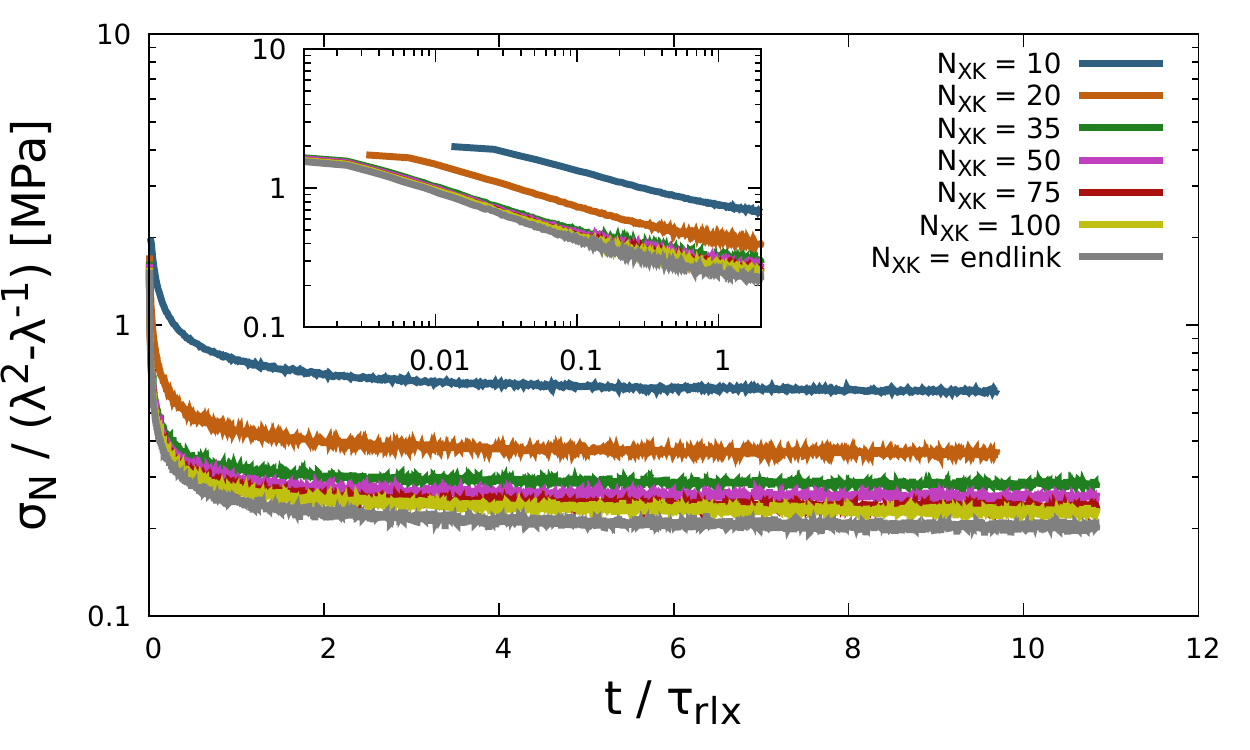}
    \caption{}
    \label{sbfgr:rlx_1D_str_rnt_vs_time}
  \end{subfigure}
  \begin{subfigure}{8.25cm}
    \includegraphics[width=\textwidth]{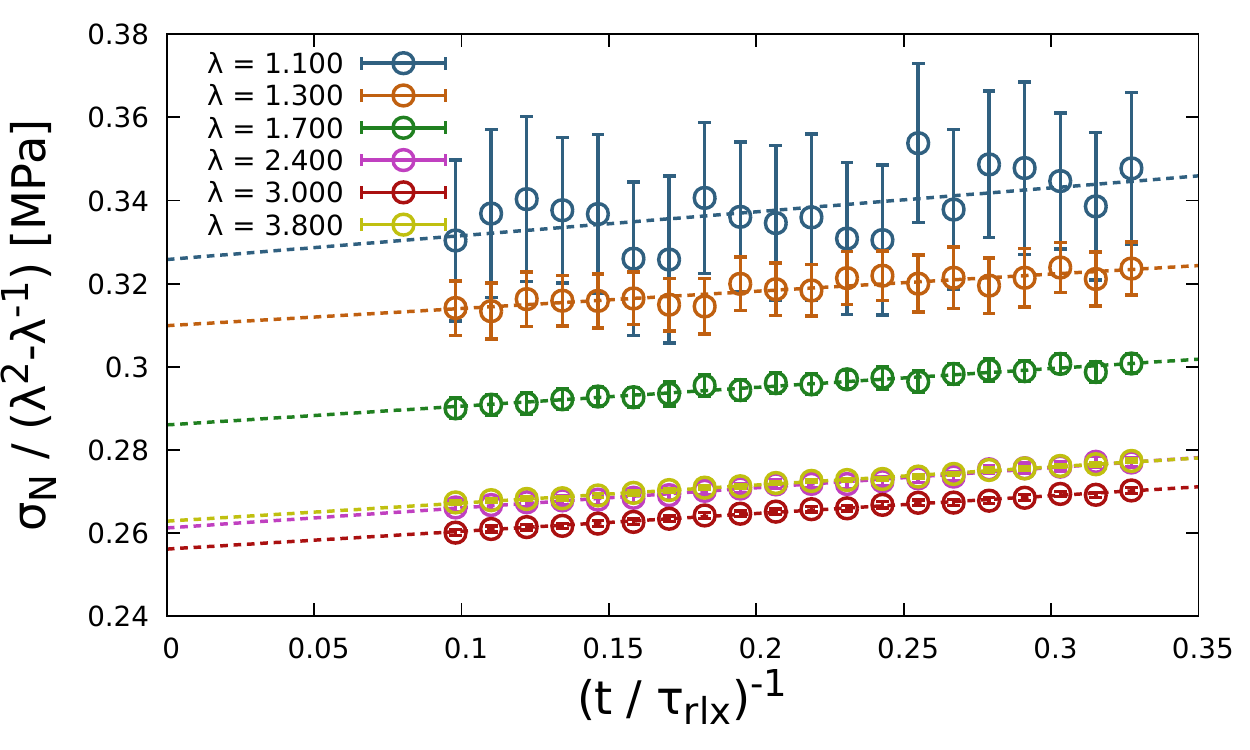}
    \caption{}
    \label{sbfgr:rlx_1D_str_rnt_vs_inv_time_Nk35}
  \end{subfigure}
  \caption{\subref{sbfgr:rlx_1D_str_rnt_vs_time}: Relaxation of the reduced stress of the networks at $\lambda=1.8$. \textbf{Inset}~shows first $2\,\tau _{rlx}$ of the relaxation. \subref{sbfgr:rlx_1D_str_rnt_vs_inv_time_Nk35}: Relaxation of the network model $N_{XK}=35$ at different deformation states plotted vs. the inverse time. The data are binned, and the bin averages and fluctuations are indicated.}
  \label{fgr:rlx_1D_str_rnt}
\end{figure}

Deformation simulations of the networks were carried out in two stages. At first, model networks were stretched uniaxially, preserving volume. Subsequently, we simulated the networks at constant strain to relax the stress. During the deformation, the~(engineering) strain rate $\dot{\varepsilon}$ was chosen large enough to save computer time and minimize relaxation effects and, secondly, small enough to avoid breaking bonds. We used $\dot{\varepsilon}=0.01\tau^{-1}$~(in LJ units, see Sect.~\ref{sbsct:KG}), which corresponds to the inverse Rouse time for a network strand of $\approx 4$ Kuhn segments~($\approx 8$ beads), occupying a spatial distance $\approx 4\,\sigma$. Hence, only short strand segments are in quasi-equilibrium upon deformation, whereas for long strand segments affine deformation response is expected. Simulations were terminated, if a bond reached length of $1.4\sigma$. For KG melts and networks in the unstrained state the topological state is preserved since there is a potential barrier of $\sim 75\,k_B\,T$ for chains to move through each other.~\cite{SukuGrestKremerEveraers2005IdentPPMesh} As bonds are stretched as a result of the network deformation, this potential barrier is progressively reduced. We estimate that the potential barrier has dropped to $\sim 30\,k_B\,T$ for a bond with length $1.4\,\sigma$, see Ref.~\cite{Note_CriticalBondLength}. Continuing the simulations beyond this point leads to spurious results as we can not be sure that entanglements are preserved.

Relaxation in elastomers has two characteristic time scales. The Rouse time~$\tau _R \sim N_{XK}^2$ is the time it takes for a network strand of length $N_{XK}$ to relax. The entanglement time~$\tau _E \sim N_{EK}^2$ is the time scale, at which the dynamics of a bead starts being affected by entanglements. Due to the exponential strand length distribution, we expect a wide spectrum of relaxation times. We supposed that the relaxation dynamics of highly cross-linked networks, $N_{XK} < N_{EK}$, was dominated by the Rouse time of short network strands, whereas the dynamics of loosely cross-linked networks, $N_{XK} > N_{EK}$, is dominated by the entanglement time. Hence, the characteristic relaxation time $\tau _{rlx}$ of a model network was estimated as the minimum between the entanglement time $\tau _E$ and the Rouse time $\tau _R$ for average network strand length. Stress relaxation was sampled for $10$--$11\,\tau _{rlx}$.

Fig.~\ref{sbfgr:rlx_1D_str_rnt_vs_time} shows the relaxation of the reduced stresses $\widetilde{\sigma}_N$ of the model networks plotted versus simulation time $t$, which is scaled by the relaxation time $\tau _{rlx}$. As clearly seen, most stress relaxation occurs for $t<\tau _{rlx}$~[see inset in Fig.~\ref{sbfgr:rlx_1D_str_rnt_vs_time}] and the subsequent dynamics is very slow. In addition, the relaxation curves in logarithmic representation are parallel and approximately linear over more than a decade. This suggests an initial powerlaw-like decay as expected from the Rouse model. Moreover, relaxation curves corresponding to the networks with $N_{XK} \ge 35$ collapse~(the inset plot). This justifies our assumption of the network dependence of the characteristic relaxation times. As expected the shear moduli are ordered by the degree of cross-linking.

Fig.~\ref{sbfgr:rlx_1D_str_rnt_vs_inv_time_Nk35} shows relaxation of the reduced normal tension of multiple deformed states of the model network $N_{XK} = 35$ as a function of the inverse time, starting from $t=3\,\tau _{rlx}$. In Fig.~\ref{sbfgr:rlx_1D_str_rnt_vs_time}, the stresses appear to be in equilibrium. However, Fig.~\ref{sbfgr:rlx_1D_str_rnt_vs_inv_time_Nk35} clearly shows that the stresses have not reached equilibrium for any deformation, and slow relaxation process is observed even after $10\,\tau _{rlx}$. To estimate the equilibrium reduced normal tension, we extrapolated the relaxation curves towards $t^{-1} \rightarrow 0$, i.e. to $t \rightarrow \infty$. As a fitting function a linear function $a + b\,t^{-1}$ was used. Due to the representation of the data as a function of the inverse time, the density of the data in the vicinity of $(t/\tau _{rlx})^{-1}=0$ is much higher than in the vicinity of $(t/\tau _{rlx})^{-1}=0.33$. Hence, the data was binned as a function of inverse time prior to fitting to correct for this. When fitting, the data within each bin was weighted by the bin variance. We observed that the relative error could be up to $7\%$ too high, if one uses the time averaged stresses for $t>3\,\tau _{rlx}$ as an estimate of the equilibrium stress rather than the present extrapolation.

%% file: 3.5.PPA.tex
\subsection{PPA: Primitive Path Analysis}
\label{sbsct:PPA}

\begin{figure}[!h]
  \begin{subfigure}{5.5cm}
    \includegraphics[width=\textwidth]{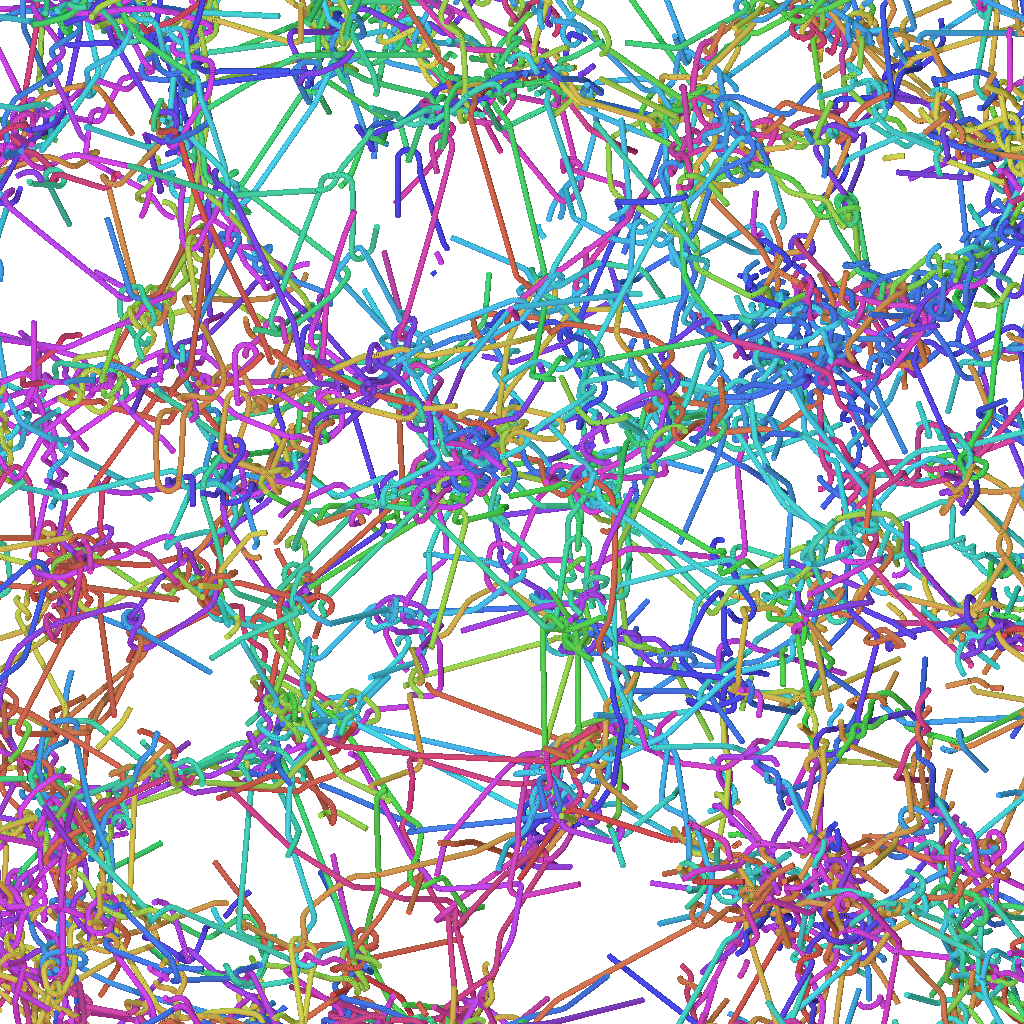}
    \caption{}
    \label{sbfgr:PPA_mesh}
  \end{subfigure}

  \begin{subfigure}{5.5cm}
    \includegraphics[width=\textwidth]{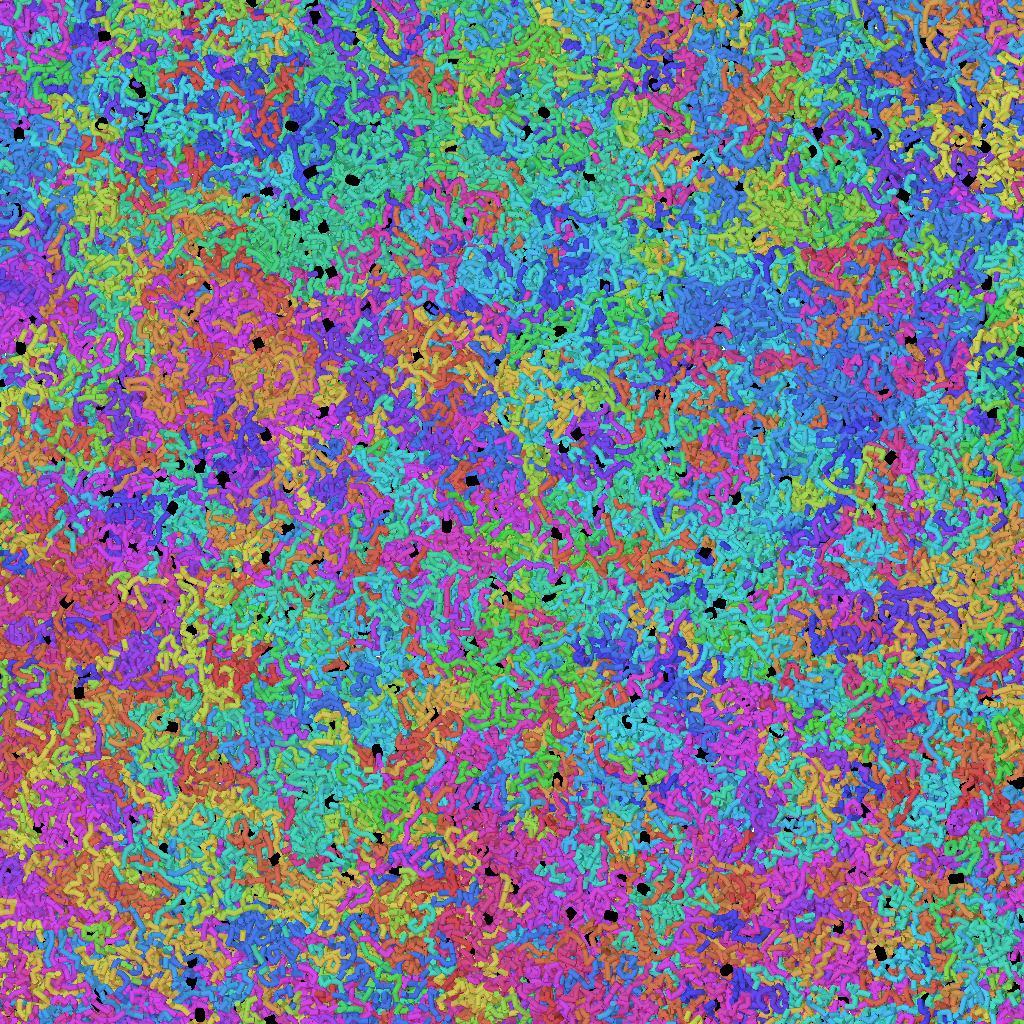}
    \caption{}
    \label{sbfgr:KG_network_NXK=35}
  \end{subfigure}
  \caption{Visualization of a thin slab of the precursor melt PPA mesh(~\subref{sbfgr:PPA_mesh}) and corresponding chain conformations of the model network $\left<N_{XK}\right>=35$~(\subref{sbfgr:KG_network_NXK=35}). Colors indicate precursor chains and in the network cross-links are represented by black bonds.}
  \label{fgr:PPA_mesh_visualization}
\end{figure}

Tube models for polymer viscoelasticity are based on the idea that thermal fluctuations of chains are localized by entanglement constraints with neighbouring chains.~\cite{DoiEdwards1986TheorPolymDyn} The central axis of the confinement tube is called primitive path. Tube models inspired the so called Primitive Path Analysis~(PPA).~\cite{EveraersSukuGrestSvaneborgSivaKremer2004RheolTopol} The analysis allows one to obtain the primitive path mesh, and in particular their average contour length $L_{pp}$. Based on $L_{pp}$, one can easily estimate the average entanglement strand length $N_{EK}$ as $\dfrac{L_{c}^2}{L_{pp}^2}$, where $L_{c}$ is the original chain contour length, and, consequently, obtain the entanglement modulus:
\begin{equation}
\label{eqn:GEPPA}
  \mathrm{G}_E = \dfrac{\rho_K\,k_B\,T}{N_{EK}}\,.
\end{equation}

Implementation of the PPA within the scope of the MD simulations is comprehensively described in Refs.~\cite{EveraersSukuGrestSvaneborgSivaKremer2004RheolTopol,SukuGrestKremerEveraers2005IdentPPMesh}~: chain ends are pinned in space, intramolecular pair interactions are switched off, intermolecular pair interactions are kept to prevent chains passing through each other, bonds are modelled as FENE springs with an arbitrary spring constant $k$, and melt is cooled down to $T=0$ to eliminate thermal fluctuations. Visualization of the precursor melt PPA mesh is shown in Fig.~\ref{sbfgr:PPA_mesh}, for comparison, we also show a network Fig.~\ref{sbfgr:KG_network_NXK=35} to illustrate the real chain structure.

We identify the result of the PPA analysis with the entanglement modulus and not the plateau modulus. The latter is reduced by $20\%$ compared to the former due to the entanglements which are lost due to chain contraction to the equilibrium contour length after the deformation of a melt.~\cite{DoiEdwards1986TheorPolymDyn} In a network no such contraction process occurs, thus making the entanglement modulus the relevant parameter.

%% file: 3.6.3PA.tex
\subsection{3PA: Phantom Primitive Path Analysis}
\label{sbsct:3PA}

\begin{figure}[!h]
  \begin{subfigure}{5.5cm}
    \includegraphics[width=\textwidth]{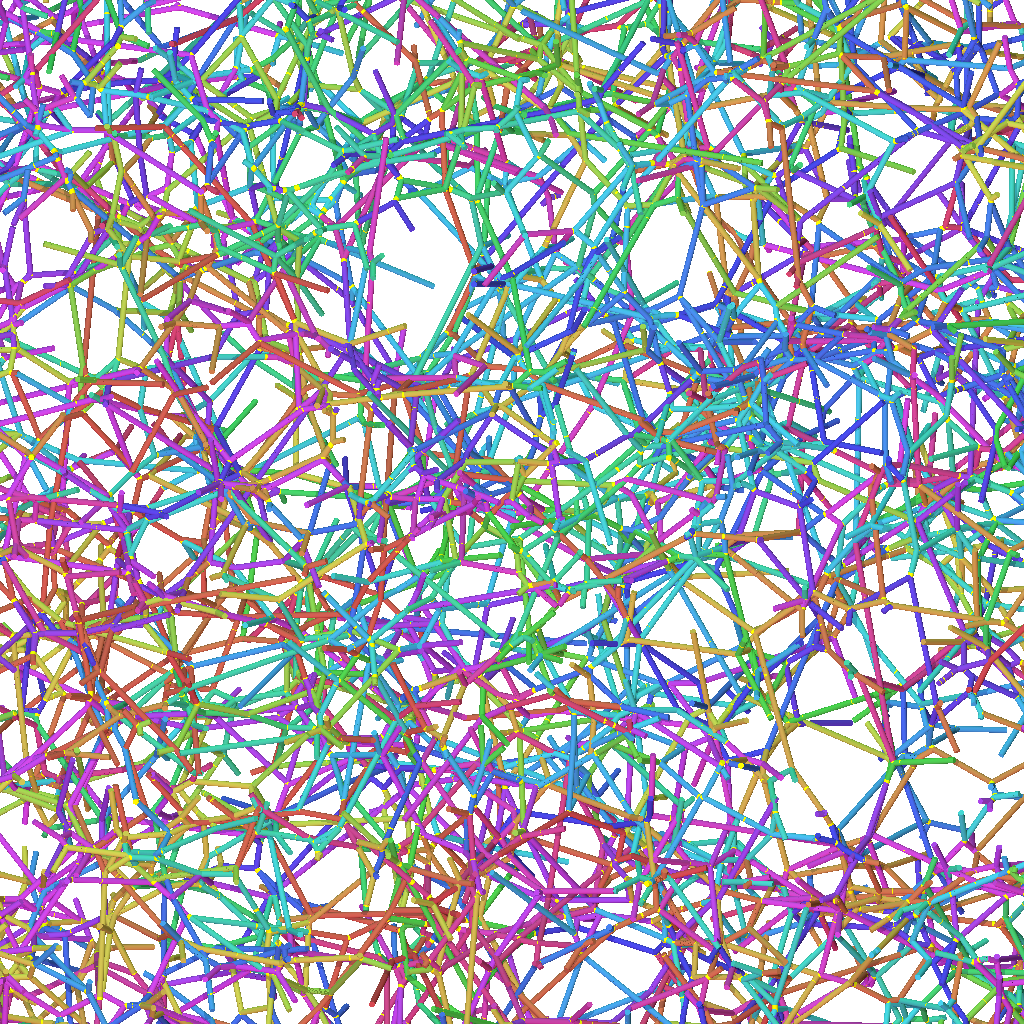}
    \caption{}
    \label{sbfgr:3PA_mesh_NXK=10}
  \end{subfigure}

  \begin{subfigure}{5.5cm}
    \includegraphics[width=\textwidth]{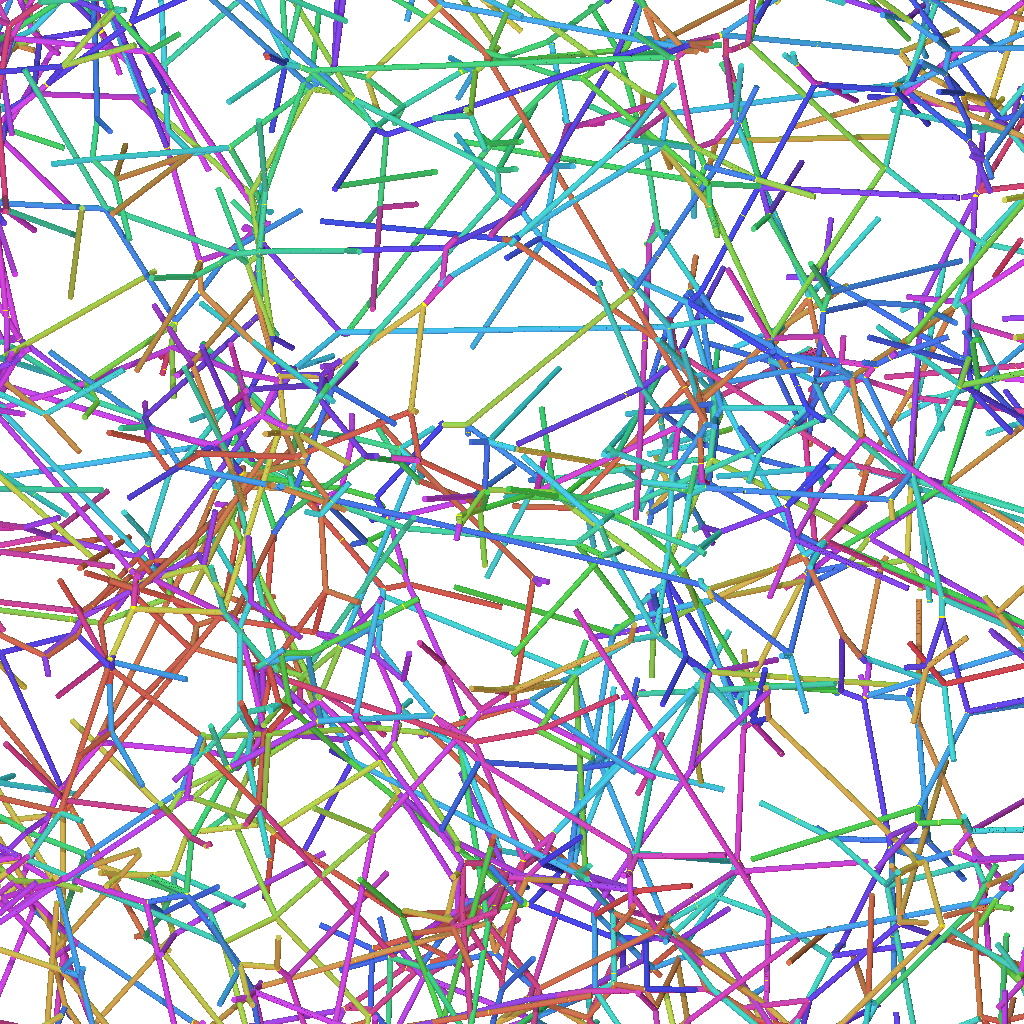}
    \caption{}
    \label{sbfgr:3PA_mesh_NXK=35}
  \end{subfigure}

  \begin{subfigure}{5.5cm}
    \includegraphics[width=\textwidth]{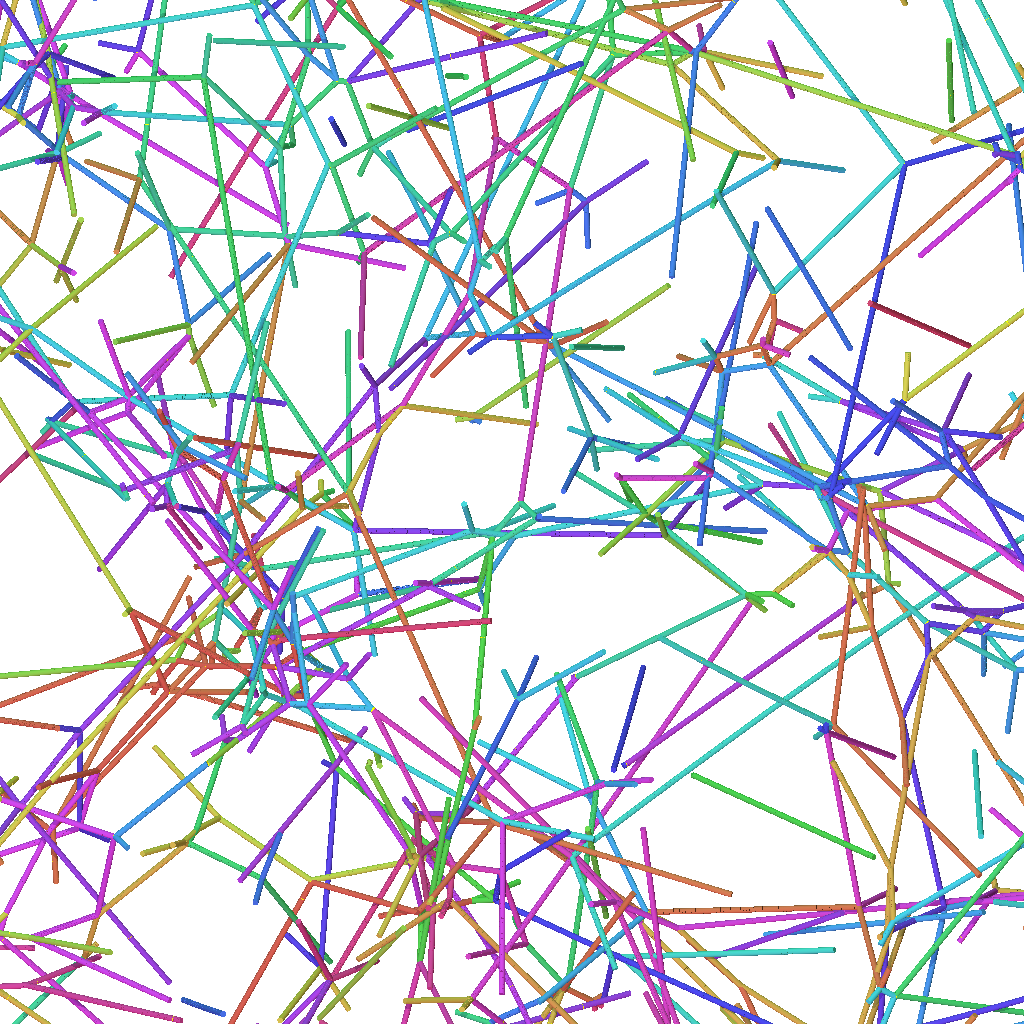}
    \caption{}
    \label{sbfgr:3PA_mesh_NXK=100}
  \end{subfigure}
  \caption{Visualization of slabs of 3PA meshes for networks with with $\left<N_{XK}\right>=10$ ( \subref{sbfgr:3PA_mesh_NXK=10}), $\left<N_{XK}\right>=35$ (\subref{sbfgr:3PA_mesh_NXK=35}), and $\left<N_{XK}\right>=100$ (\subref{sbfgr:3PA_mesh_NXK=100}). Slabs and chain colors as in Fig. \protect\ref{fgr:PPA_mesh_visualization}}
  \label{fgr:3PA_mesh_visualization}
\end{figure}

It would be natural to ask if we can estimate the cross-link contribution to the shear modulus by performing an analysis similar to the PPA, but applied to the model networks. The theoretical rational for how to formulate such an analysis is provided by the phantom network model. The Phantom Primitive Path Analysis~(3PA)~\cite{SvaneborgEveraersGrestCurro2008StressContributions} allows us to measure the cross-link modulus by deforming phantom meshes.

The 3PA analysis proceeds similar to PPA analysis, except that all pair interactions between beads are switched off. The model network is cooled down to $T=0$ to eliminate thermal fluctuations, hence, it is converted into the mechanical equilibrium state of the corresponding phantom model~--~the phantom mesh. To eliminate finite extensibility effects, we replaced FENE bonds by harmonic bonds. Instead of analyzing the network topology, we measure the stress tensor of the deformed phantom meshes. Since thermal fluctuations in the phantom model are strain independent, they do not contribute to the stress. Hence, the stress-strain behaviour of the phantom mesh is sufficient to estimate the cross-link contribution to the network modulus.

The theoretical derivation of the phantom model assumes a local tree-like structure of the network.~\cite{EdwardsViglis1986EffOfEntanglRubbElast, RubinPanyuk1997NFFDfrmElstctPlmNtwk} Recently, there has been several studies of the effect of loops on the phantom modulus estimate, see e.g. Refs.~\cite{
ZhongWangKawamotoOlsenJohnson2016RENT,
Lang2018,
Panyukov2019,
LinWangJohnsonOlsen2019ElasticityRealGaussPhtnmNtwrk}.
The phantom modulus estimate based on a network analysis excludes such loop contributions, however, since the 3PA analysis exactly generates deformed phantom model ground states for our actual network structures, the 3PA modulus estimate does include loop contributions.

Since the 3PA force field differs from the KG force field, the 3PA stress has to be converted to an equivalent KG stress value.~\cite{SvaneborgEveraersGrestCurro2008StressContributions} Recall the expression for the entropic stress tensor of a single polymer strand~(see Eq.~(4.129) in Ref.~\cite{DoiEdwards1986TheorPolymDyn}):
\begin{equation}
\label{eqn:entropic_stress}
  \mathbf{\sigma}^{KG} = -\dfrac{3\,k_B\,T}{N_K\,l_K^2}\,\mathbf{R}\otimes\mathbf{R}\,,
\end{equation}
where $\mathbf{R}$ is the strand end-to-end vector. Whereas, in the mechanical equilibrium state the same network strand has a virial stress:
\begin{equation}
\label{eqn:3PA_stress}
  \mathbf{\sigma}^{3PA} = -\dfrac{k^{3PA}}{N_b}\,\mathbf{R}\otimes\mathbf{R}\,,
\end{equation}
where $k^{3PA}$ is the spring constant used in 3PA simulations, which was chosen arbitrarily as $k^{3PA}=100\,\varepsilon\,\sigma^{-2}$, and $N_b$ is number of beads in the strand. Eliminating tensor $\mathbf{R}\,\mathbf{R}$ from Eqs.~(\ref{eqn:entropic_stress}),~(\ref{eqn:3PA_stress}) and converting from Kuhn to bead units as $N_K\,l_K^2 = C_{\infty}\,N_b\,l_b^2$, one obtains:
\begin{equation}
\label{eqn:entropic_nt_via_3PA_stress}
  \mathbf{\sigma}^{KG} = \mathbf{\sigma}^{3PA}\,\dfrac{3\,k_B\,T}{k^{3PA}\,l_b\,l_K}\,,
\end{equation}
here $l_b$ is KG bond length, $C_{\infty}$ is polymer specific characteristic ratio~(see Sect.~\ref{sbsct:Mapping}).

Compared to generation of stress-strain data for the full KG networks, the 3PA analysis provides the cross-link moduli with much less computational effort, since the method requires only a single energy minimization. Visualization of some of the resulting phantom meshes are shown in Fig.~\ref{fgr:3PA_mesh_visualization}. We observe that all strands forms straight lines connecting cross-links, and each cross-link position is determined by the force balance of all connected strands. We observe a more and more dense mesh as the density of cross-links increase, in particular, we note the qualitative similarity between the phantom mesh with $N_{XK}=35$ and the primitive path mesh shown in Fig.~\ref{sbfgr:PPA_mesh}.

%% file: 4.0.Results.tex
\section{Results and discussion}
\label{sct:Results}

In the present section, we show the results of the analysis of the our stress-strain simulation data within the scope of the MR model~[Sect.~\ref{sbsct:MR_res}]. Next, we show the results of the PPA methods, compare estimations of the shear modulus given by two different methods and discuss the microscopic origin of the MR model parameters~[Sect.~\ref{sbsct:PPA_res}]. Finally, we fit microscopic elasticity models to our simulation stress-strain data, using either the full range of the simulation data or that range, where finite extensibility effects are expected to be negligible~[Sect.~\ref{sbsct:Micro_res}].

%% file: 4.1.MR.tex
\subsection{MR analysis}
\label{sbsct:MR_res}

\begin{figure}[!h]
  \begin{subfigure}[!h]{8.25cm}
    \includegraphics[width=\textwidth]{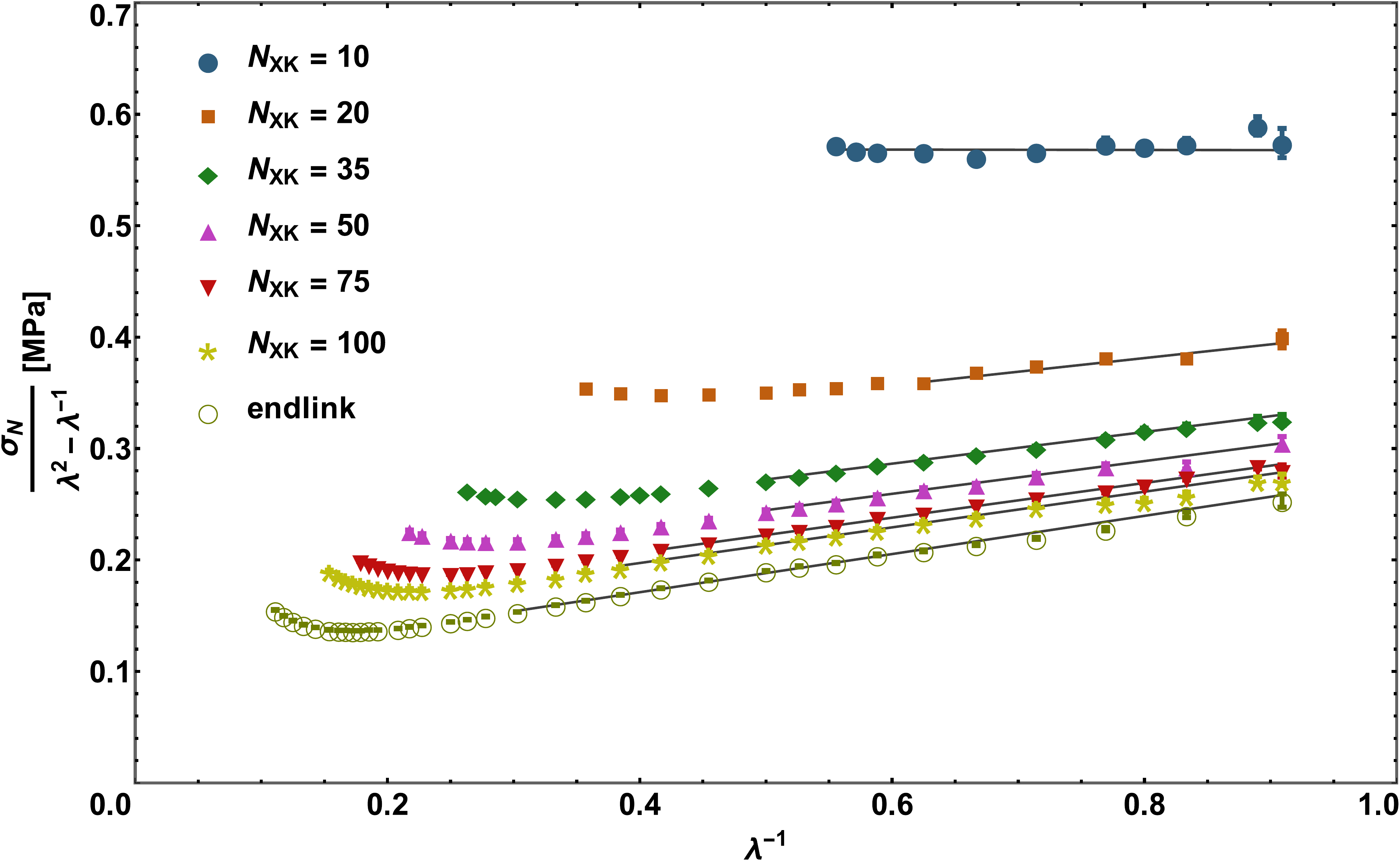}
    \caption{}
    \label{sbfgr:KG_MR_rnt_vs_iElF}
  \end{subfigure}
  \hfill
  \begin{subfigure}[!h]{8.25cm}
    \includegraphics[width=\textwidth]{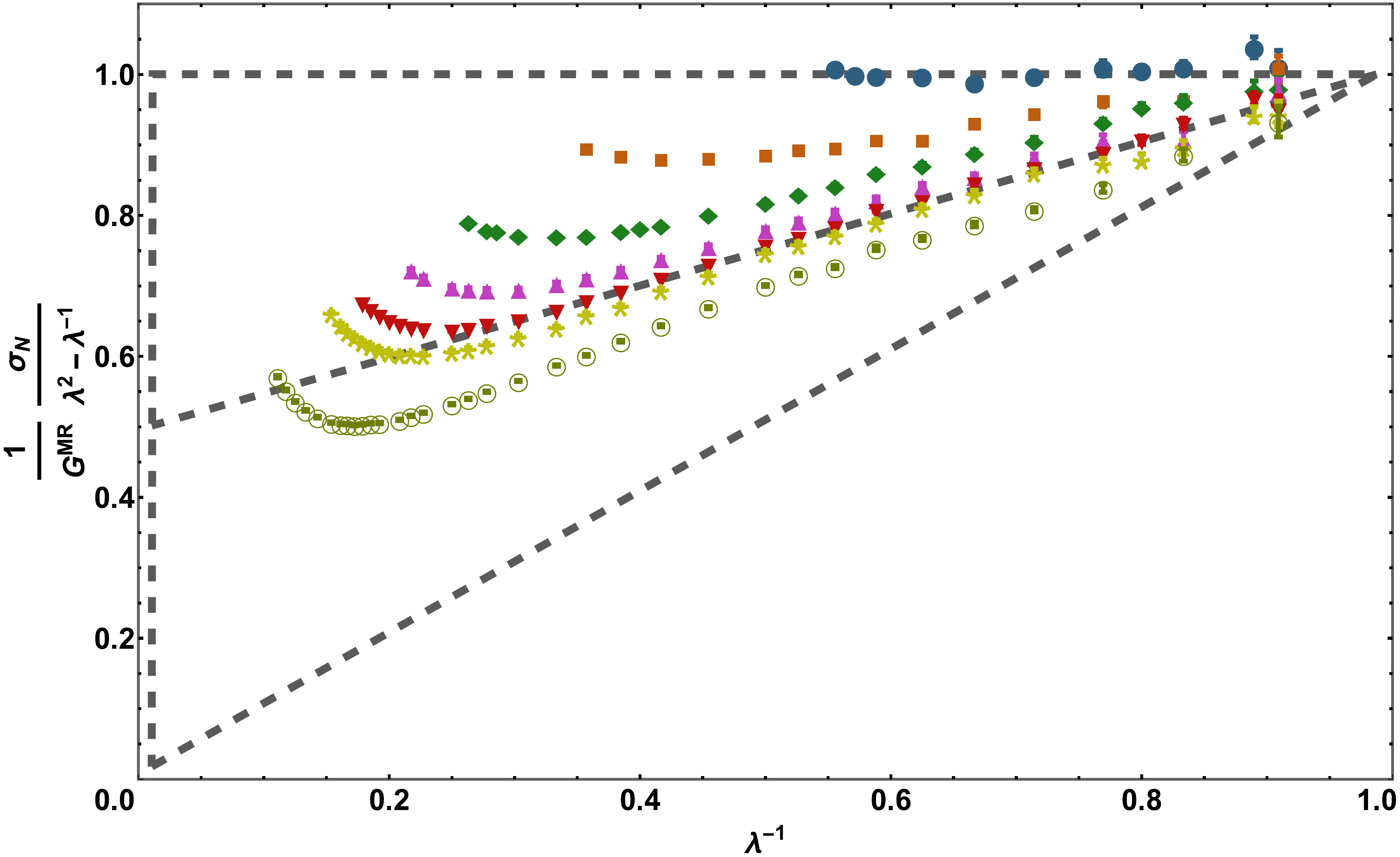}
    \caption{}
    \label{sbfgr:KG_MR_nrm_rnt_vs_iElF}
  \end{subfigure}
  \hfill
  \begin{subfigure}[!h]{8.25cm}
    \includegraphics[width=\textwidth]{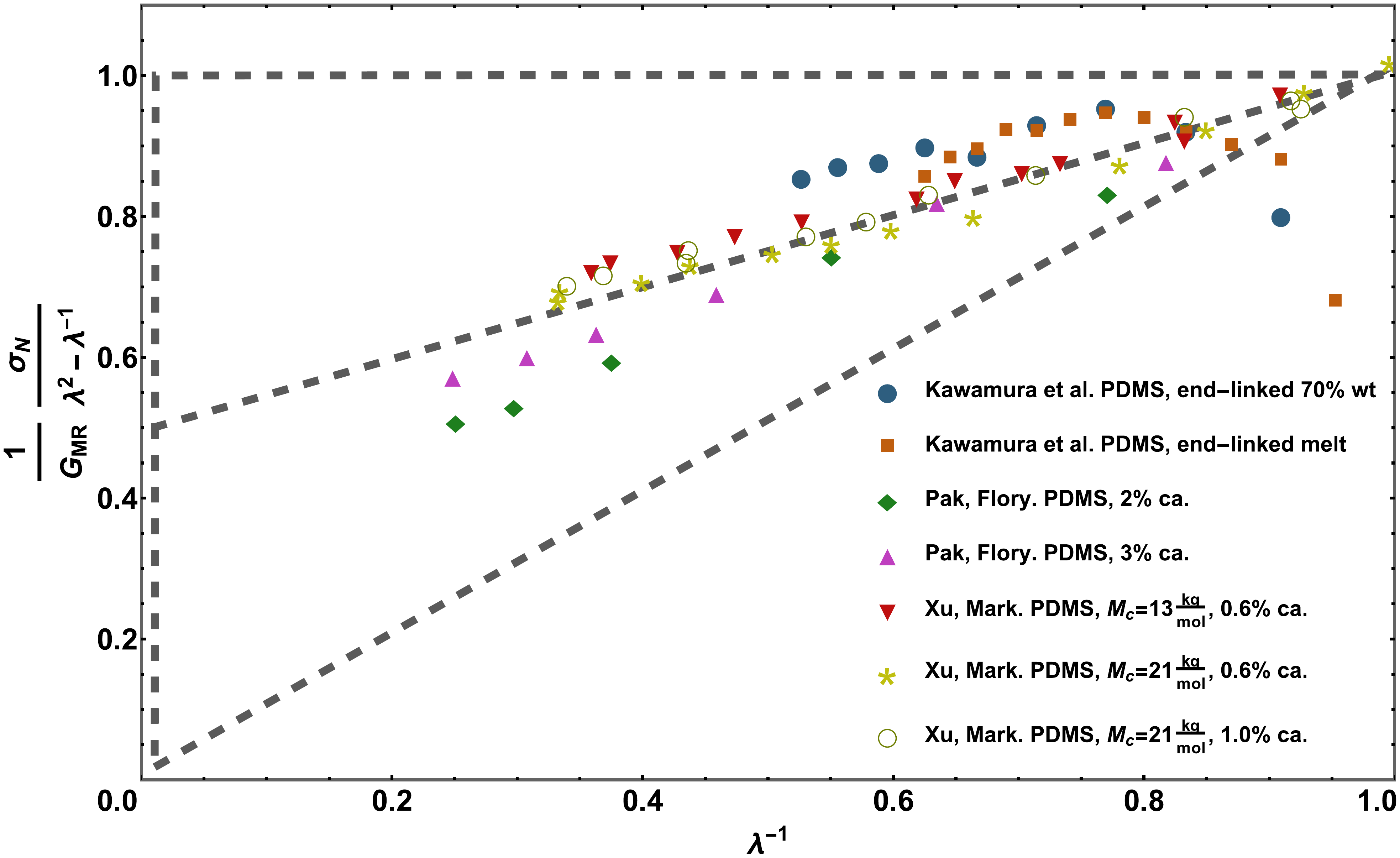}
    \caption{}
    \label{sbfgr:data}
  \end{subfigure}
  \caption{\subref{sbfgr:KG_MR_rnt_vs_iElF}: Mooney-Rivlin plot of the reduced normal tension $\widetilde{\sigma}_N$ vs. the inverse elongation $\lambda^{-1}$. Black solid lines denote linear fits of the stress-strain data. \subref{sbfgr:KG_MR_nrm_rnt_vs_iElF}: Normalized Mooney-Rivlin plot, where the stress data are normalized by the shear modulus estimates. Grey thick dashed lines define the triangle, showing the cross-over from the cross-link to the entanglement dominated stress-strain behaviour in accordance with the MR model. \subref{sbfgr:data}: Experimental data for PDMS, shown in the normalized Mooney-Rivlin representation.}
  \label{fgr:KG_MR_rnt}
\end{figure}

Fig.~\ref{sbfgr:KG_MR_rnt_vs_iElF} shows stress data on uniaxial stretching simulations of our model networks in Mooney-Rivlin representation. Each data point is the result of the extrapolation procedure to infinite time presented in Sect.~\ref{sbsct:SetUp}. The stress response increases with increasing cross-link density. Moreover, a stress upturn is observed for large deformations, which is due to finite extensibility effects. We read off the elongations $\lambda^{MR}_{max}$, where the stress upturn starts. The results are summarized in Table~\ref{tbl:Res}. For the data where the finite extensibility effects are negligible, we observe a linear relation between the reduced normal tension $\widetilde{\sigma}_N$ and the inverse elongation $\lambda^{-1}$ as predicted by the MR model~(\ref{eqn:MR_rnt}). This allowed us to perform linear fitting to the stress-strain data. To avoid finite extensibility effects, only the data with $\varepsilon < \varepsilon_{max}^{MR}/2$ were fitted, where $\varepsilon=\lambda-1$ is deformation, $\varepsilon_{max}^{MR}$ is the deformation, corresponding to the stress upturn. We have performed the present MR analysis and model fits below for several criteria, gauging which data points are affected by finite extensibility effects. Including too few data points causes statistical uncertainties in the fits, while including too many data points cause finite-extensibility artifacts. We choose the criterion $\varepsilon < \varepsilon_{max}^{MR}/2$, since it was observed to produce both robust and accurate fit results. The black lines in Fig.~\ref{sbfgr:KG_MR_rnt_vs_iElF} illustrate that range of the data, and the coloured symbols in Fig.~\ref{fgr:Micro_fit_lin_uni_rnt_vs_iElF} justify our choice. We observe linear behaviour in the former and an excellent collapse in the latter, hence showing that finite extensibility effects are not expected to influence our conclusions. The statistical errors of the measured stress-strain data were accounted for when performing the fits, though the error bars are comparable to or smaller than the symbols. According to Eq.~(\ref{eqn:MR_rnt}), extrapolation of the linear fitting function towards $\lambda^{-1} \rightarrow 1$ provides an estimate for the shear modulus $\mathrm{G}^{MR}$, whereas extrapolation towards $\lambda^{-1} \rightarrow 0$ gives estimate for the $\mathrm{C}_1$ model parameter. The resulting values for $\mathrm{C}_1$, $\mathrm{C}_2$ and $\mathrm{G}_{MR}$ are presented in Table~\ref{tbl:Res}.

Fig.~\ref{sbfgr:KG_MR_nrm_rnt_vs_iElF} shows the simulation data in a reduced Mooney-Rivlin representation, where the data is normalized by the shear moduli estimates obtained via the MR analysis. In this representation, all stress-strain data are enclosed by the triangular domain $(0;\,0)$, $(0;\,1)$, $(1;\,1)$~(shown as the grey dashed line). This allows one to directly visually gauge the relative importance of the $\mathrm{C}_1$, $\mathrm{C}_2$ contributions to the shear modulus. Horizontal stress-strain data are dominated by the $\mathrm{C}_1$ term, wheras diagonal stress-strain data would be dominated by the $\mathrm{C}_2$ term. The middle dashed line corresponds to an even balance between the $\mathrm{C}_1$ and $\mathrm{C}_2$ contributions. It is clearly seen that our the most strongly cross-linked model network with $4$ cross-links between each entanglement on average is mostly dominated by the $\mathrm{C}_1$ term. The $\mathrm{C}_2$ term contribution progressively increases for weaker cross-linked networks. The stress-strain data for the end-linked network is below the middle line, indicating that the $\mathrm{C}_2$ term is slightly larger than the $\mathrm{C}_1$ term in this case. The end-linked network has on average $56$ entanglements between cross-links and is completely dominated by entanglements effects.

Fig.~\ref{sbfgr:data} shows experimental data for PDMS~\cite{KawamuraUrayamaKohjiya2001MltXDfrmOfELPolymerNtwrkI, PakFlory1979Stress1DStrPDMS, XuMark1990} also in the normalized Mooney-Rivlin form. The experimental papers provide no estimates for the strand length after cross-linking. However, since the networks were formed via end-linking, we expect the network strand lengths to be comparable to the chain length of the precursor melt. The precursor melt used by Kawamura et al.~\cite{KawamuraUrayamaKohjiya2001MltXDfrmOfELPolymerNtwrkI} had chains of $150$ Kuhn segments, while Xu et al.~\cite{XuMark1990} used a precursor melt with $42-68$ Kuhn segments. Consequently, the resulting experimental results are expected to be in the entanglement dominated regime. No special attempts to find or select experimental data matching our model networks have been made. Nevertheless, we observe that the experimental data roughly falls into the same triangular region as our simulation data. This suggests that the range of network cross-links density we have used is relevant for comparison to experimental data and the conclusions, which we draw based on our model networks, are applicable to the experimental systems. Additionally, the reliability of the MR model for the description of uniaxial stretching of rubbery materials is justified. We note some scatter for small deformations (data of Kawamura et al.~\cite{KawamuraUrayamaKohjiya2001MltXDfrmOfELPolymerNtwrkI}). The Mooney-Rivlin representation of the stress-strain data significantly amplifies any experimental error close to the unstrained state, since it is difficult to measure small stresses precisely as well as to determine the length of the sample exactly in the vicinity of the unstrained state. For comparison, elongation of model networks is exactly defined in our simulations, but we have a significantly reduced signal-to-noise ratio for stresses at small deformations.

%% file: 4.2.PP.tex
\subsection{PPA and 3PA analysis}
\label{sbsct:PPA_res}

\begin{table}[!h]
  \centering
  \caption{Network characterization. Fraction of ''good'' strands $\alpha_S^{good}$; the phantom network model estimates for the cross-link modulus $\mathrm{G}^{ph}$, Eq.~(\ref{eqn:Phnt_ntwrk_GX_org}); estimates of the cross-link modulus $\mathrm{G}_X$ obtained via the 3PA analysis; elongation values $\lambda^{MR}_{max}$ at the upturn of the stress data in the Mooney represenation~(plot in Fig.~\ref{sbfgr:KG_MR_rnt_vs_iElF}); estimates of the MR model parameters $\mathrm{C}_1^{MR}$, $\mathrm{C}_2^{MR}$ and the shear modulus $\mathrm{G}^{MR}$ obtained by means of the MR analysis of the simulation data; estimates of the shear modulus provided by the PPA methods $\mathrm{G}_X + \mathrm{G}_E$. EL stands for "end-linked". For comparison, the entanglement modulus of the precursor melt is $\mathrm{G}_E = 0.226\,\mathrm{MPa}$.}
  \label{tbl:Res}
  {\footnotesize
  \begin{tabular}{|c|c|c|c|c|c|c|c|c|}
    \hline
    $N_{XK}$ &
    $\alpha_S^{good}$ &
    $\mathrm{G}^{ph}\,\left[\mathrm{MPa}\right]$ &
    $\mathrm{G}_X\,\left[\mathrm{MPa}\right]$ &
    $\lambda^{MR}_{max}$ &
    $\mathrm{C}_1^{MR}\,\left[\mathrm{MPa}\right]$ &
    $\mathrm{C}_2^{MR}\,\left[\mathrm{MPa}\right]$ &
    $\mathrm{G}^{MR}\,\left[\mathrm{MPa}\right]$ &
    $\mathrm{G}_X + \mathrm{G}_E\,\left[\mathrm{MPa}\right]$ \\
    \hline
    $10$  & $0.8648$ & $0.2980$ & $0.3154$ & $1.8$ & $0.2846$ & $-0.0007$ & $0.5677 \pm 0.0209$ & $0.5418$ \\
    \hline
    $20$  & $0.8611$ & $0.1410$ & $0.1443$ & $2.2$ & $0.1415$ & $0.0614$  & $0.4058 \pm 0.0247$ & $0.3707$ \\
    \hline
    $35$  & $0.8582$ & $0.0795$ & $0.0793$ & $3.0$ & $0.1006$ & $0.0710$  & $0.3433 \pm 0.0035$ & $0.3057$ \\
    \hline
    $50$  & $0.8539$ & $0.0549$ & $0.0545$ & $3.3$ & $0.0856$ & $0.0734$  & $0.3181 \pm 0.0050$ & $0.2809$ \\
    \hline
    $75$  & $0.8496$ & $0.0359$ & $0.0344$ & $4.0$ & $0.0723$ & $0.0778$  & $0.3002 \pm 0.0024$ & $0.2608$ \\
    \hline
    $100$ & $0.8485$ & $0.0268$ & $0.0254$ & $4.4$ & $0.0667$ & $0.0800$  & $0.2932 \pm 0.0042$ & $0.2519$ \\
    \hline
    EL    & $0.8321$ & $0.0013$ & $0.0010$ & $5.6$ & $0.0513$ & $0.0857$  & $0.2739 \pm 0.0039$ & $0.2275$ \\
    \hline
  \end{tabular}}
\end{table}

\begin{figure}[!h]
  \includegraphics[width=8.25cm]{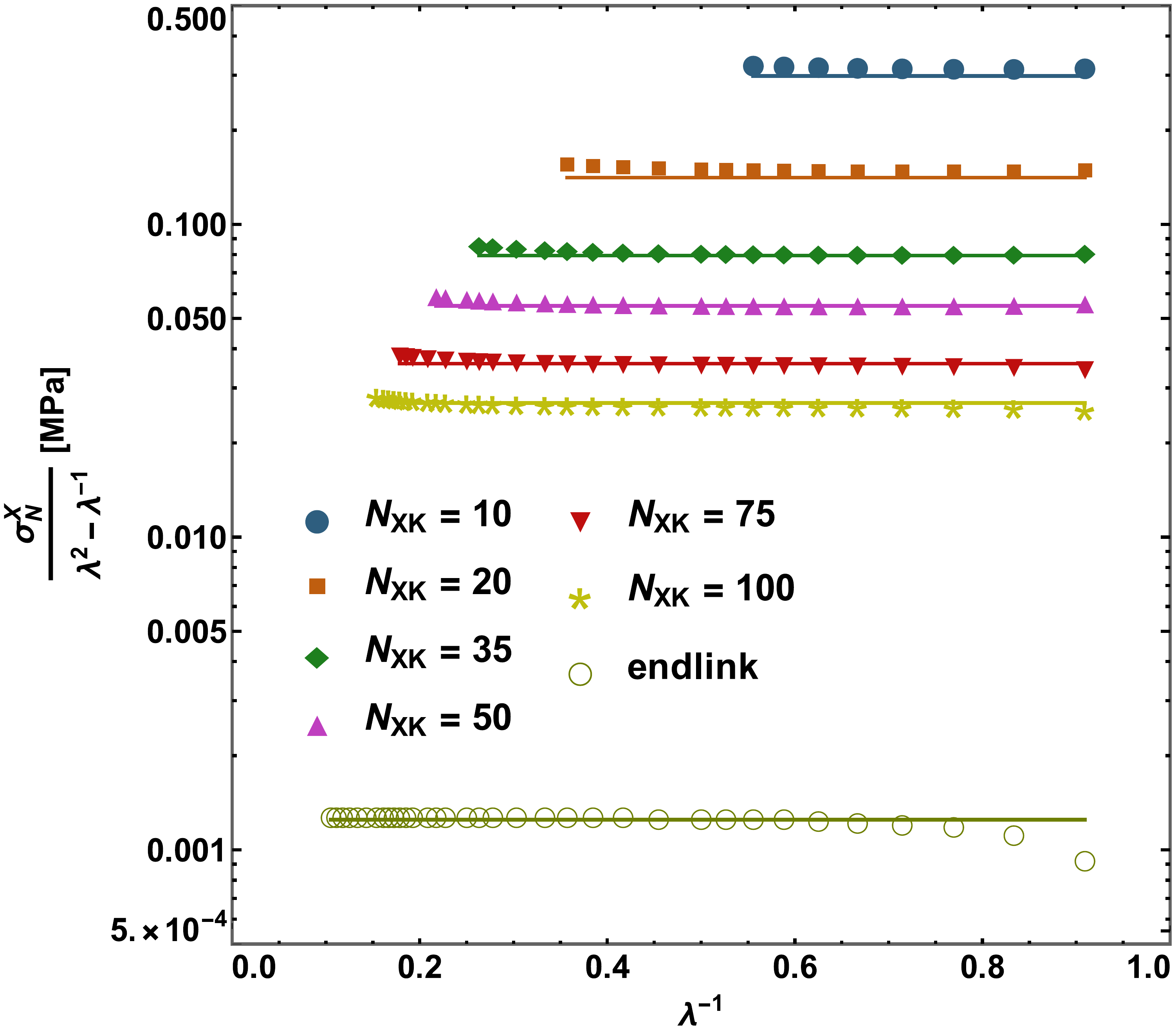}
  \caption{Mooney-Rivlin plot of the 3PA reduced normal tension $\widetilde{\sigma}_N^X$. Lines are parameter free predictions of the phantom network model, Eq.~(\ref{eqn:Phnt_ntwrk_GX_org}), using as input the data from the network characterization, see Sect.~\ref{sbsct:Characterization}.}
  \label{fgr:3PA_rnt_vs_iElF}
\end{figure}

\begin{figure}[!h]
  \includegraphics[width=8.25cm]{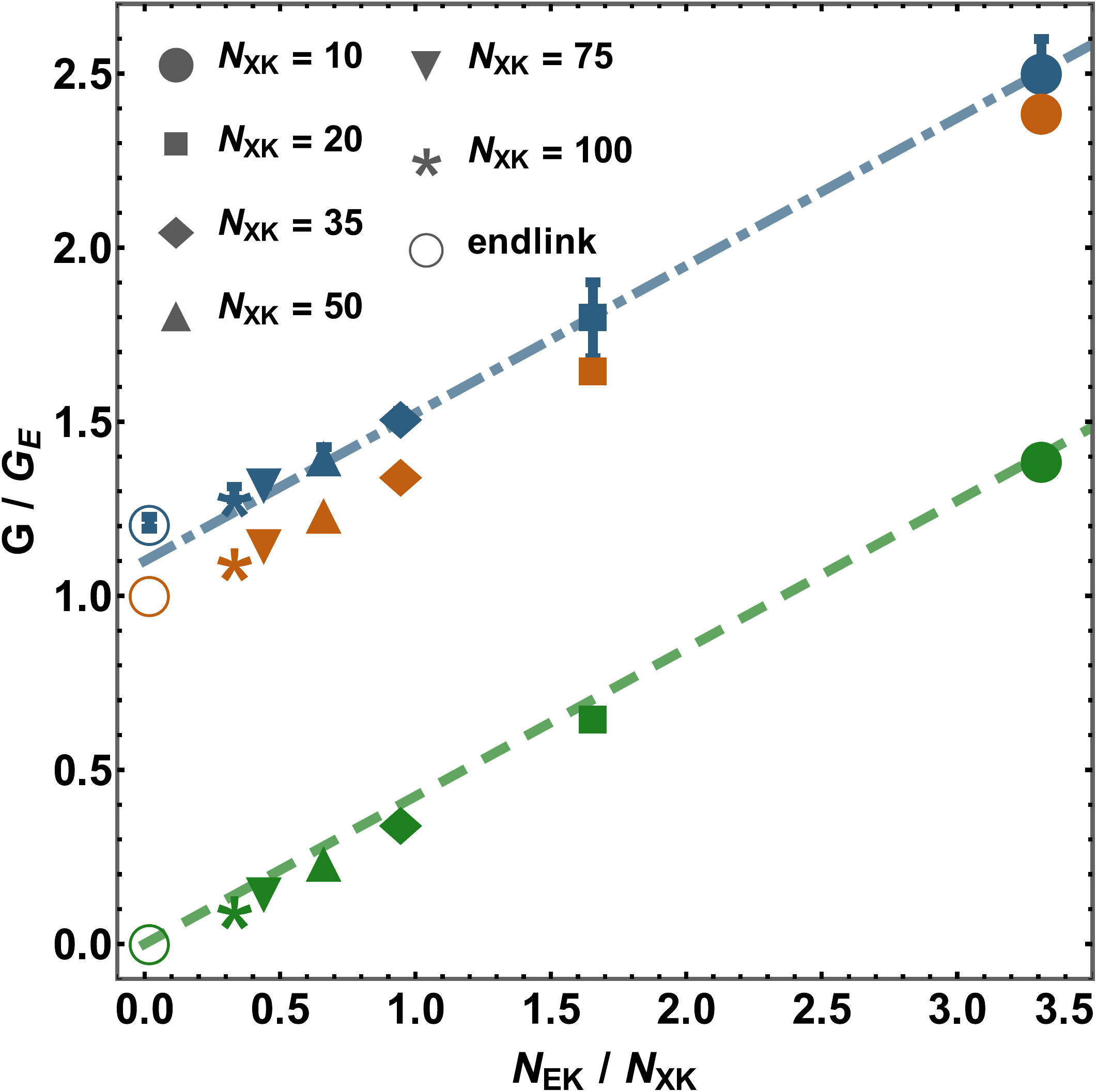}
  \caption{Comparison of the shear moduli estimates from the MR analysis, $\mathrm{G}^{MR}$~(blue markers), and from the Primitive Path methods: the cross-link modulus $\mathrm{G}_X$~(green markers), $\mathrm{G}_X + \mathrm{G}_E$~(orange markers). For the discussion of the dashed and the dot-dashed lines, see the text.}
  \label{fgr:GMR_vs_G3PA}
\end{figure}

\begin{figure}[!h]
  \includegraphics[width=8.25cm]{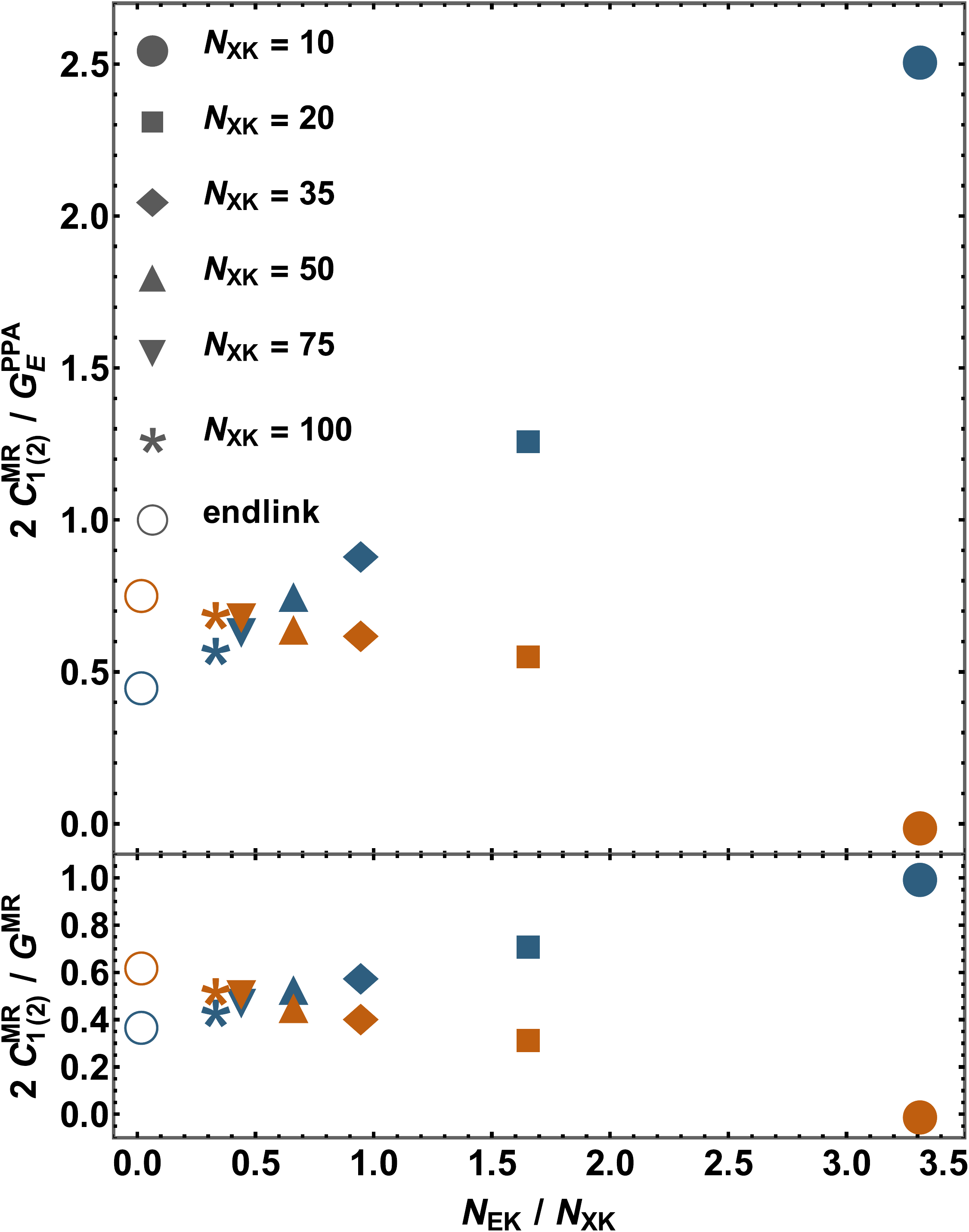}
  \caption{Identification of the MR modulus contributions $2\,\mathrm{C}_1$, $2\,\mathrm{C}_2$ in relation to the entanglement modulus $\mathrm{G}_E$ as function of the network structure. The bottom plot shows the same data relatively to the MR estimates for the shear modulus $\mathrm{G}^{MR}$. Blue markers denote the $\mathrm{C}_1$ parameter, orange markers show the $\mathrm{C}_2$ parameter.}
  \label{fgr:MR_vs_PPA_Coeffs}
\end{figure}

To complement the MR analysis, we performed PPA analysis of the precursor melt and 3PA analysis of the deformed model networks. PPA analysis of the precursor melt provides an estimate of $N_{EK} \approx 33.1$. Using the mapping from PDMS-KG model to SI units, see Sect.~\ref{sbsct:Mapping}, we calculate the entanglement modulus $\mathrm{G}_E = 0.23\,\mathrm{MPa}$, which is in a good agreement with the literature values of $N_{EK}=31.1$ and $\mathrm{G}_E=0.25\,\mathrm{MPa}$.~\cite{FettersLohseColby2007ChainDimEntaSpac} Comparing the MR modulus estimate of the end-linked network to the entanglement modulus, we obtain fairly good agreement $\mathrm{G}^{MR}(EL)/\mathrm{G}_E \approx 1.2$, which is discussed below.

Fig.~\ref{fgr:3PA_rnt_vs_iElF} shows Mooney-Rivlin representation of the stress-strain data provided by 3PA analysis of the deformed networks. The phantom network model predicts that the reduced stress is independent of elongation, and this is in excellent agreement with the results of the 3PA analysis. From the reduced stress plateaus, one can directly read off the cross-link modulus $\mathrm{G}_X$. For comparison, the figure also shows parameter free predictions of $\mathrm{G}^{ph}$ (Eq.~(\ref{eqn:Phnt_ntwrk_GX_org})). To evaluate $\mathrm{G}^{ph}$, we used the densities of good strands and cross-links obtained from the network analysis, see Eq.~\ref{eqn:Phnt_ntwrk_GX_org}. We observe that phantom model estimates are in excellent agreement with the 3PA analysis results. 3PA method implements Hamiltonian of the phantom network model exactly for a given network, hence, the resulting cross-link modulus estimate $\mathrm{G}_X$ is completely independent of any assumptions used to derive the phantom network modulus Eqs.~(\ref{eqn:Phnt_ntwrk_GX_org}, \ref{eqn:Phnt_ntwrk_GX_prt}). Numerical values of the moduli $\mathrm{G}^{ph}$ and $\mathrm{G}_X$ are summarized in Table~\ref{tbl:Res}. We note that loop contributions~\cite{Lang2018, Panyukov2019} to the modulus are only presented in $\mathrm{G}_X$. Perhaps, this is why $\mathrm{G}_X$ is slightly larger than the corresponding phantom modulus for the most cross-linked network.

Having obtained \emph{model independent} estimates of the cross-link moduli of all networks and the entanglement modulus of the precursor melt, one can ask how they are related to the network structure, e.g., to the quantitative relation between entanglement and network strands. Fig.~\ref{fgr:GMR_vs_G3PA} shows a Langley plot of the results of the 3PA analysis as blue symbols. The phantom network model Eq.~(\ref{eqn:Phnt_ntwrk_GX_prt}) with functionality $f=4$ predicts $\mathrm{G}_X/\mathrm{G}_E = 0.5\,N_{EK}/N_{XK}$, whereas we observe $\mathrm{G}_X/\mathrm{G}_E = 0.42\,N_{EK}/N_{XK}$~(dashed green line). This is perfectly consistent with the number of good strands we obtained from the network analysis, suggesting a $15\%$ reduction of the cross-link modulus due to the presence of loops. We also show the linear relation $\mathrm{G}/\mathrm{G}_E=1.10+0.42\,N_{EK}/N_{XK}$~(blue dot-dashed line), which is observed to be in excellent agreement with the moduli obtained from MR analysis. The choice of prefactor will be discussed below.

Many theories assume that the effects of entanglements and cross-links are additive, see e.g. Refs.~\cite{RubinPanyuk1997NFFDfrmElstctPlmNtwk, KaliskeHeinrich1999ExtendedTubeModel, RubinPanyuk2002ElastPolymNetw, DavidsonGoulbourne2013NffNtwkModel, XiangZhongWangMaoYuQu2018GnrlModelSoftElast}. We are not aware of any theoretical arguments or proofs for why this assumption is valid. For instance, the double tube model of Mergell and Everaers~\cite{MergellEveraers2001TubeModels} produces a non-additive relation for the modulus. The predictions of the double tube model are obtained by means of a statistical mechanical approach based on a constraint mode Hamiltonian, and hence are thermodynamically consistent.

Having \emph{independent} estimates for the network shear modulus as well as the cross-link and the entanglement moduli, we can test the equality between the $\mathrm{G}_{MR}$ estimate on one hand and $\mathrm{G}_X + \mathrm{G}_E$ on the other hand. To our knowledge, this is perhaps the first time this relation has been tested directly. Fig.~\ref{fgr:GMR_vs_G3PA} shows the shear moduli estimates obtained via the MR model analysis as well as the sum of cross-link and entanglement moduli. We observe that the MR modulus estimates are consistent with but slightly larger than the sum. The deviation is in the range $5-20\%$ and is decreasing with increasing density of cross-links. We note that the statistical error of the MR shear moduli estimates is smaller than $3\%$. We also emphasise that our analysis is based on the entanglement modulus of the precursor melt, which is $20$\% larger than the corresponding plateau modulus, which is often reported in the literature. Hence, a larger deviation would have been observed in the latter case. We attribute this systematic deviation to the capture of entanglements during cross-linking. Langley proposed that the network modulus comprised the cross-link modulus and entanglements contribution multiplied by Langley trapping factor.~\cite{Langley1968} Our results are consistent with a Langley trapping factor $T_e<0.2$ essentially \emph{independent} of the degree of cross-linking.

The discussion of the physical interpretation of the $\mathrm{C}_1$, $\mathrm{C}_2$ terms has a long history in the literature, see, e.g., Refs.~\cite{
Mark1975MRConstants,
MarkSullivan1977ModelNetworksOfELPDMSI,
SharafMark1994Interpretation,
Treloar1973ElastPropRub,
Treloar1974ElastPropRub,
Treloar1974MechRubElast}.
Early on, it was recognized that the $\mathrm{C}_2$ term goes to zero, when the polymer network is swollen~\cite{
GumbrellMullinsRivlin1953,
Treloar1975PhysRubElasticity,
HanHorkaMcKenna1999},
and the breakdown of the neo-Hookean theory~(assuming $\mathrm{C}_2=0$) in the intermediate range of strains was attributed to entanglements.~\cite{
Flory1976,
RoncaAllegra1975,
BoyerMiller1987} Moreover, the tube theories discussed in Sect.~\ref{sbsct:Micro} also suggests that the parameter $\mathrm{C}_1$ is related to both cross-links and entanglements, while the parameter $\mathrm{C}_2$ is only related to entanglements. The notable exceptions being the Warner-Edwards tube theory~\cite{
DeamEdwards1976ThrRubElast,
WarnerEdwards1978NeutronScatter,
MergellEveraers2001TubeModels},
which has a strain independent tube diameter and reduces to a phantom-like stress-strain response, and the double tube model~\cite{MergellEveraers2001TubeModels}, where the $\mathrm{C}_2$ term depends on both the cross-link and the entanglement moduli $\mathrm{G}_X$, $\mathrm{G}_E$. To our knowledge, the most recent contribution to this long discussion is the paper of Schl{\"o}gl et al.~\cite{SchloglTrutschelChasseRiessSaalwachter2014}. The authors performed NMR experiments on dry and swollen samples to measure critical molecular weights. Swelling is believed to minimize the entanglement effects, allowing the cross-link modulus to be estimated experimentally. Moreover, the correlations between the critical molecular weights and the MR parameters were studied. The analysis showed that $\mathrm{C}_1$ was related to cross-links and entanglements, while $\mathrm{C}_2$ were related to entanglements only.

Fig.~\ref{fgr:MR_vs_PPA_Coeffs} shows the MR model parameters $\mathrm{C}_1$, $\mathrm{C}_2$ in units of the entanglement modulus. We observe a roughly linear increase of the $\mathrm{C}_1$ parameter with a slope of $0.63$ and a concomitant linear decrease of the $\mathrm{C}_2$ parameter with a slope of $-0.23$ with increasing cross-link density. The former is expected as the phantom model alone would predict $2\,\mathrm{C}_1/\mathrm{G}_E=\mathrm{G}_X/\mathrm{G}_E=0.42\,N_{EK}/N_{XK}$. Our data suggest transfer of the entanglement contributions progressively from $\mathrm{C}_2$ to $\mathrm{C}_1$ as the network is progressively cross-linked causing the increase of the slope as observed. This is similar to a Langley trapping factor, but acting between the two MR parameters and not affecting the resulting shear modulus. The bottom plot shows the MR model parameters $\mathrm{C}_1$, $\mathrm{C}_2$ in units of the shear modulus $\mathrm{G}^{MR}$ and, hence, how the two terms are balanced as function of network structure. In the entanglement dominated limit $N_{EK}/N_{XK}\rightarrow 0$, we observe that the entanglement modulus is distributed $40\%$, $60\%$ between the $\mathrm{C}_1$, $\mathrm{C}_2$ parameters, respectively. At the point $N_{EK} = 0.5\,N_{XK}$, the contributions are approximately equal. Finally, in the limit $N_{EK} \gg N_{XK}$, the $\mathrm{C}_2 \approx 0$ and $2\,\mathrm{C}_1 \approx \mathrm{G}^{MR}$. The fact that the network strand length $N_{XK}$ should be roughly twice the entanglement strand length $N_{EK}$ is a consequence of the definitions of the entanglement modulus $\mathrm{G}_E\,N_{EK}=\rho_K\,k_B\,T$, and the phantom modulus $0.5\,\mathrm{G}_X,N_{XK}=\rho_K\,k_B\,T$, where the prefactor comes from the phantom model for a $4$-functional network. Hence, with these definitions $N_{EK}=0.5\,N_{XK}$ is required for the entanglement and the cross-link moduli to match. Assuming entanglements are binary and, hence, corresponding to a $4$-functional cross-links, suggests that a better definition of the entanglement modulus would be the one that includes the phantom network model prefactor, such that $N_{EK}$ and $N_{XK}$ would be defined on an equal footing.~\cite{Everaers2012TopoVSRheo}

Both models of Rubinstein and Panyukov~\cite{RubinPanyuk1997NFFDfrmElstctPlmNtwk, RubinPanyuk2002ElastPolymNetw}, the extended tube model of Kaliske and Heinrich~\cite{KaliskeHeinrich1999ExtendedTubeModel}, the non-affine network model of Davidson and Goulbourne~\cite{DavidsonGoulbourne2013NffNtwkModel}, and the general constitutive model of Xiang et al.~\cite{XiangZhongWangMaoYuQu2018GnrlModelSoftElast}, predict that the entanglement modulus $\mathrm{G}_E$ is a constant fraction of the $\mathrm{C}_2$ parameter, see Eqs.~(\ref{eqn:Micro2MR_NffT},~\ref{eqn:Micro2MR_SlT},~\ref{eqn:Micro2MR_ExT},~\ref{eqn:Micro2MR_DG},~\ref{eqn:Micro2MR_XI}), respectively. This is not consistent with the data shown in Fig.~\ref{fgr:MR_vs_PPA_Coeffs}. We note that the decreasing $\mathrm{C}_2$ parameter is predicted by the Double Tube theory~\cite{MergellEveraers2001TubeModels}, see Eq.~(\ref{eqn:Micro2MR_DT}). However, the model also predicts that both quantities $2\,\mathrm{C}_1$, $2\,\mathrm{C}_2$ converge to the same value in the limit $N_{EK} / N_{XK} \rightarrow 0$, which is not what we observe.

%% file: 4.3.0.Micro.tex
\subsection{Comparison of microscopic models}
\label{sbsct:Micro_res}

In Sect.~\ref{sbsct:PPA_res}, we presented our results for the cross-link and entanglement moduli, and qualitatively discussed their relations to the total shear modulus as well as to the MR parameters.

Many theories predict a universal form of the stress-strain relation, when subtracting the cross-link modulus from the reduced normal tension and normalizing by the entanglement modulus. In Sect.~\ref{sbsbsct:Universality}, we check this assumption \emph{independently} of any model.

Another interesting question is whether the microscopic models, where the entanglement and the cross-link model fit parameters are identified with the Primitive Path estimates, are able to predict the simulation stress-strain data. This is the most stringent consistency test possible for the models, since there are no free parameters, and, hence, no way for a fit to hide systematic errors. In~\ref{sbsbsct:Universality}, we apply this test to the stress-strain data not affected by finite extensibility effects.

For models failing such a test, one can fit the moduli prefactors and check how accurate the models are in estimating the shear, the cross-link and entanglement moduli from the stress-strain data. In Sect.~\ref{sbsbsct:Fit_linear_data}, we apply this test to the models not taking finite extensbility into account, while in Sect.~\ref{sbsbsct:Fit_full_data} we perform the analysis for the models that include finite extensibility effects. Moreover, we compare the fitted estimates for the cross-link and the entanglement moduli to the independent benchmarks provided by the Primitive Path methods.

We emphasize that here and below $\mathrm{G}_E$ and $\mathrm{G}_X$ refer to the PPA estimate of the entanglement modulus of the precursor melt and 3PA estimate of the network cross-link modulus, respectively. We fit the models as they are formulated in Sect.~\ref{sbsct:Micro}. The model fit parameters are referred to as $\mathrm{G}_E^{model}$ and $\mathrm{G}_X^{model}$ where necessary to avoid confusion. We identify the model parameters with our moduli estimates without attempting to include any model specific prefactors or microscopic parameters used to define the cross-link and entanglement moduli. All models that assume additivity of the entanglement and the cross-link moduli obey $\mathrm{G}_X^{model} + \mathrm{G}_E^{model} = 2\,\left(\mathrm{C}_1 + \mathrm{C}_2\right) = \mathrm{G}_{MR}$, and in this respect are compared to our simulation data on an equal footing.

%% file: 4.3.1.Universality.tex
\subsubsection{Universal representation of the stress-strain data}
\label{sbsbsct:Universality}

\begin{figure}[!h]
  \includegraphics[width=8.25cm]{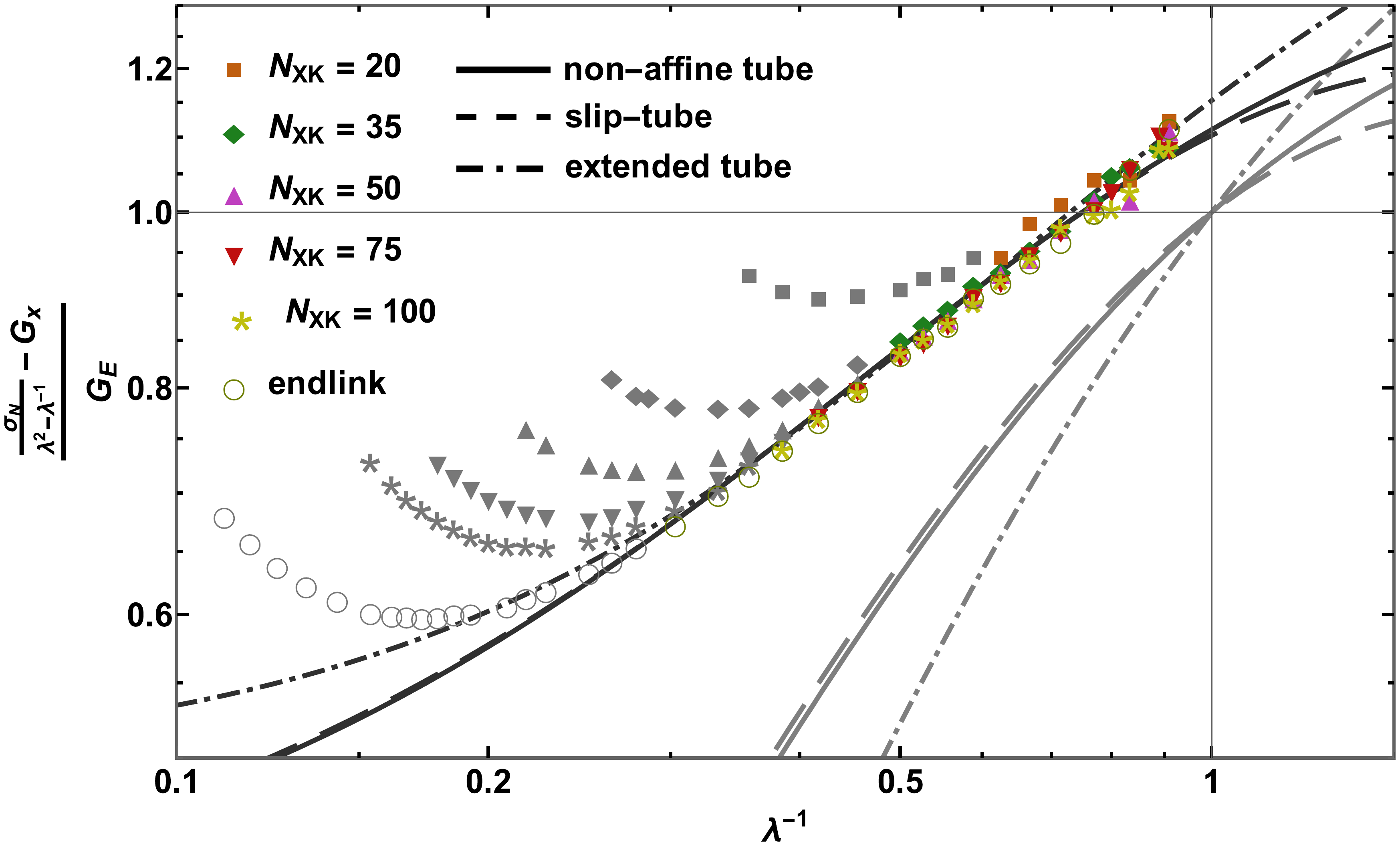}
  \caption{Universal representation of the simulation stress-strain data along with the parameter free predictions (gray lines) and fits of the microscopic elasticity models (black lines). Solid, large dashed and dot-dashed lines illustrate the non-affine tube, the slip tube and the extended tube models, respectively. Correspondence of symbols colouring and markers to the model networks is the same as in Fig.~\ref{sbfgr:KG_MR_rnt_vs_iElF}, gray symbols denote stress data discarded due to finite extensibility effects.}
  \label{fgr:Micro_fit_lin_uni_rnt_vs_iElF}
\end{figure}

The non-affine tube model, the slip tube model and the extended tube model predict reduced normal tensions in the form $\widetilde{\sigma}_N(\lambda)= \mathrm{G}_X + \mathrm{G}_E\,\varphi(\lambda)$, where $\varphi(\lambda)$ is a model specific universal function. Consequently, $\varphi(\lambda)$ can be isolated as:~\cite{Kluppel1992}
\begin{equation*}
 \varphi(\lambda) = \dfrac{\widetilde{\sigma}(\lambda)_N - \mathrm{G}_X}{\mathrm{G}_E}\,,
\end{equation*}
independently of the network structure.

In Fig.~\ref{fgr:Micro_fit_lin_uni_rnt_vs_iElF}, we plot the simulation stress-strain data in this reduced form using the cross-link and the entanglement moduli provided independently by the Primitive Path methods. We observe an excellent collapse of that range of the data, which is not affected by finite extensibility. The most strongly cross-linked network $N_K=10$ does not fall on the universal curve~(these data are not shown), and small deviations are observed for $N_K=20$. We attribute these effects to the onset of glassy dynamics, where entropic elasticity theory is not applicable. For the remaining data range, obviously, no collapse can be expected, since finite extensibility effects depend on the specific chain length distribution of a given network.

To observe universality of experimental data, often, a theory is fitted to stress-strain data and, afterwards, the experimental data are plotted using the fitted cross-link and the entanglement moduli, see, e.g. Ref.~\cite{RubinPanyuk2002ElastPolymNetw}. This essentially forces the data to collapse around the theoretical prediction. Here our aim is to compare the theories to our simulation stress-strain data and not vise versa. We emphasise that the observed collapse is \emph{independent} of any microscopic elasticity theory.

Having \emph{model independent} estimates of the cross-link and the entanglement moduli, one can ask whether the microscopic theories can make parameter free predictions of the simulation stress-strain data. The gray lines show the default, parameter free predictions for the strain dependent entanglement contributions $\varphi(\lambda)$ of the non-affine tube model, the slip tube model and the extended tube model. These predictions correspond to the naive identification of the model parameters $\mathrm{G}_X^{model}=\mathrm{G}_X$, $\mathrm{G}_E^{model}=\mathrm{G}_E$ and they are observed to be in poor agreement with the simulation data. The models assume additivity of entanglement and cross-link effects, and hence by construction fall $5-20\%$ short of predicting the correct shear modulus. Furthermore, the slope of the universal functions is observed to be too large. If the models correctly predicted the shear modulus, we could attempt to their stress-strain predictions by shifting part of the entanglement modulus into the cross-link modulus, thereby changing the slope, while keeping the intercept at $\lambda=1$ fixed. The plot suggests, this is not sufficient, and both a scaling~(entanglement modulus) and a shift~(cross-link modulus) should be adjusted simultaneously for the models to agree with the simulation data.

Since the direct identification of the fit parameters of the microscopic models with the Primitive Path estimates did not work, we approximated the collapsed stress-strain data in the universal representation by two-parameter fits $\gamma_X + \gamma_E\,\varphi(\lambda)$, where $\varphi(\lambda)$ is the model specific function~(shown as the black lines) and $\gamma_X$, $\gamma_E$ are prefactors given by $\mathrm{G}_X^{model}=\gamma_X\,\mathrm{G}_X$, $\mathrm{G}_E^{model}=\gamma_E\,\mathrm{G}_E$. As a result, we obtained an excellent agreement between the theories and the collapsed stress-strain data. Fit qality is quantified by the reduced chi-square $\chi_{\nu}^2$, and we obtained $\chi_{\nu}^2$ values $6.53$, $6.66$, $7.54$ for the non-affine tube, the slip tube and the extended tube models, respectively. As an empirical parameterization of our stress-strain data, we ''recalibrated'' non-affine tube model to obtain the universal function $\varphi(\lambda)=0.39+0.72/\left(\lambda-\lambda^{1/2}+1\right)$, which is in excellent agreement with the simulation data.

%% file: 4.3.2.FitLinearData.tex
\subsubsection{Fitting the data not affected by finite extensibility}
\label{sbsbsct:Fit_linear_data}

\begin{figure}[!h]
  \begin{subfigure}[!h]{8.2cm}
    \includegraphics[width=\textwidth]{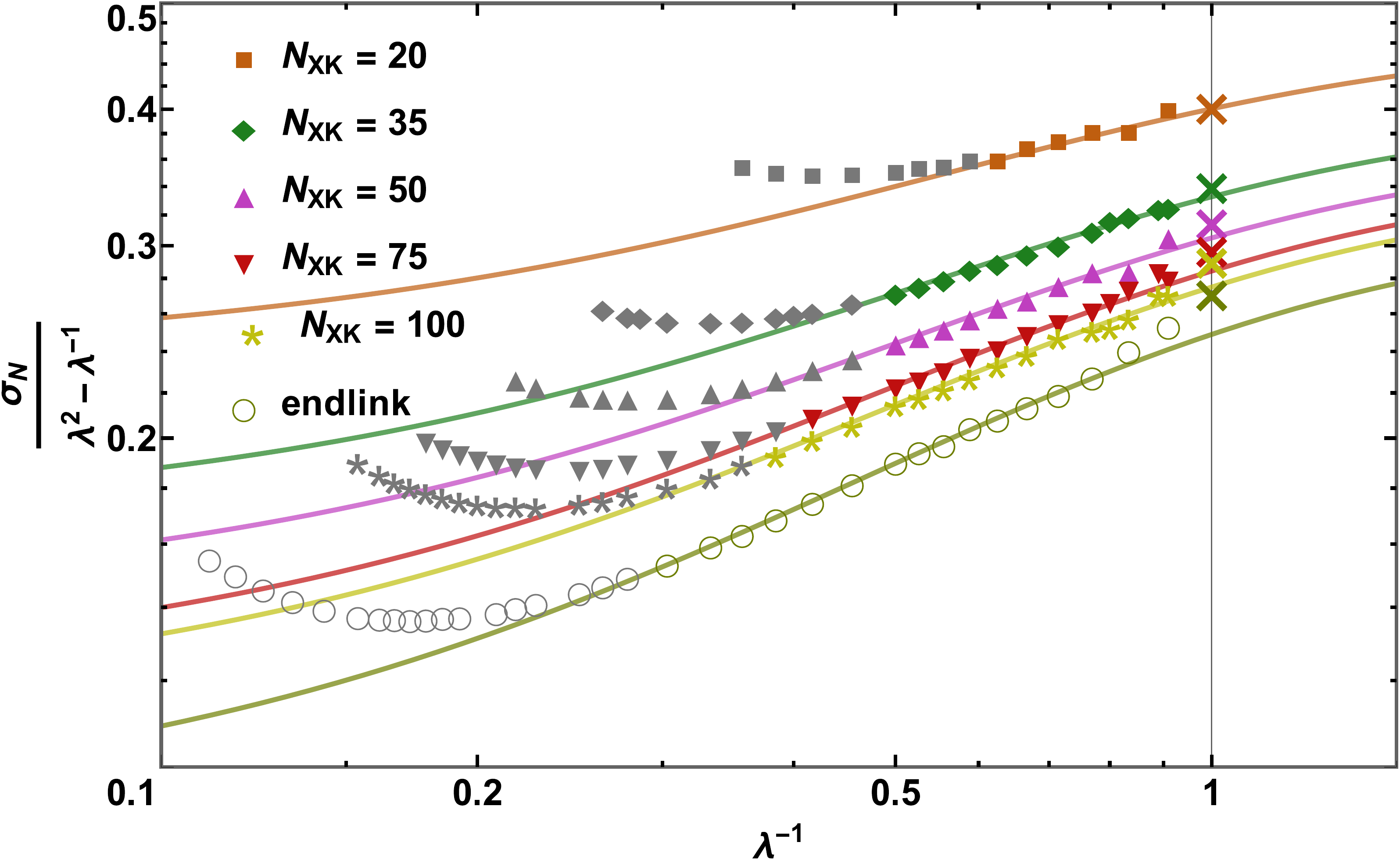}
    \caption{non-affine tube model}
    \label{sbfgr:2p_fit_lin_NffT}
  \end{subfigure}
  \begin{subfigure}{8.2cm}
    \includegraphics[width=\textwidth]{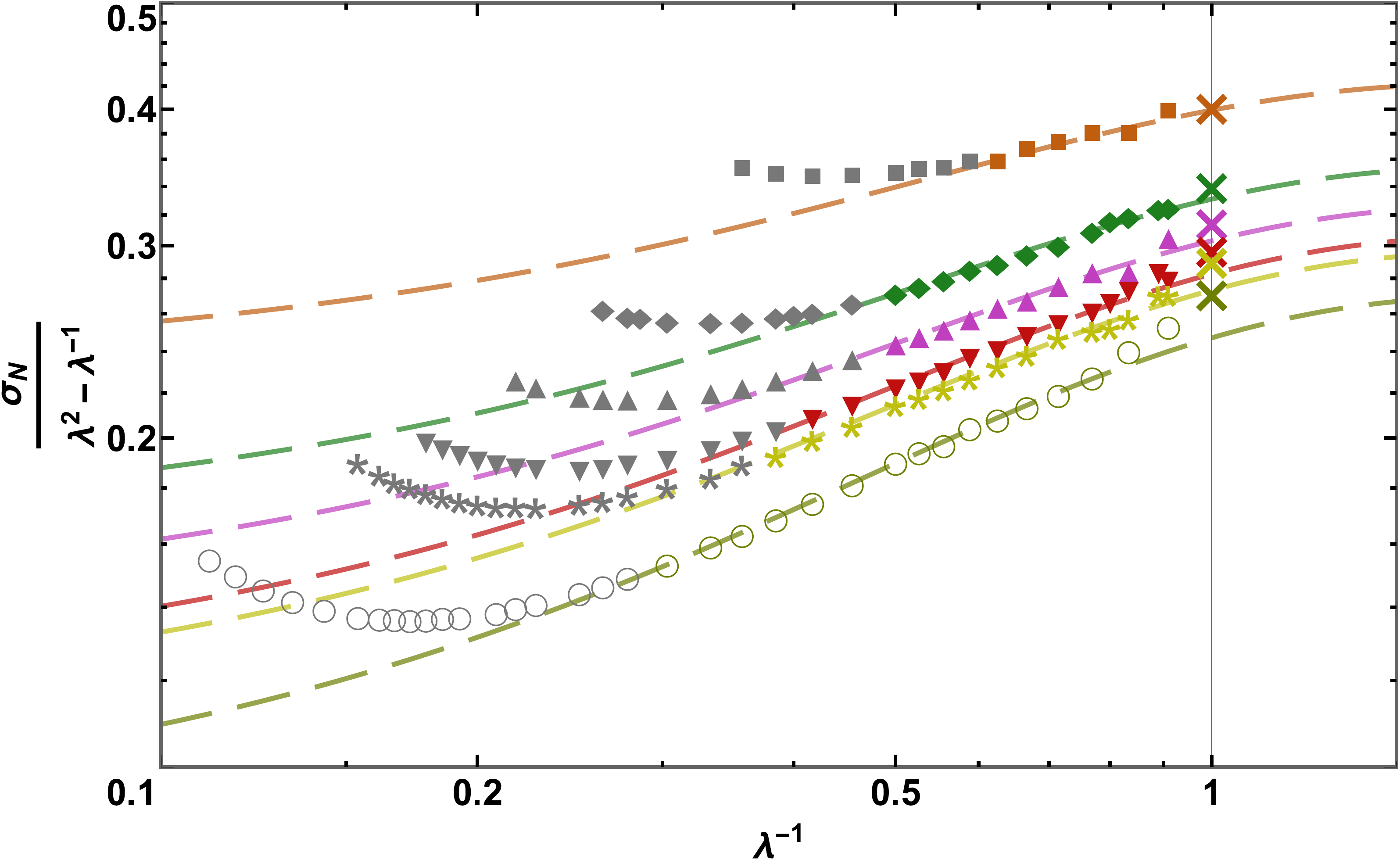}
    \caption{slip tube model}
    \label{sbfgr:2p_fit_lin_SlT}
  \end{subfigure}
  \begin{subfigure}{8.2cm}
    \includegraphics[width=\textwidth]{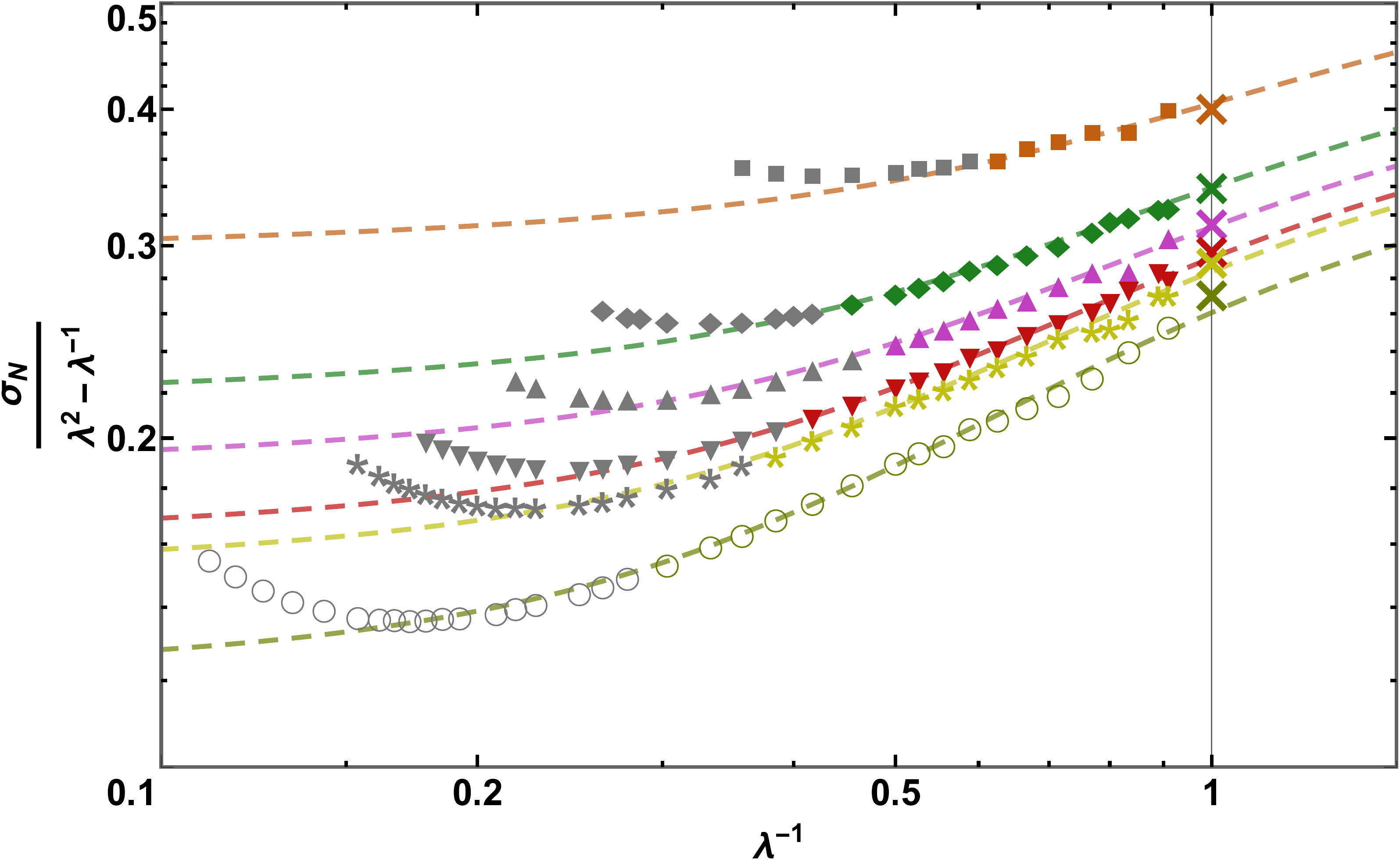}
    \caption{double tube model}
    \label{sbfgr:2p_fit_lin_DT}
  \end{subfigure}
  \begin{subfigure}{8.2cm}
    \includegraphics[width=\textwidth]{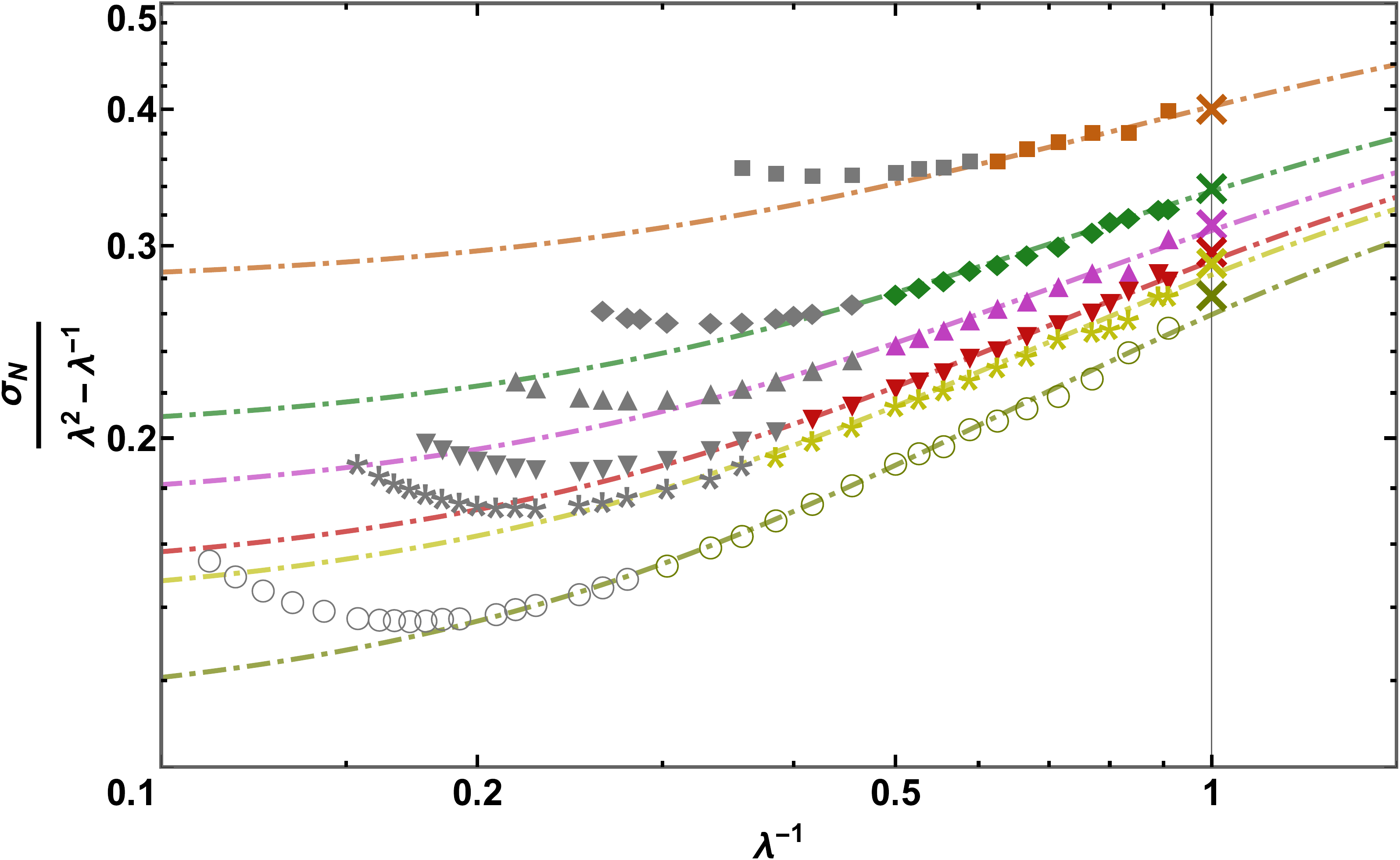}
    \caption{extended tube model}
    \label{sbfgr:2p_fit_lin_ExT}
  \end{subfigure}
  \caption{Fits of the microscopic elasticity theories to the simulation stress-strain data not affected by finite extensibility. Coloured crosses located at the vertical line $\lambda^{-1}=1$ indicate the MR estimates of the shear modulus.}
  \label{fgr:Micro_fit_lin_rnt_vs_iElF}
\end{figure}

\begin{figure}[!h]
  \begin{subfigure}{5.5cm}
    \includegraphics[width=\textwidth]{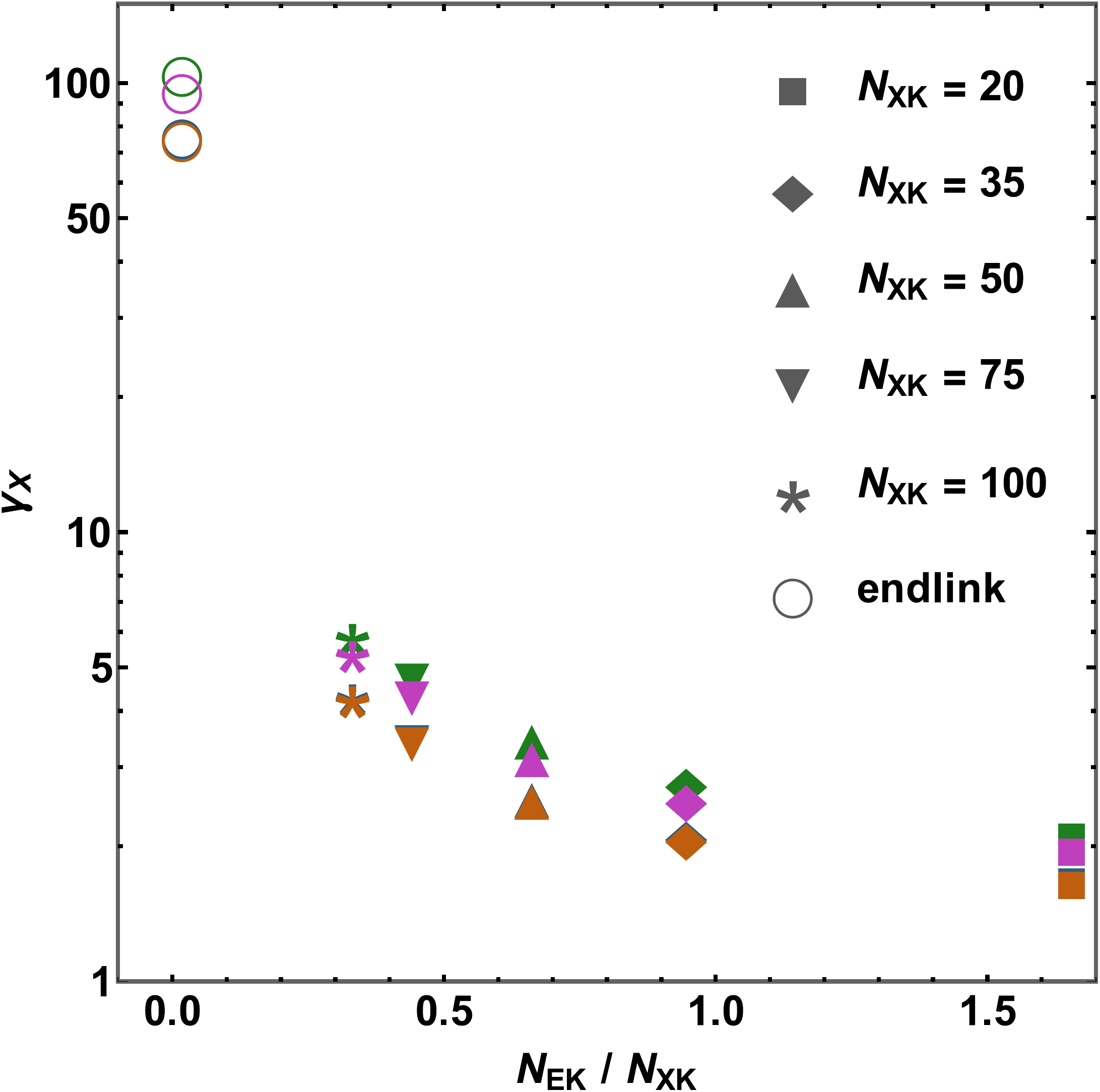}
    \caption{}
    \label{sbfgr:2p_fit_lin_gammaX}
  \end{subfigure}

  \begin{subfigure}{5.5cm}
    \includegraphics[width=\textwidth]{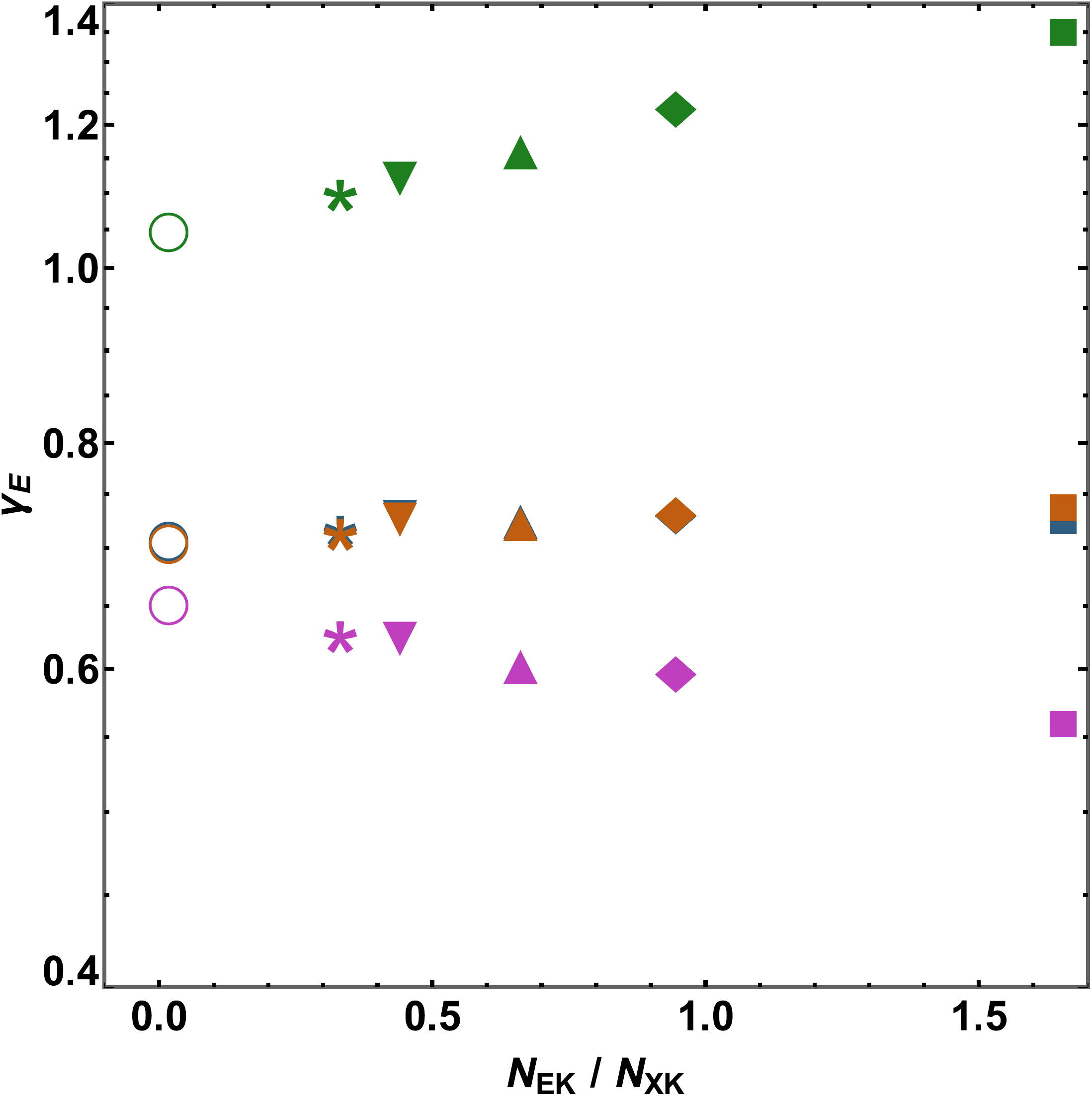}
    \caption{}
    \label{sbfgr:2p_fit_lin_gammaE}
  \end{subfigure}
  \caption{Fitting parameters \subref{sbfgr:2p_fit_lin_gammaX}: $\gamma_X$, \subref{sbfgr:2p_fit_lin_gammaE}: $\gamma_E$. Blue, orange, green and purple symbols show the optimal values for the non-affine tube, the slip tube, the double tube and the extended tube models, respectively.}
  \label{fgr:Micro_fit_lin_parameters}
\end{figure}

We fit the microscopic elasticity theories that do not account for finite extensibility effects to the corresponding stress-strain data. For the fitting, two parameters $\gamma_X$, $\gamma_E$ were introduced as multipliers of $\mathrm{G}_X$, $\mathrm{G}_E$, respectively. Fitted data are illustrated by coloured symbols, whereas gray symbols indicate the data discarded due to significant finite extensibility effects. We chose the non-affine tube model~\cite{RubinPanyuk1997NFFDfrmElstctPlmNtwk}, the slip tube model~\cite{RubinPanyuk2002ElastPolymNetw}, and the double tube model~\cite{MergellEveraers2001TubeModels}, because they do not capture finite extensibility. The extended tube model~\cite{KaliskeHeinrich1999ExtendedTubeModel} has two additional parameters $\beta$, which roughly accounts for network defects, and $\delta$ for finite extensibility. We set $\beta=1$ as our model networks are defect free in that sense that they do not have dangling ends (see Sect.~\ref{sbsct:Micro}), and we set $\delta=0$, since we are discarding simulation data affected by finite extensibility. The non-affine network model of Davidson and Goulbourne~\cite{DavidsonGoulbourne2013NffNtwkModel} and the general constitutive model of Xiang et al.~\cite{XiangZhongWangMaoYuQu2018GnrlModelSoftElast} were not fitted, since in the limit, where finite extensibility effects can ne neglected, the models reduce to the non-affine tube model and to the extended tube model with $\delta=0$, respectively.

The model fits are shown in Fig.~\ref{fgr:Micro_fit_lin_rnt_vs_iElF}. We observe an excellent agreement with the stress-strain data. The reduced chi-square $\chi_{\nu}^2$ was $2.51$, $2.70$, $3.54$ and $2.63$ for the non-affine tube, the slip tube, the double tube and the extended tube model fits, respectively, hence, all fits are of comparable quality. The fits provided new estimates for the shear modulus, which was equal to $\gamma_X\,\mathrm{G}_X + \gamma_E\,\mathrm{G}_E$ for the non-affine tube, the slip tube and the extended tube models and was equal to:
\begin{equation*}
  \left(\left(\gamma_X\,\mathrm{G}_X\right)^2 + 2\,\left(\gamma_E\,\mathrm{G}_E\right)^2\right)\,\left(\sqrt{\left(\gamma_X\,\mathrm{G}_X\right)^2 + 4\,\left(\gamma_E\,\mathrm{G}_E\right)^2}\right)^{-1/2}
\end{equation*}
for the double tube model. The shear modulus estimates can be seen in Fig.~\ref{fgr:Micro_fit_lin_rnt_vs_iElF} as intersections of the coloured fitting curves with the vertical line $x=1$. They are all observed to be in good agreement with the MR estimates shown as big coloured crosses.

In Fig.~\ref{fgr:Micro_fit_lin_parameters}, optimal values for the fit parameters $\gamma_X$, $\gamma_E$ are shown. All model fits show that $\gamma_X$ decreases strongly with increasing cross-link density and is located within the range $2-6$ for almost all model networks. The only exception is the end-linked network, for which we observe that $\gamma_X$ has huge values of the order $10^2$. We attribute this to the tiny cross-link modulus $\mathrm{G}_X$ of the end-linked model network, which is used for the normalization. Consequently, though the data for the end-linked can be fitted with high accuracy by any microscopic model, none of them provides a reasonable accurate estimate for the cross-link modulus. At the same time, all fits are consistent with $\gamma_E$ in the range $0.5-1.4$. The double tube model fit shows the growth of $\gamma_E$ with increasing cross-link density, whereas all other model fits demonstrate a roughyl constant or weakly decreasing trend. Note that the fitting results of the slip tube and the non-affine models are identical. As noted in Sect.~\ref{sbsct:Micro}, the slip tube model allows for the chain contour length redistribution between parallel and perpendicular tube sections upon deformation. This is expected to important, especially for the weakly cross-linked networks. Nevertheless, this model feature seems to have no influence on the model fit results.

The microscopic models introduce cross-link and entanglement localization effects using different mathematical approximations, hence, one can not expect the corresponding fit parameters to produce exactly the same values. We observe agreement on a scaling level only with a $O(1)$ prefactor. By fitting the models to the stress-strain data for completely characterized systems, where the cross-link and entanglement moduli are known independently of any microscopic model, we have in essense performed a calibration of these models to a Primitive Path reference standard. Consequently, from the fit parameters, one can attempt to estimate the correct cross-link and the entanglement moduli as $\gamma_X\,\mathrm{G}_X$, $\gamma_E\,\mathrm{G}_E$.

%% file: 4.3.3.FitFullData.tex
\subsubsection{Fitting full range of the data}
\label{sbsbsct:Fit_full_data}

\begin{figure}[!h]
  \begin{subfigure}{8.25cm}
    \includegraphics[width=\textwidth]{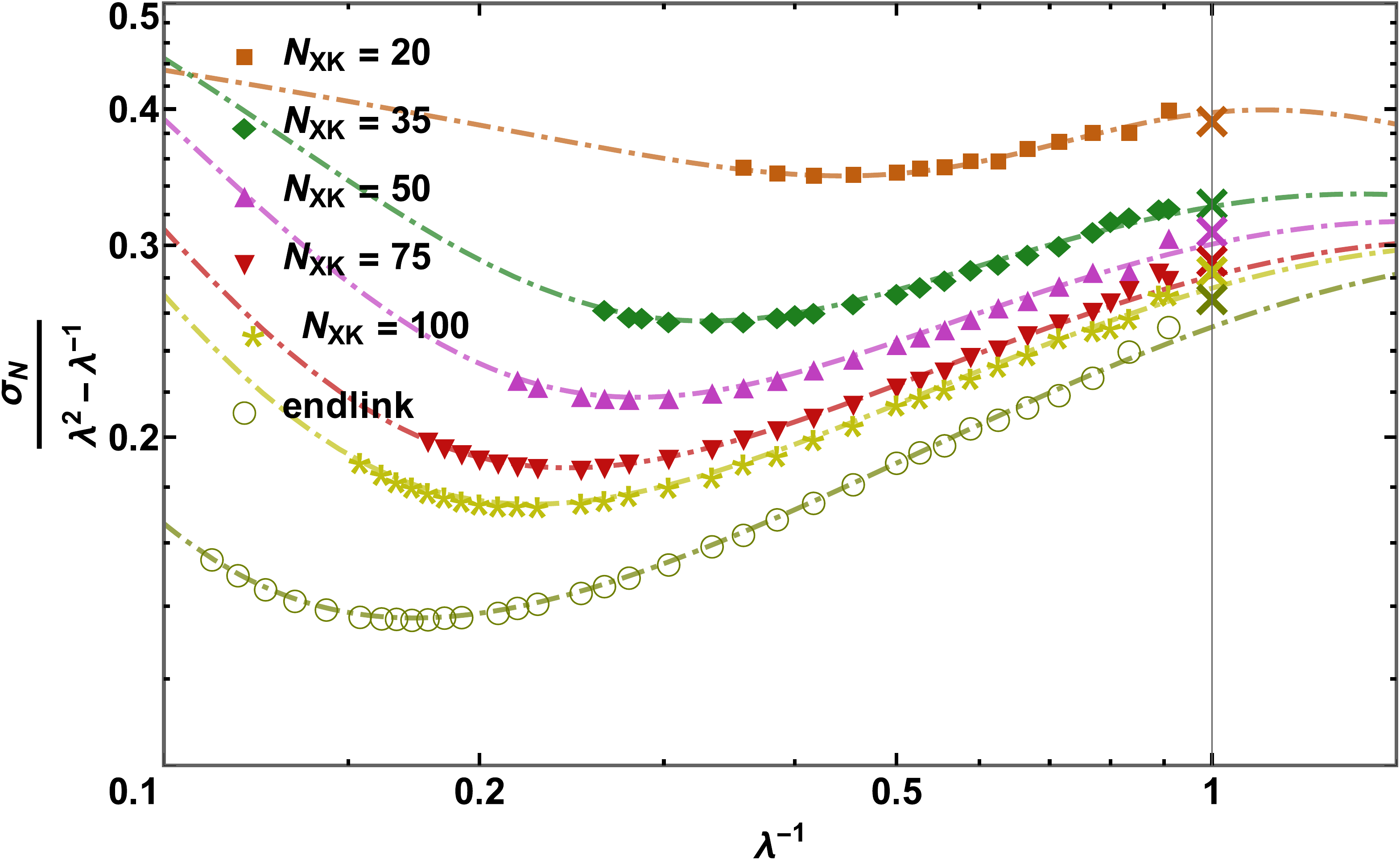}
    \caption{extended tube model of Kaliske, Heinrich}
    \label{sbfgr:3p_fit_nlin_ExT}
  \end{subfigure}
  \begin{subfigure}{8.25cm}
    \includegraphics[width=\textwidth]{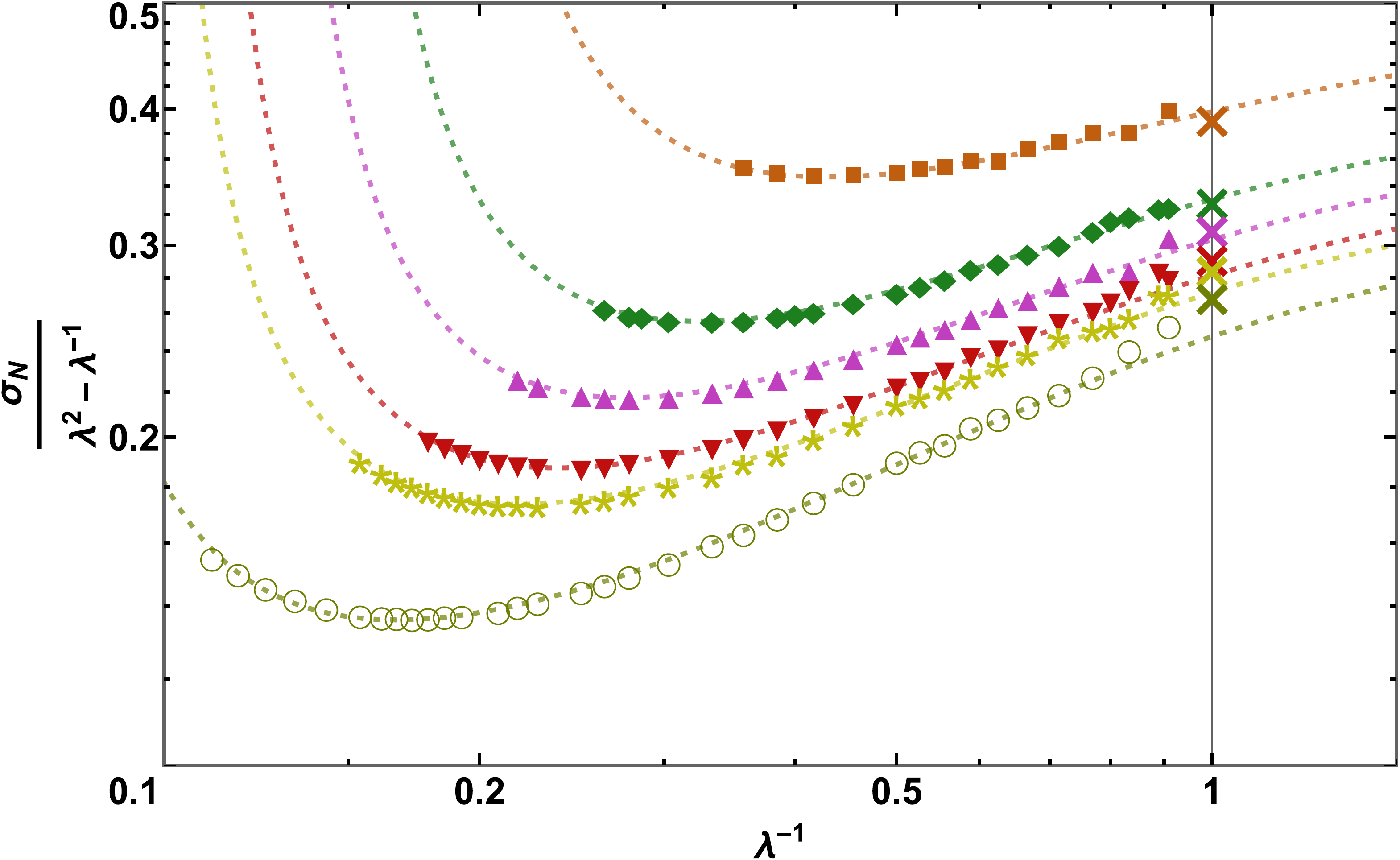}
    \caption{non-affine network model of Davidson, Goulbourne}
    \label{sbfgr:3p_fit_nlin_DG}
  \end{subfigure}
  \begin{subfigure}{8.25cm}
    \includegraphics[width=\textwidth]{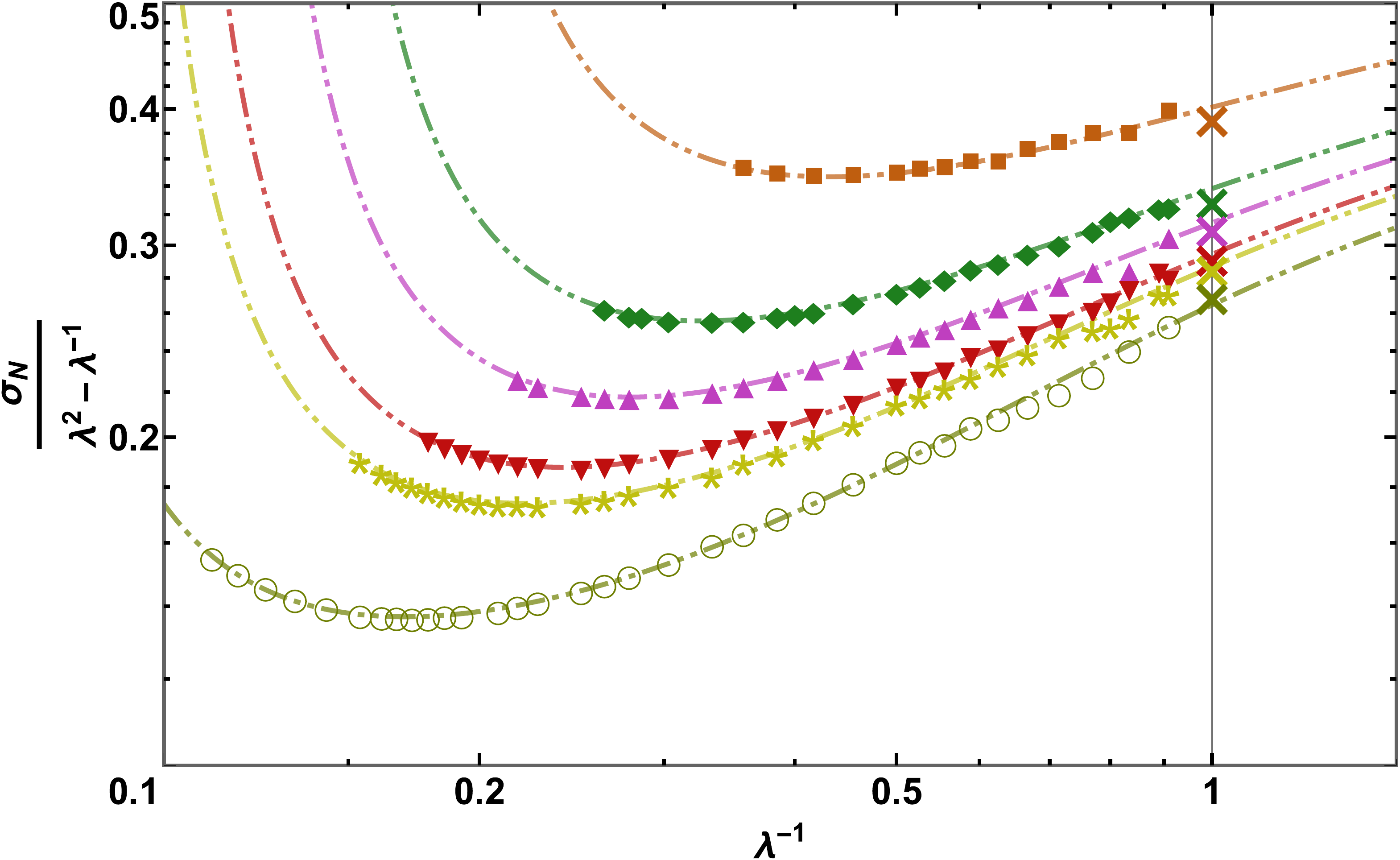}
    \caption{general constitutive model of Xiang et al.}
    \label{sbfgr:3p_fit_nlin_XI}
  \end{subfigure}
  \caption{Fits of the microscopic elasticity theories to the full range of the simulation stress-strain data. Coloured crosses located at the vertical line $\lambda^{-1}=1$ indicate the MR estimates of the shear modulus.}
  \label{fgr:Micro_fit_nlin_rnt_vs_iElF}
\end{figure}

\begin{figure}[!h]
  \begin{subfigure}{5.5cm}
    \includegraphics[width=\textwidth]{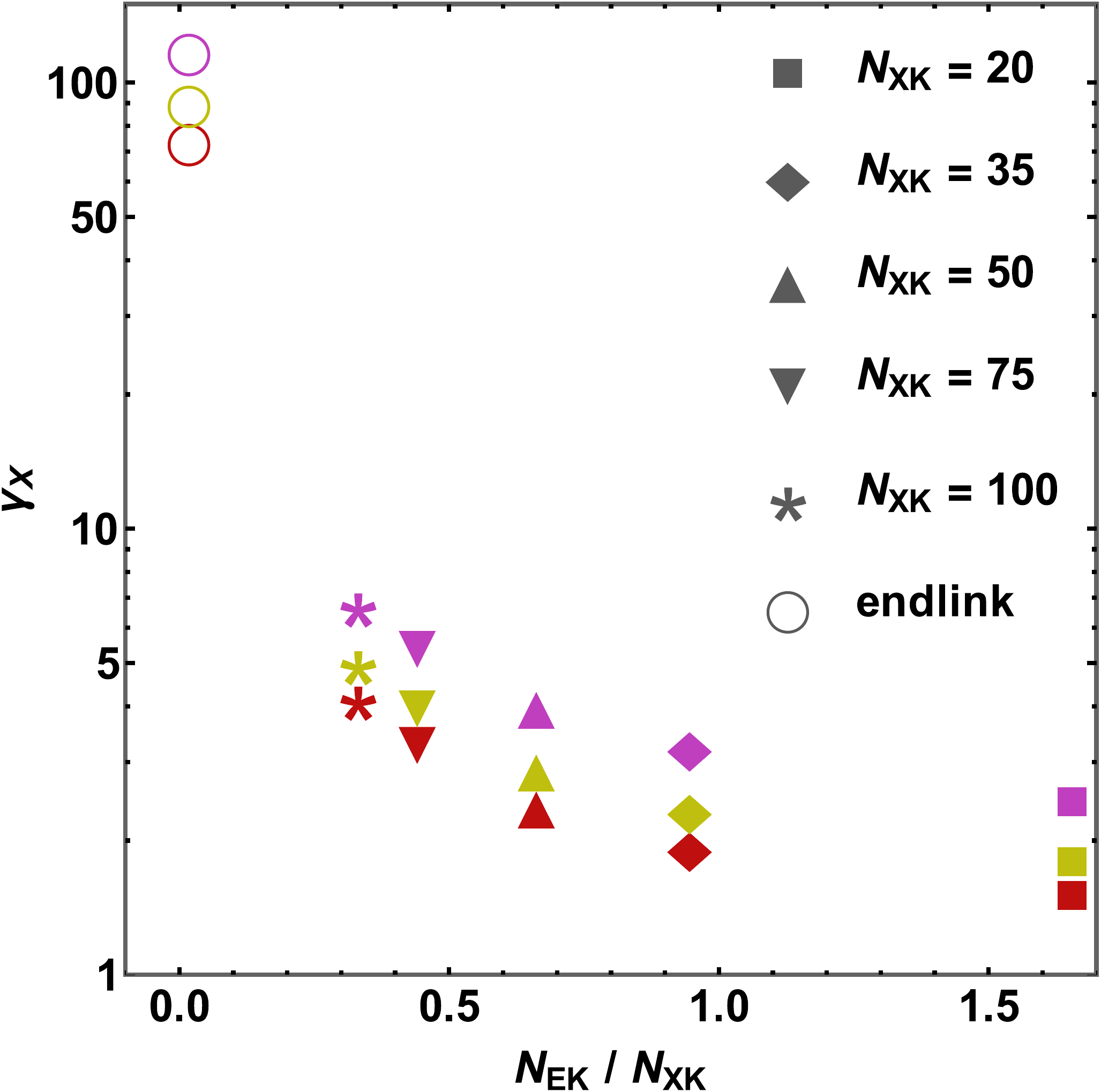}
    \caption{}
    \label{sbfgr:3p_fit_nlin_gammaX}
  \end{subfigure}

  \begin{subfigure}{5.5cm}
    \includegraphics[width=\textwidth]{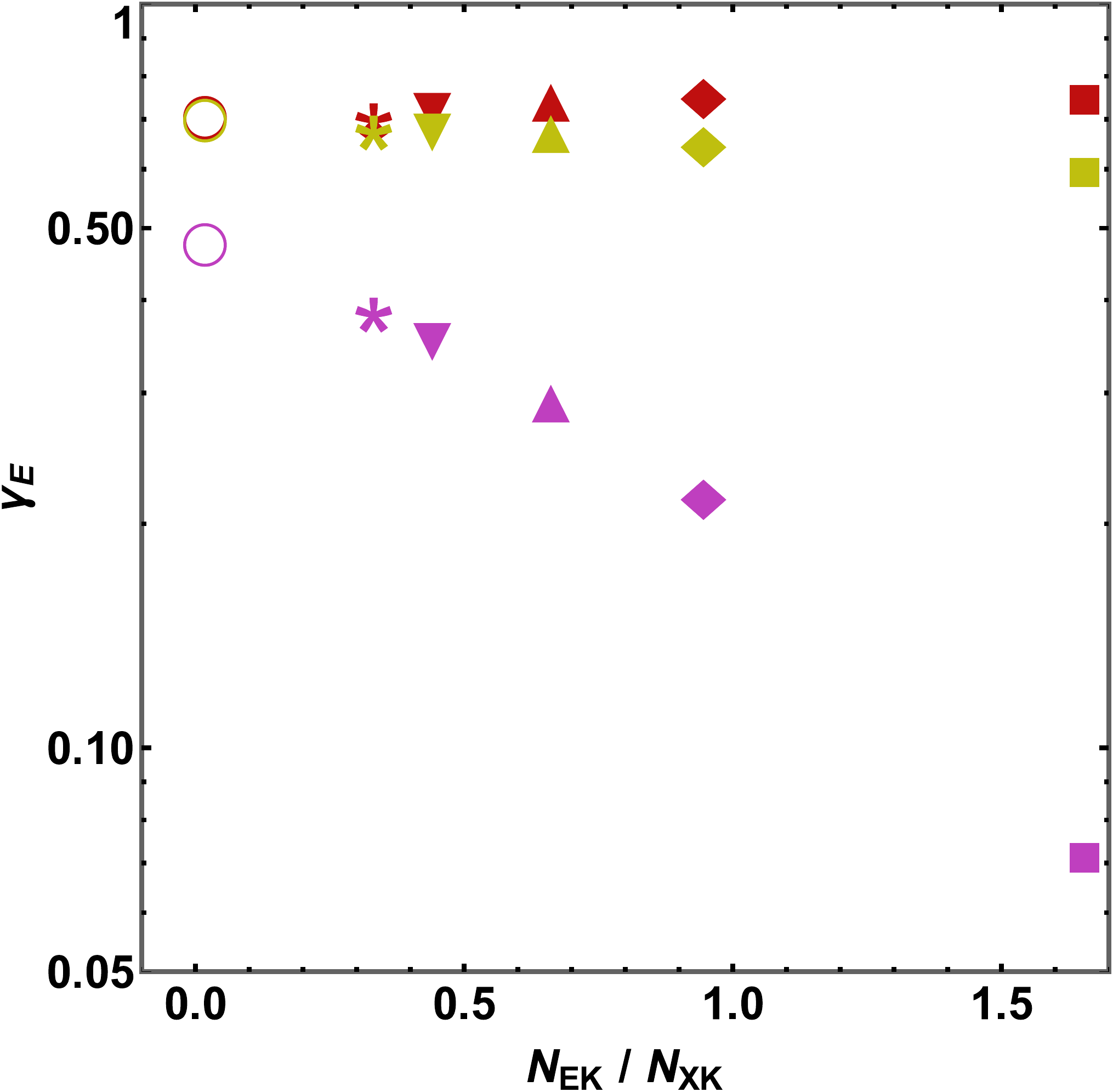}
    \caption{}
    \label{sbfgr:3p_fit_nlin_gammaE}
  \end{subfigure}

  \begin{subfigure}{5.5cm}
    \includegraphics[width=\textwidth]{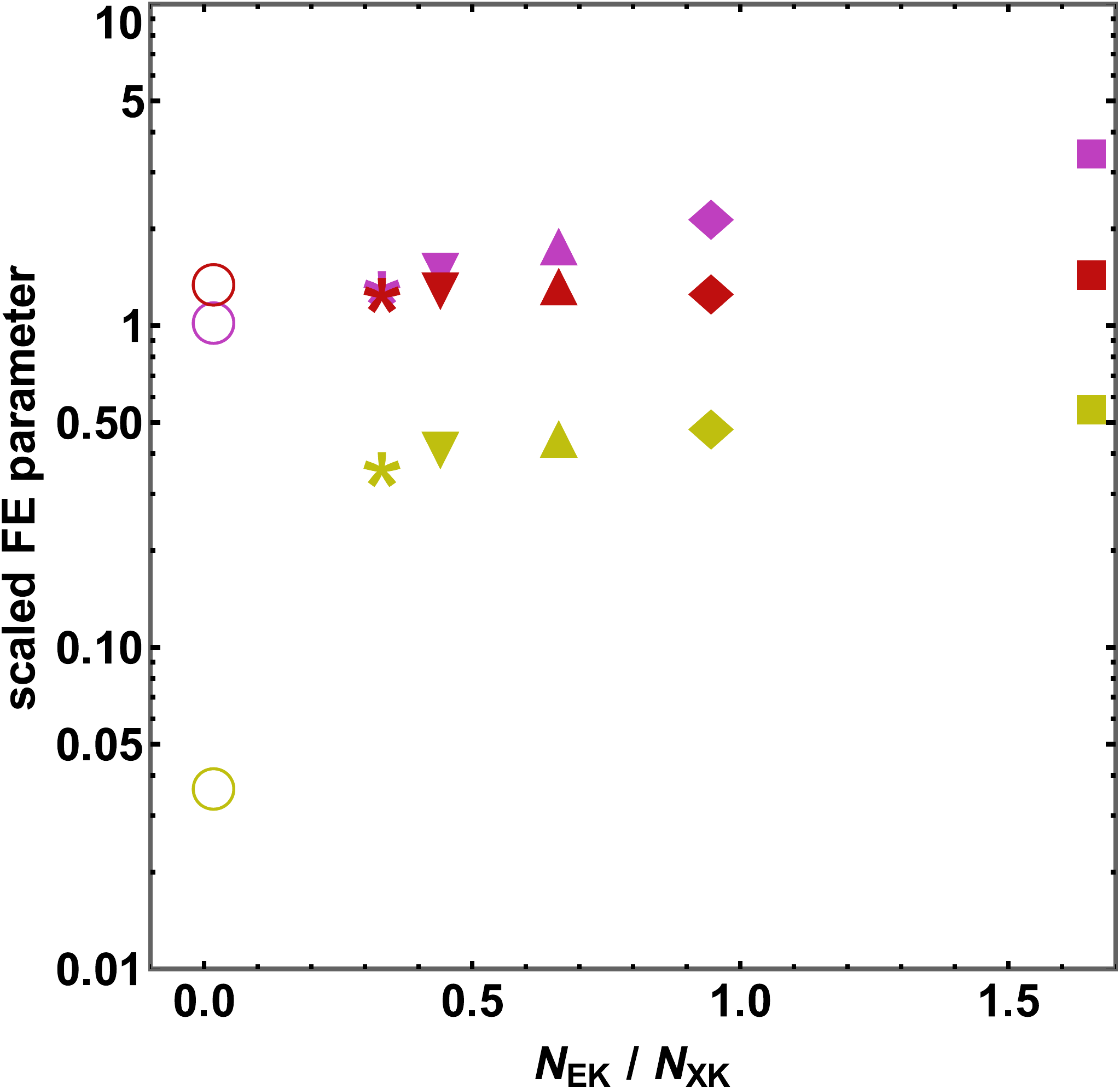}
    \caption{}
    \label{sbfgr:3p_fit_nlin_FE}
  \end{subfigure}
  \caption{Fitting parameters \subref{sbfgr:3p_fit_nlin_gammaX}: $\gamma_X$, \subref{sbfgr:3p_fit_nlin_gammaE}: $\gamma_E$, \subref{sbfgr:3p_fit_nlin_FE}: finite extensibility parameter. Purple, red and yellow symbols show the optimal values of the fitting parameters for the extended tube model of Kaliske and Heinrich, the non-affine network model of Davidson and Goulbourne and the general constitutive model of Xiang et al., respectively. The plot in Panel~\subref{sbfgr:3p_fit_nlin_FE} shows parameters responsible for the finite extensibility, $\alpha = \delta\,\sqrt{N_{EK}}$ for the extended tube model, $\lambda_{max}^{DG} / \lambda^{MR}_{max}$ for the Davidson and Goulbourne model, $N_{XK}^{XIA} / N_{XK}$ for the model of Xiang et al.}
  \label{fgr:Micro_fit_nlin_parameters}
\end{figure}

The KG polymer model includes finite extensibility effects due to the FENE potential used to model the bonds. At large macroscopic deformations, the upturn of the simulation stress-strain data induced by finite extensibility is observed, see Fig.~\ref{sbfgr:KG_MR_rnt_vs_iElF}.

In Fig.~\ref{fgr:Micro_fit_nlin_rnt_vs_iElF}, we show fits of the extended tube model of Kaliske and Heinrich~\cite{KaliskeHeinrich1999ExtendedTubeModel}, the non-affine network model of Davidson and Goulbourne~\cite{DavidsonGoulbourne2013NffNtwkModel}, and the general constitutive model of Xiang et al.~\cite{XiangZhongWangMaoYuQu2018GnrlModelSoftElast} to the full range of simulation data. The fitting was performed in a similar way as in Sect.~\ref{sbsbsct:Fit_linear_data}. In addition to the fit parameters $\gamma_X$, $\gamma_E$, additional model specific parameters responsible for finite extensibility were also fitted. These are $\delta$, $\lambda_{max}$, $N_{XK}$ for Kaliske and Heinrich, Davidson and Goulbourne, Xiang et al. models, respectively. We set $\beta=1$ for the extended tube model as before. The resulting fits~(illustrated as coloured lines) are observed to be in perfect agreement with the simulation data. The reduced chi-square values are $2.95$, $7.61$ and $3.39$ for the extended tube, the non-affine network and the general constitutive models, respectively. While the model fits are able to reproduce the stress upturn at large deformations, they also provide estimates for the shear modulus~(intersections of the coloured fitting curves with the vertical line $x=1$ in Fig.~\ref{fgr:Micro_fit_lin_rnt_vs_iElF}), which are in good quantitative agreement with the MR estimates $\mathrm{G}^{MR}$~(shown as big coloured crosses).

In Fig.~\ref{fgr:Micro_fit_nlin_parameters}, the fit parameters are shown. Similarly to Sect.~\ref{sbsbsct:Fit_linear_data}, for all model fits, the value of $\gamma_X$ varies within the range $1.5-6.9$, it decreases with increasing cross-link density and has a huge value of the order $10^2$ for the end-linked model network. For the fit parameter $\gamma_E$, the extended tube model fit shows a strong decrease from $0.5$ down to $0.07$. At the same time, both the non-affine network model of Davidson and Goulbourne and the general constitutive model of Xiang et al. demonstrate almost constant values of $\gamma_E \approx 0.6-0.7$. For the extended tube model, we plot $\alpha = \delta\,\sqrt{N_{EK}}$ instead of $\delta$. In Ref.~\cite{KaliskeHeinrich1999ExtendedTubeModel}, the parameter $\alpha$ is introduced as a measure of the network chains inextensibility, and it is claimed that $\alpha \in (0;\,1)$. However, the model fit shows that $\alpha$ rather varies within range $1-3.5$. For the non-affine network model, we plot the ratio $\lambda_{max}^{DG} / \lambda^{MR}_{max}$. We observe that $\lambda_{max}^{DG} / \lambda^{MR}_{max} \approx 1.3$ for all model networks. For the general constitutive model, we observe that the ratio $N_{XK}^{Xiang} / \left<N_{XK}\right>$ is within the range $0.3-0.5$. The exception is the end-linked network, where $N_{XK}^{Xiang} / \left<N_{XK}\right> \approx 0.04$, presumably, due to large value of the average network strand length $\left<N_{XK}\right> \approx 1850$.

%% file: 5.Conclusion.tex
\section{Conclusion}
\label{sct:Conclusion}

In this work, we studied the relation between the macroscopic mechanical properties and microscopic structure of Kremer-Grest~(KG) model polymer networks. We used a PDMS-KG polymer model~\cite{SvaneborgkEveraers2020KGModel, EveraersKarimiFleckHojdisSvaneborg2020KGMap} designed to match the Kuhn number of PDMS and, hence, to reproduce, e.g. the packing length, the entanglement length and the shear modulus of PDMS. We generated well equilibrated precursor melts, linked the chain ends and then proceeded to cross-link random bead pairs. The resulting model networks were free of dangling ends, but contained loops mostly due to intramolecular cross-linking. The networks had different number of cross-links, hence, ranging from cross-link to entanglement dominated elastic response. We characterized the networks in terms of strand length and cross-link functionality distributions.

We performed Molecular Dynamics simulations of uniaxial stretching of the model networks and carefully estimated the equilibrium stresses in the deformed states to obtain stress-strain curves. We invested in excess of $200$ core years of computer time on stress relaxation of the deformed networks states.

We applied Primitive Path methods~\cite{EveraersSukuGrestSvaneborgSivaKremer2004RheolTopol, SvaneborgEveraersGrestCurro2008StressContributions} to estimate the entanglement modulus  of the precursor melt and the cross-link moduli of the networks. The entanglement modulus was found to be in a good agreement with the literature data.~\cite{FettersLohseColby2007ChainDimEntaSpac} A variation of the PPA analysis named Phantom Primitive Path Analysis~(3PA)~\cite{SvaneborgEveraersGrestCurro2008StressContributions} was presented, and we applied it to the deformed model networks to estimate the cross-link moduli. The results turned out to be in excellent agreement with the phantom network model estimates based on network analysis.

To estimate the shear moduli, we applied the Mooney-Rivlin~(MR) model. Mooney-Rivlin representation of the data clearly demonstrated linear dependency of the reduced normal tension on the inverse elongation that justifies the applicability of the MR model analysis to estimation of the network moduli.

We observed that the balance of the MR model parameters $\mathrm{C}_1$, $\mathrm{C}_2$ depends on the level of the network cross-linking. The most strongly cross-linked network was accurately described only by the $\mathrm{C}_1$ parameter. With increasing average network strand length, the influence of the $\mathrm{C}_2$ parameter concomitantly increased, and, at the point $N_{EK} = 0.5\,N_{XK}$, the contributions of $\mathrm{C}_1$, $\mathrm{C}_2$ to the shear modulus are approximately equal. For the most weakly cross-linked network, the relative contributions from the $\mathrm{C}_1$, $\mathrm{C}_2$ terms to the shear modulus were observed to be $40\% / 60\%$, respectively.

We compared the MR modulus estimates with the sum of the cross-link and entanglement moduli and observed good agreement. The MR moduli were systematically $5-20\%$ larger than the sums. This partially supports the assumption often used in microscopic theories of elasticity of polymer networks that the effects of cross-links and entanglements are additive, see e.g. Refs.~\cite{RubinPanyuk1997NFFDfrmElstctPlmNtwk, RubinPanyuk2002ElastPolymNetw, KaliskeHeinrich1999ExtendedTubeModel}. We attribute this deviation to the entanglement trapping. Langley~\cite{Langley1968} proposed a trapping factor $\mathrm{T}_e$, where a fraction of the entanglement modulus contribute to the total modulus of a polymer network. Our data are consistent with a tiny trapping factor $\mathrm{T}_e \leq 0.2$ independent of network structure.

Plotting the simulation stress-strain data in a universal form, i.e. subtracting the cross-link modulus and normalizing by the entanglement modulus, we observed a collapse of the data to a univeral curve. The collapse only occurs for stress data not affected by finite extensibility, which is system specific. This collapse was already anticipated from several microscopic models, see e.g. Refs.~\cite{
EdwardsViglis1986EffOfEntanglRubbElast,
RubinPanyuk1997NFFDfrmElstctPlmNtwk,
KaliskeHeinrich1999ExtendedTubeModel,
RubinPanyuk2002ElastPolymNetw}.
We note that the observed collapse is independent of any microscopic model assumptions, since the cross-link and entanglement moduli were obtained via the Primitive Path methods and not via fitting models to the simulation data. As a calibration standard, we obtained an empirical estimate for the universal function of $\varphi(\lambda)=0.38+0.73/\left(\lambda-\lambda^{1/2}+1\right)$ for PDMS networks that describe our data well.

We tested the quality of parameter free predictions of the non-affine tube model~\cite{RubinPanyuk1997NFFDfrmElstctPlmNtwk}, the slip tube model~\cite{RubinColby2003PolymerPhysics}, and the extended tube model~\cite{KaliskeHeinrich1999ExtendedTubeModel}. For this purpose, we identified the model parameters with the entanglement and the cross-link moduli estimates obtained by the Primitive Path methods and compared the resulting stress-strain predictions to the simulation data. All the models failed in description the data. Therefore, we proceeded to fit the models as well as the double tube model~\cite{MergellEveraers2001TubeModels} to the simulation data, which were not affected by finite extensibility. We performed two-parameter fitting by allowing the prefactors of the entanglement and the cross-link moduli to vary. We observed an excellent agreement between the resulting model fits and the stress-strain data and obtained accurate estimates of the shear moduli. However, obtained estimates for the cross-link and the entanglement moduli varied significantly when compared to the benchmarks provided by the Primitive Path methods. We did not observe quantitative agreement, but the prefactors agreed within a $O(1)$ scaling. The entanglement prefactor $\gamma_E$ was in the range of $0.5-1.4$ and was roughly constant for all models. On the other hand, the cross-link prefactor $\gamma_X$ varied within the range $2-6$ and showed a strong decreasing trend with increasing density of cross-links.

Finally, we fitted the extended tube model, the non-affine network model~\cite{DavidsonGoulbourne2013NffNtwkModel} and the general constitutive model~\cite{XiangZhongWangMaoYuQu2018GnrlModelSoftElast} to the full range of the simulation stress-strain data. Each model has an additional parameter to account for finite extensibility effects, bringing the number of fit parameters to three. We observed excellent fits including the stress upturns at large deformations due to finite extensibility effects. Moreover, obtained estimates for the shear moduli were in close agreement with the MR estimates. Nevertheless, the model estimates for the cross-link modulus were $1.5-7$ too large compared to the estimates provided by the 3PA analysis. At the same time, the model estimates for the entanglement modulus were within $0.5-2$ range of the PPA analysis estimate. Again, we observed a significant decreasing trend of the cross-link prefactors $\gamma_X$ with increasing cross-link density. The entanglement prefactors $\gamma_E$ were roughly constant, except for the extended tube model, where it decreased strongly with increasing cross-link density. The finite extensibility parameters were compared to our estimates and were observed to be $O(1)$ and roughly constant. The model that provided the best fit was the non-affine network model of Davidson and Goulbourne, it showed the smallest prefactor for the cross-link moduli, a nearly network independent prefactor for the entanglement moduli, and also a network independent scaling factor for the finite extensibility effects. The reduced chi-square varied from $2.5$ to $7.5$ for all fits and, hence, the models describe the simulation data equally accurately. We note that, in essence, we performed a calibration of the model parameters to an independent standard offered by the results of the Primitive Path analysis methods applied to our model PDMS networks.

From a computational perspective, the Primitive Path methods have the advantage as computationally they are orders of magnitude cheaper than generation of an equilibrated stress-strain data for the MR analysis. For the latter, one needs a large number of well equilibrated deformed states of a model network, where the data is not affected by finite extensibility effects. Moremover, time consuming simulations are required for accurate estimation of the equilibrium stresses due to the slow relaxation dynamics. For instance, the virial stress tensor is a strongly fluctuating quantity for KG melts due to the hard interaction potentials. The good agreement between the MR estimates and the Primitive Path estimates of the shear moduli suggests that the latter provide a computationally effective alternative for the moduli estimation.

The present results are based only on an analysis of PDMS model KG polymer networks, which represents a single data point in terms of entanglement length/modulus. We are currently generating networks for other polymer models and will also be cross-linking networks in the swollen state to generalize the present results to the situation where the entanglement length also is systematically varied. We are also performing multiple mode deformations of the present networks to provide an even more stringent test data for future theories of rubber elasticity. We hope that our simulation data can offer a benchmark as the reference data for future development of the models that attempt to describe both entanglement and cross-link contributions to polymer network elasticity.

%% file: Main.bbl
\providecommand{\latin}[1]{#1}
\providecommand*\mcitethebibliography{\thebibliography}
\csname @ifundefined\endcsname{endmcitethebibliography}
  {\let\endmcitethebibliography\endthebibliography}{}
\begin{mcitethebibliography}{98}
\providecommand*\natexlab[1]{#1}
\providecommand*\mciteSetBstSublistMode[1]{}
\providecommand*\mciteSetBstMaxWidthForm[2]{}
\providecommand*\mciteBstWouldAddEndPuncttrue
  {\def\EndOfBibitem{\unskip.}}
\providecommand*\mciteBstWouldAddEndPunctfalse
  {\let\EndOfBibitem\relax}
\providecommand*\mciteSetBstMidEndSepPunct[3]{}
\providecommand*\mciteSetBstSublistLabelBeginEnd[3]{}
\providecommand*\EndOfBibitem{}
\mciteSetBstSublistMode{f}
\mciteSetBstMaxWidthForm{subitem}{(\alph{mcitesubitemcount})}
\mciteSetBstSublistLabelBeginEnd
  {\mcitemaxwidthsubitemform\space}
  {\relax}
  {\relax}

\bibitem[Strobl(1997)]{Strobl1997PhysPolym}
Strobl,~G.~R. \emph{The physics of polymers}, 2nd ed.; Springer-Verlag: Berlin
  Heidelberg, 1997\relax
\mciteBstWouldAddEndPuncttrue
\mciteSetBstMidEndSepPunct{\mcitedefaultmidpunct}
{\mcitedefaultendpunct}{\mcitedefaultseppunct}\relax
\EndOfBibitem
\bibitem[Mark(2007)]{Mark2007PhysPropertPolymHandbook}
Mark,~J.~E. \emph{Physical properties of polymers handbook}, 2nd ed.;
  Springer-Verlag: New York, 2007\relax
\mciteBstWouldAddEndPuncttrue
\mciteSetBstMidEndSepPunct{\mcitedefaultmidpunct}
{\mcitedefaultendpunct}{\mcitedefaultseppunct}\relax
\EndOfBibitem
\bibitem[Edwards(1967)]{Edwards1967StatMechTopologyI}
Edwards,~S. Statistical mechanics with topological constraints: {I}.
  \emph{Proc. Phys. Soc.} \textbf{1967}, \emph{91}, 513--519\relax
\mciteBstWouldAddEndPuncttrue
\mciteSetBstMidEndSepPunct{\mcitedefaultmidpunct}
{\mcitedefaultendpunct}{\mcitedefaultseppunct}\relax
\EndOfBibitem
\bibitem[Edwards(1967)]{Edwards1967StatMechPolymMat}
Edwards,~S. The statistical mechanics of polymerized material. \emph{Proc.
  Phys. Soc.} \textbf{1967}, \emph{92}, 9--16\relax
\mciteBstWouldAddEndPuncttrue
\mciteSetBstMidEndSepPunct{\mcitedefaultmidpunct}
{\mcitedefaultendpunct}{\mcitedefaultseppunct}\relax
\EndOfBibitem
\bibitem[Deam and Edwards(1976)Deam, and Edwards]{DeamEdwards1976ThrRubElast}
Deam,~R.; Edwards,~S. The theory of rubber elasticity. \emph{Philos. Trans. R.
  Soc. A} \textbf{1976}, \emph{280}, 317--353\relax
\mciteBstWouldAddEndPuncttrue
\mciteSetBstMidEndSepPunct{\mcitedefaultmidpunct}
{\mcitedefaultendpunct}{\mcitedefaultseppunct}\relax
\EndOfBibitem
\bibitem[Khokhlov and Nechaev(1990)Khokhlov, and
  Nechaev]{KhokhlovNechaev1990PolymChainArrObstac}
Khokhlov,~A.; Nechaev,~S. Polymer chain in an array of obstacles. \emph{Phys.
  Lett. A} \textbf{1990}, \emph{12}, 156--160\relax
\mciteBstWouldAddEndPuncttrue
\mciteSetBstMidEndSepPunct{\mcitedefaultmidpunct}
{\mcitedefaultendpunct}{\mcitedefaultseppunct}\relax
\EndOfBibitem
\bibitem[Kuhn(1936)]{Kuhn1936}
Kuhn,~W. Beziehungen zwischen {M}olek{\"u}lgr{\"o}{\ss}e, statistischer
  {M}olek{\"u}lgestalt und elastischen {E}igenschaften hochpolymerer {S}toffe.
  \emph{Kolloid-Zeitschrift} \textbf{1936}, \emph{76}, 258--271\relax
\mciteBstWouldAddEndPuncttrue
\mciteSetBstMidEndSepPunct{\mcitedefaultmidpunct}
{\mcitedefaultendpunct}{\mcitedefaultseppunct}\relax
\EndOfBibitem
\bibitem[Kuhn and Gr{\"u}n(1942)Kuhn, and Gr{\"u}n]{KuhnGrun1942}
Kuhn,~V.~W.; Gr{\"u}n,~F. Beziehungen zwischen elastischen {K}onstanten und
  {D}ehnungsdoppelbrechung hochelastischer {S}toffe. \emph{Kolloid-Zeitschrift}
  \textbf{1942}, \emph{3}, 248--271\relax
\mciteBstWouldAddEndPuncttrue
\mciteSetBstMidEndSepPunct{\mcitedefaultmidpunct}
{\mcitedefaultendpunct}{\mcitedefaultseppunct}\relax
\EndOfBibitem
\bibitem[Kuhn(1946)]{Kuhn1946}
Kuhn,~W. Dependence of the average transversal on the longitudinal dimensions
  of statistical coils formed by chain molecules. \emph{J. Polym. Sci.}
  \textbf{1946}, \emph{1}, 380--388\relax
\mciteBstWouldAddEndPuncttrue
\mciteSetBstMidEndSepPunct{\mcitedefaultmidpunct}
{\mcitedefaultendpunct}{\mcitedefaultseppunct}\relax
\EndOfBibitem
\bibitem[Wall(1942)]{Wall1942StatTDRubI}
Wall,~F. Statistical thermodynamics of rubber. \emph{J. Chem. Phys.}
  \textbf{1942}, \emph{10}, 132--134\relax
\mciteBstWouldAddEndPuncttrue
\mciteSetBstMidEndSepPunct{\mcitedefaultmidpunct}
{\mcitedefaultendpunct}{\mcitedefaultseppunct}\relax
\EndOfBibitem
\bibitem[Wall(1942)]{Wall1942StatTDRubII}
Wall,~F. Statistical thermodynamics of rubber. {II}. \emph{J. Chem. Phys.}
  \textbf{1942}, \emph{10}, 485--488\relax
\mciteBstWouldAddEndPuncttrue
\mciteSetBstMidEndSepPunct{\mcitedefaultmidpunct}
{\mcitedefaultendpunct}{\mcitedefaultseppunct}\relax
\EndOfBibitem
\bibitem[Wall(1943)]{Wall1943StatTDRubIII}
Wall,~F. Statistical thermodynamics of rubber. {III}. \emph{J. Chem. Phys.}
  \textbf{1943}, \emph{11}, 527--530\relax
\mciteBstWouldAddEndPuncttrue
\mciteSetBstMidEndSepPunct{\mcitedefaultmidpunct}
{\mcitedefaultendpunct}{\mcitedefaultseppunct}\relax
\EndOfBibitem
\bibitem[Flory and Rehner(1943)Flory, and
  Rehner]{FloryRehner1943StatMechXlnkdPolymNtwkI}
Flory,~P.; Rehner,~J. Statistical mechanics of cross-linked polymer networks.
  {I}. {R}ubberlike elasticity. \emph{J. Chem. Phys.} \textbf{1943}, \emph{11},
  512--520\relax
\mciteBstWouldAddEndPuncttrue
\mciteSetBstMidEndSepPunct{\mcitedefaultmidpunct}
{\mcitedefaultendpunct}{\mcitedefaultseppunct}\relax
\EndOfBibitem
\bibitem[Treloar(1943)]{Treloar1943ElasticityOfNtwkI}
Treloar,~L. The elasticity of a network of long-chain molecules. {I}.
  \emph{Trans. Faraday Soc.} \textbf{1943}, \emph{39}, 36--41\relax
\mciteBstWouldAddEndPuncttrue
\mciteSetBstMidEndSepPunct{\mcitedefaultmidpunct}
{\mcitedefaultendpunct}{\mcitedefaultseppunct}\relax
\EndOfBibitem
\bibitem[James and Guth(1943)James, and Guth]{JamesGuth1943ThrElstPropRub}
James,~H.~M.; Guth,~E. Theory of the elastic properties of rubber. \emph{J.
  Chem. Phys.} \textbf{1943}, \emph{11}, 455--481\relax
\mciteBstWouldAddEndPuncttrue
\mciteSetBstMidEndSepPunct{\mcitedefaultmidpunct}
{\mcitedefaultendpunct}{\mcitedefaultseppunct}\relax
\EndOfBibitem
\bibitem[James(1947)]{James1947StatPropNetwFlexChains}
James,~H.~M. Statistical properties of networks of flexible chains. \emph{J.
  Chem. Phys.} \textbf{1947}, \emph{15}, 651--668\relax
\mciteBstWouldAddEndPuncttrue
\mciteSetBstMidEndSepPunct{\mcitedefaultmidpunct}
{\mcitedefaultendpunct}{\mcitedefaultseppunct}\relax
\EndOfBibitem
\bibitem[James and Guth(1947)James, and Guth]{JamesGuth1947}
James,~H.~M.; Guth,~E. Theory of the increase in rigidity of rubber during
  cure. \emph{J. Chem. Phys.} \textbf{1947}, \emph{15}, 669--683\relax
\mciteBstWouldAddEndPuncttrue
\mciteSetBstMidEndSepPunct{\mcitedefaultmidpunct}
{\mcitedefaultendpunct}{\mcitedefaultseppunct}\relax
\EndOfBibitem
\bibitem[James and Guth(1949)James, and Guth]{JamesGuth1949}
James,~H.; Guth,~E. Simple presentation of network theory of rubber, with a
  discussion of other theories. \emph{J. Polym. Sci.} \textbf{1949}, \emph{14},
  153--182\relax
\mciteBstWouldAddEndPuncttrue
\mciteSetBstMidEndSepPunct{\mcitedefaultmidpunct}
{\mcitedefaultendpunct}{\mcitedefaultseppunct}\relax
\EndOfBibitem
\bibitem[Flory(1976)]{Flory1976}
Flory,~P.~J. Statistical thermodynamics of random networks. \emph{Proc. R. Soc.
  A} \textbf{1976}, \emph{351}, 351--380\relax
\mciteBstWouldAddEndPuncttrue
\mciteSetBstMidEndSepPunct{\mcitedefaultmidpunct}
{\mcitedefaultendpunct}{\mcitedefaultseppunct}\relax
\EndOfBibitem
\bibitem[Treloar(1975)]{Treloar1975PhysRubElasticity}
Treloar,~L. \emph{The physics of rubber elasticity}, 3rd ed.; Clarendon Press:
  Oxford, 1975\relax
\mciteBstWouldAddEndPuncttrue
\mciteSetBstMidEndSepPunct{\mcitedefaultmidpunct}
{\mcitedefaultendpunct}{\mcitedefaultseppunct}\relax
\EndOfBibitem
\bibitem[Rubinstein and Panyukov(1997)Rubinstein, and
  Panyukov]{RubinPanyuk1997NFFDfrmElstctPlmNtwk}
Rubinstein,~M.; Panyukov,~S. Nonaffine deformation and elasticity of polymer
  networks. \emph{Macromolecules} \textbf{1997}, \emph{30}, 8036--8044\relax
\mciteBstWouldAddEndPuncttrue
\mciteSetBstMidEndSepPunct{\mcitedefaultmidpunct}
{\mcitedefaultendpunct}{\mcitedefaultseppunct}\relax
\EndOfBibitem
\bibitem[Rubinstein and Panyukov(2002)Rubinstein, and
  Panyukov]{RubinPanyuk2002ElastPolymNetw}
Rubinstein,~M.; Panyukov,~S. Elasticity of polymer networks.
  \emph{Macromolecules} \textbf{2002}, \emph{35}, 6670--6686\relax
\mciteBstWouldAddEndPuncttrue
\mciteSetBstMidEndSepPunct{\mcitedefaultmidpunct}
{\mcitedefaultendpunct}{\mcitedefaultseppunct}\relax
\EndOfBibitem
\bibitem[Kaliske and Heinrich(1999)Kaliske, and
  Heinrich]{KaliskeHeinrich1999ExtendedTubeModel}
Kaliske,~M.; Heinrich,~G. An extended tube model for rubber elasticity:
  {S}tatistical-mechanical theory and finite element implementation.
  \emph{Rubber Chem. Technol.} \textbf{1999}, \emph{72}, 602--632\relax
\mciteBstWouldAddEndPuncttrue
\mciteSetBstMidEndSepPunct{\mcitedefaultmidpunct}
{\mcitedefaultendpunct}{\mcitedefaultseppunct}\relax
\EndOfBibitem
\bibitem[Mergell and Everaers(2001)Mergell, and
  Everaers]{MergellEveraers2001TubeModels}
Mergell,~B.; Everaers,~R. Tube models for rubber-elastic systems.
  \emph{Macromolecules} \textbf{2001}, \emph{34}, 5675--5686\relax
\mciteBstWouldAddEndPuncttrue
\mciteSetBstMidEndSepPunct{\mcitedefaultmidpunct}
{\mcitedefaultendpunct}{\mcitedefaultseppunct}\relax
\EndOfBibitem
\bibitem[Davidson and Goulbourne(2013)Davidson, and
  Goulbourne]{DavidsonGoulbourne2013NffNtwkModel}
Davidson,~J.~D.; Goulbourne,~N. A nonaffine network model for elastomers
  undergoing finite deformations. \emph{J. Mech. Phys. Solids} \textbf{2013},
  \emph{61}, 1784--1797\relax
\mciteBstWouldAddEndPuncttrue
\mciteSetBstMidEndSepPunct{\mcitedefaultmidpunct}
{\mcitedefaultendpunct}{\mcitedefaultseppunct}\relax
\EndOfBibitem
\bibitem[Xiang \latin{et~al.}(2018)Xiang, Zhong, Wang, Mao, Yu, and
  Qu]{XiangZhongWangMaoYuQu2018GnrlModelSoftElast}
Xiang,~Y.; Zhong,~D.; Wang,~P.; Mao,~G.; Yu,~H.; Qu,~S. A general constitutive
  model of soft elastomers. \emph{J. Mech. Phys. Solids} \textbf{2018},
  \emph{117}, 110--122\relax
\mciteBstWouldAddEndPuncttrue
\mciteSetBstMidEndSepPunct{\mcitedefaultmidpunct}
{\mcitedefaultendpunct}{\mcitedefaultseppunct}\relax
\EndOfBibitem
\bibitem[Gottlieb and Gaylord(1983)Gottlieb, and
  Gaylord]{GottliebGaylord1983I1D}
Gottlieb,~M.; Gaylord,~R.~J. Experimental tests of entanglement models of
  rubber elasticity: 1. {U}niaxial extension-compression. \emph{Polymer}
  \textbf{1983}, \emph{24}, 1644--1646\relax
\mciteBstWouldAddEndPuncttrue
\mciteSetBstMidEndSepPunct{\mcitedefaultmidpunct}
{\mcitedefaultendpunct}{\mcitedefaultseppunct}\relax
\EndOfBibitem
\bibitem[Gottlieb and Gaylord(1984)Gottlieb, and
  Gaylord]{GottliebGaylord1984IISwelling}
Gottlieb,~M.; Gaylord,~R.~J. Experimental tests of entanglement models of
  rubber elasticity. 2. {S}welling. \emph{Macromolecules} \textbf{1984},
  \emph{17}, 2024--2030\relax
\mciteBstWouldAddEndPuncttrue
\mciteSetBstMidEndSepPunct{\mcitedefaultmidpunct}
{\mcitedefaultendpunct}{\mcitedefaultseppunct}\relax
\EndOfBibitem
\bibitem[Gottlieb and Gaylord(1987)Gottlieb, and
  Gaylord]{GottliebGaylord1987III2D}
Gottlieb,~M.; Gaylord,~R.~J. Experimental tests of entanglement models of
  rubber elasticity. 3. {B}iaxial deformations. \emph{Macromolecules}
  \textbf{1987}, \emph{20}, 130--138\relax
\mciteBstWouldAddEndPuncttrue
\mciteSetBstMidEndSepPunct{\mcitedefaultmidpunct}
{\mcitedefaultendpunct}{\mcitedefaultseppunct}\relax
\EndOfBibitem
\bibitem[Higgs and Gaylord(1990)Higgs, and Gaylord]{HiggsGaylord1990}
Higgs,~P.; Gaylord,~R.~J. Slip-links, hoops and tubes: tests of entanglement
  models of rubber elasticity. \emph{Polymer} \textbf{1990}, \emph{31},
  70--74\relax
\mciteBstWouldAddEndPuncttrue
\mciteSetBstMidEndSepPunct{\mcitedefaultmidpunct}
{\mcitedefaultendpunct}{\mcitedefaultseppunct}\relax
\EndOfBibitem
\bibitem[Urayama \latin{et~al.}(2001)Urayama, Kawamura, and
  Kohjiya]{KawamuraUrayamaKohjiya2001MltXDfrmOfELPolymerNtwrkII}
Urayama,~K.; Kawamura,~T.; Kohjiya,~S. Multiaxial deformations of end-linked
  poly~(dimethylsiloxane) networks. {II}. {E}xperimental tests of molecular
  entanglement models of rubber elasticity. \emph{Macromolecules}
  \textbf{2001}, \emph{34}, 8261--8269\relax
\mciteBstWouldAddEndPuncttrue
\mciteSetBstMidEndSepPunct{\mcitedefaultmidpunct}
{\mcitedefaultendpunct}{\mcitedefaultseppunct}\relax
\EndOfBibitem
\bibitem[Mooney(1940)]{Mooney1940Origin}
Mooney,~M. A theory of large elastic deformation. \emph{J. Appl. Phys.}
  \textbf{1940}, \emph{11}, 582--592\relax
\mciteBstWouldAddEndPuncttrue
\mciteSetBstMidEndSepPunct{\mcitedefaultmidpunct}
{\mcitedefaultendpunct}{\mcitedefaultseppunct}\relax
\EndOfBibitem
\bibitem[Rivlin(1948)]{Rivlin1948LrgLstDfrmI}
Rivlin,~R. Large elastic deformations of isotropic materials. {I}.
  {F}undamental concepts. \emph{Philos. Trans. R. Soc. A} \textbf{1948},
  \emph{240}, 459--490\relax
\mciteBstWouldAddEndPuncttrue
\mciteSetBstMidEndSepPunct{\mcitedefaultmidpunct}
{\mcitedefaultendpunct}{\mcitedefaultseppunct}\relax
\EndOfBibitem
\bibitem[Rivlin(1948)]{Rivlin1948LrgLstDfrmII}
Rivlin,~R. Large elastic deformations of isotropic materials. {II}. {S}ome
  uniqueness theorems for pure, homogeneous deformation. \emph{Philos. Trans.
  R. Soc. A} \textbf{1948}, \emph{240}, 491--508\relax
\mciteBstWouldAddEndPuncttrue
\mciteSetBstMidEndSepPunct{\mcitedefaultmidpunct}
{\mcitedefaultendpunct}{\mcitedefaultseppunct}\relax
\EndOfBibitem
\bibitem[Rivlin(1948)]{Rivlin1948LrgLstDfrmIII}
Rivlin,~R. Large elastic deformations of isotropic materials. {III}. {S}ome
  simple problems in cylindrical polar co-ordinates. \emph{Philos. Trans. R.
  Soc. A} \textbf{1948}, \emph{240}, 509--525\relax
\mciteBstWouldAddEndPuncttrue
\mciteSetBstMidEndSepPunct{\mcitedefaultmidpunct}
{\mcitedefaultendpunct}{\mcitedefaultseppunct}\relax
\EndOfBibitem
\bibitem[Rivlin(1948)]{Rivlin1948LrgLstDfrmIV}
Rivlin,~R. Large elastic deformations of isotropic materials. {IV}. {F}urther
  developments of the general theory. \emph{Philos. Trans. R. Soc. A}
  \textbf{1948}, \emph{241}, 379--397\relax
\mciteBstWouldAddEndPuncttrue
\mciteSetBstMidEndSepPunct{\mcitedefaultmidpunct}
{\mcitedefaultendpunct}{\mcitedefaultseppunct}\relax
\EndOfBibitem
\bibitem[Rivlin and Saunders(1951)Rivlin, and
  Saunders]{RivlinSaunders1951LrgLstDfrmVII}
Rivlin,~R.; Saunders,~D. Large elastic deformations of isotropic materials.
  {VII}. {E}xperiments on the deformation of rubber. \emph{Philos. Trans. R.
  Soc. A} \textbf{1951}, \emph{243}, 251--288\relax
\mciteBstWouldAddEndPuncttrue
\mciteSetBstMidEndSepPunct{\mcitedefaultmidpunct}
{\mcitedefaultendpunct}{\mcitedefaultseppunct}\relax
\EndOfBibitem
\bibitem[Mark(1975)]{Mark1975MRConstants}
Mark,~J. The constants $2\,\mathrm{C}_1$ and $2\,\mathrm{C}_2$ in
  phenomenological elasticity theory and their dependence on experimental
  variables. \emph{Rubber Chem. Technol.} \textbf{1975}, \emph{48},
  495--512\relax
\mciteBstWouldAddEndPuncttrue
\mciteSetBstMidEndSepPunct{\mcitedefaultmidpunct}
{\mcitedefaultendpunct}{\mcitedefaultseppunct}\relax
\EndOfBibitem
\bibitem[Mark and Sullivan(1977)Mark, and
  Sullivan]{MarkSullivan1977ModelNetworksOfELPDMSI}
Mark,~J.; Sullivan,~J. Model networks of end-linked polydimethylsiloxane
  chains. {I}. {C}omparisons between experimental and theoretical values of the
  elastic modulus and the equilibrium degree of swelling. \emph{J. Chem. Phys.}
  \textbf{1977}, \emph{66}, 1006--1011\relax
\mciteBstWouldAddEndPuncttrue
\mciteSetBstMidEndSepPunct{\mcitedefaultmidpunct}
{\mcitedefaultendpunct}{\mcitedefaultseppunct}\relax
\EndOfBibitem
\bibitem[Sharaf and Mark(1994)Sharaf, and Mark]{SharafMark1994Interpretation}
Sharaf,~M.; Mark,~J. Interpretation of the small-strain moduli of model
  networks of polydimethylsiloxane. \emph{J. Chem. Phys.} \textbf{1994},
  \emph{35}, 740--751\relax
\mciteBstWouldAddEndPuncttrue
\mciteSetBstMidEndSepPunct{\mcitedefaultmidpunct}
{\mcitedefaultendpunct}{\mcitedefaultseppunct}\relax
\EndOfBibitem
\bibitem[Schl{\"o}gl \latin{et~al.}(2014)Schl{\"o}gl, Trutschel, Chass{\'e},
  Riess, and Saalw{\"a}chter]{SchloglTrutschelChasseRiessSaalwachter2014}
Schl{\"o}gl,~S.; Trutschel,~M.-L.; Chass{\'e},~W.; Riess,~G.;
  Saalw{\"a}chter,~K. Entanglement effects in elastomers: macroscopic vs
  microscopic properties. \emph{Macromolecules} \textbf{2014}, \emph{47},
  2759--2773\relax
\mciteBstWouldAddEndPuncttrue
\mciteSetBstMidEndSepPunct{\mcitedefaultmidpunct}
{\mcitedefaultendpunct}{\mcitedefaultseppunct}\relax
\EndOfBibitem
\bibitem[Svaneborg \latin{et~al.}(2016)Svaneborg, Karimi-Varzaneh, Hojdis,
  Fleck, and
  Everaers]{SvaneborgKarimiHojdisFleckEveraers2016MultiscaleApproach}
Svaneborg,~C.; Karimi-Varzaneh,~H.~A.; Hojdis,~N.; Fleck,~F.; Everaers,~R.
  Multiscale approach to equilibrating model polymer melts. \emph{Phys. Rev. E}
  \textbf{2016}, \emph{94}, 032502\relax
\mciteBstWouldAddEndPuncttrue
\mciteSetBstMidEndSepPunct{\mcitedefaultmidpunct}
{\mcitedefaultendpunct}{\mcitedefaultseppunct}\relax
\EndOfBibitem
\bibitem[Zhang \latin{et~al.}(2014)Zhang, Moreira, Stuehn, Daoulas, and
  Kremer]{ZhangMoreiraStuehnDaoulasKremer2014EquilHierarchStrat}
Zhang,~G.; Moreira,~L.~A.; Stuehn,~T.; Daoulas,~K.~C.; Kremer,~K. Equilibtation
  of high molecular weight polymer melts: a hierarchical strategy. \emph{ACS
  Macro Lett.} \textbf{2014}, \emph{3}, 198--203\relax
\mciteBstWouldAddEndPuncttrue
\mciteSetBstMidEndSepPunct{\mcitedefaultmidpunct}
{\mcitedefaultendpunct}{\mcitedefaultseppunct}\relax
\EndOfBibitem
\bibitem[Auhl \latin{et~al.}(2003)Auhl, Everaers, Grest, Kremer, and
  Plimpton]{AuhlEveraersGrestKremerPlimpton2003EQLongChainPlmMltInCmpSim}
Auhl,~R.; Everaers,~R.; Grest,~G.~S.; Kremer,~K.; Plimpton,~S.~J. Equilibration
  of long chain polymer melts in computer simulations. \emph{J. Chem. Phys.}
  \textbf{2003}, \emph{119}, 12718--12728\relax
\mciteBstWouldAddEndPuncttrue
\mciteSetBstMidEndSepPunct{\mcitedefaultmidpunct}
{\mcitedefaultendpunct}{\mcitedefaultseppunct}\relax
\EndOfBibitem
\bibitem[Everaers \latin{et~al.}(2020)Everaers, Karimi-Varzaneh, Fleck, Hojdis,
  and Svaneborg]{EveraersKarimiFleckHojdisSvaneborg2020KGMap}
Everaers,~R.; Karimi-Varzaneh,~H.~A.; Fleck,~F.; Hojdis,~N.; Svaneborg,~C.
  {K}remer-{G}rest models for commodity polymer melts: linking theory,
  experiment and simulation at the {K}uhn scale. \emph{Macromolecules}
  \textbf{2020}, \emph{53}, 1901--1916\relax
\mciteBstWouldAddEndPuncttrue
\mciteSetBstMidEndSepPunct{\mcitedefaultmidpunct}
{\mcitedefaultendpunct}{\mcitedefaultseppunct}\relax
\EndOfBibitem
\bibitem[Lang \latin{et~al.}(2005)Lang, G{\"o}ritz, and
  Kreitmeier]{LangGoritzKreitmeier2005IntraReactions}
Lang,~M.; G{\"o}ritz,~D.; Kreitmeier,~S. Intramolecular reactions in randomly
  end-linked polymer networks and linear (co)polymerizations.
  \emph{Macromolecules} \textbf{2005}, \emph{38}, 2515--2523\relax
\mciteBstWouldAddEndPuncttrue
\mciteSetBstMidEndSepPunct{\mcitedefaultmidpunct}
{\mcitedefaultendpunct}{\mcitedefaultseppunct}\relax
\EndOfBibitem
\bibitem[Everaers \latin{et~al.}(2004)Everaers, Sukumaran, Grest, Svaneborg,
  Sivasubramanian, and
  Kremer]{EveraersSukuGrestSvaneborgSivaKremer2004RheolTopol}
Everaers,~R.; Sukumaran,~S.~K.; Grest,~G.~S.; Svaneborg,~C.;
  Sivasubramanian,~A.; Kremer,~K. Rheology and microscopic topology of
  entangled polymeric liquids. \emph{Science} \textbf{2004}, \emph{303},
  823--826\relax
\mciteBstWouldAddEndPuncttrue
\mciteSetBstMidEndSepPunct{\mcitedefaultmidpunct}
{\mcitedefaultendpunct}{\mcitedefaultseppunct}\relax
\EndOfBibitem
\bibitem[Svaneborg \latin{et~al.}(2008)Svaneborg, Everaers, Grest, and
  Curro]{SvaneborgEveraersGrestCurro2008StressContributions}
Svaneborg,~C.; Everaers,~R.; Grest,~G.~S.; Curro,~J.~G. Connectivity and
  entanglement stress contributions in strained polymer networks.
  \emph{Macromolecules} \textbf{2008}, \emph{41}, 4920--4928\relax
\mciteBstWouldAddEndPuncttrue
\mciteSetBstMidEndSepPunct{\mcitedefaultmidpunct}
{\mcitedefaultendpunct}{\mcitedefaultseppunct}\relax
\EndOfBibitem
\bibitem[Kuhn(1934)]{Kuhn1934}
Kuhn,~W. {\"U}ber die {G}estalt fadenf{\"o}rmiger Molek{\"u}le in {L}{\"o}sung.
  \emph{Kolloid-Zeitschrift} \textbf{1934}, \emph{68}, 2--15\relax
\mciteBstWouldAddEndPuncttrue
\mciteSetBstMidEndSepPunct{\mcitedefaultmidpunct}
{\mcitedefaultendpunct}{\mcitedefaultseppunct}\relax
\EndOfBibitem
\bibitem[Flory(1953)]{Flory1953PrincipPolymerChem}
Flory,~P.~J. \emph{Principles of polymer chemistry}, 1st ed.; Cornell
  University Press: Ithaca, New York, 1953\relax
\mciteBstWouldAddEndPuncttrue
\mciteSetBstMidEndSepPunct{\mcitedefaultmidpunct}
{\mcitedefaultendpunct}{\mcitedefaultseppunct}\relax
\EndOfBibitem
\bibitem[Flory(1949)]{Flory1949ConfigRealPolyChains}
Flory,~P.~J. The configuration of real polymer chains. \emph{J. Chem. Phys.}
  \textbf{1949}, \emph{17}, 303--310\relax
\mciteBstWouldAddEndPuncttrue
\mciteSetBstMidEndSepPunct{\mcitedefaultmidpunct}
{\mcitedefaultendpunct}{\mcitedefaultseppunct}\relax
\EndOfBibitem
\bibitem[Prince E.~Rouse(1953)]{Rouse1953Origin}
Prince E.~Rouse,~J. A theory of the linear viscoelastic properties of dilute
  solutions of coiling polymers. \emph{J. Chem. Phys.} \textbf{1953},
  \emph{21}, 1272--1280\relax
\mciteBstWouldAddEndPuncttrue
\mciteSetBstMidEndSepPunct{\mcitedefaultmidpunct}
{\mcitedefaultendpunct}{\mcitedefaultseppunct}\relax
\EndOfBibitem
\bibitem[Kawamura \latin{et~al.}(2001)Kawamura, Urayama, and
  Kohjiya]{KawamuraUrayamaKohjiya2001MltXDfrmOfELPolymerNtwrkI}
Kawamura,~T.; Urayama,~K.; Kohjiya,~S. Multiaxial deformations of end-linked
  poly~(dimethylsiloxane) networks. {I}. {P}henomenological approach to strain
  energy density function. \emph{Macromolecules} \textbf{2001}, \emph{34},
  8252--8260\relax
\mciteBstWouldAddEndPuncttrue
\mciteSetBstMidEndSepPunct{\mcitedefaultmidpunct}
{\mcitedefaultendpunct}{\mcitedefaultseppunct}\relax
\EndOfBibitem
\bibitem[Graessley(1975)]{Graessley1975}
Graessley,~W.~W. Elasticity and chain dimensions in {G}aussian networks.
  \emph{Macromolecules} \textbf{1975}, \emph{8}, 865--868\relax
\mciteBstWouldAddEndPuncttrue
\mciteSetBstMidEndSepPunct{\mcitedefaultmidpunct}
{\mcitedefaultendpunct}{\mcitedefaultseppunct}\relax
\EndOfBibitem
\bibitem[Edwards and Viglis(1988)Edwards, and
  Viglis]{EdwardsViglis1988TubeModelThrRubbElast}
Edwards,~S.; Viglis,~T. The tube model theory of rubber elasticity. \emph{Rep.
  Prog. Phys.} \textbf{1988}, \emph{51}, 243--297\relax
\mciteBstWouldAddEndPuncttrue
\mciteSetBstMidEndSepPunct{\mcitedefaultmidpunct}
{\mcitedefaultendpunct}{\mcitedefaultseppunct}\relax
\EndOfBibitem
\bibitem[Zhong \latin{et~al.}(2016)Zhong, Wang, Kawamoto, Olsen, and
  Johnson]{ZhongWangKawamotoOlsenJohnson2016RENT}
Zhong,~M.; Wang,~R.; Kawamoto,~K.; Olsen,~B.~D.; Johnson,~J.~A. Quantifying the
  impact of molecular defects on polymer network elasticity. \emph{Science}
  \textbf{2016}, \emph{353}, 1264--1268\relax
\mciteBstWouldAddEndPuncttrue
\mciteSetBstMidEndSepPunct{\mcitedefaultmidpunct}
{\mcitedefaultendpunct}{\mcitedefaultseppunct}\relax
\EndOfBibitem
\bibitem[Lang(2018)]{Lang2018}
Lang,~M. Elasticity of phantom model networks with cyclic defects. \emph{ACS
  Macro Lett.} \textbf{2018}, \emph{7}, 536--539\relax
\mciteBstWouldAddEndPuncttrue
\mciteSetBstMidEndSepPunct{\mcitedefaultmidpunct}
{\mcitedefaultendpunct}{\mcitedefaultseppunct}\relax
\EndOfBibitem
\bibitem[Panyukov(2019)]{Panyukov2019}
Panyukov,~S. Loops in polymer networks. \emph{Macromolecules} \textbf{2019},
  \emph{52}, 4145--4153\relax
\mciteBstWouldAddEndPuncttrue
\mciteSetBstMidEndSepPunct{\mcitedefaultmidpunct}
{\mcitedefaultendpunct}{\mcitedefaultseppunct}\relax
\EndOfBibitem
\bibitem[Lin \latin{et~al.}(2019)Lin, Wang, Johnson, and
  Olsen]{LinWangJohnsonOlsen2019ElasticityRealGaussPhtnmNtwrk}
Lin,~T.-S.; Wang,~R.; Johnson,~J.~A.; Olsen,~B.~D. Revisiting the elasticity
  theory for real gaussian phantom networks. \emph{Macromolecules}
  \textbf{2019}, \emph{52}, 1685--1694\relax
\mciteBstWouldAddEndPuncttrue
\mciteSetBstMidEndSepPunct{\mcitedefaultmidpunct}
{\mcitedefaultendpunct}{\mcitedefaultseppunct}\relax
\EndOfBibitem
\bibitem[Edwards and Viglis(1986)Edwards, and
  Viglis]{EdwardsViglis1986EffOfEntanglRubbElast}
Edwards,~S.; Viglis,~T. The effect of entanglements in rubber elasticity.
  \emph{Polymer} \textbf{1986}, \emph{27}, 483--492\relax
\mciteBstWouldAddEndPuncttrue
\mciteSetBstMidEndSepPunct{\mcitedefaultmidpunct}
{\mcitedefaultendpunct}{\mcitedefaultseppunct}\relax
\EndOfBibitem
\bibitem[Warner and Edwards(1978)Warner, and
  Edwards]{WarnerEdwards1978NeutronScatter}
Warner,~M.; Edwards,~S. Neutron scattering from strained polymer networks.
  \emph{J. Phys. A} \textbf{1978}, \emph{11}, 1649--1655\relax
\mciteBstWouldAddEndPuncttrue
\mciteSetBstMidEndSepPunct{\mcitedefaultmidpunct}
{\mcitedefaultendpunct}{\mcitedefaultseppunct}\relax
\EndOfBibitem
\bibitem[Cohen(1991)]{Cohen1991}
Cohen,~A. A {P}ad{\'e} approximant to the inverse {L}angevin function.
  \emph{Rheol. Acta} \textbf{1991}, \emph{30}, 270--273\relax
\mciteBstWouldAddEndPuncttrue
\mciteSetBstMidEndSepPunct{\mcitedefaultmidpunct}
{\mcitedefaultendpunct}{\mcitedefaultseppunct}\relax
\EndOfBibitem
\bibitem[Heinrich and Straube(1984)Heinrich, and Straube]{HeinrichStraube1984}
Heinrich,~G.; Straube,~E. On the strength and deformation dependence of the
  tube-like topological constraints of polymer networks, melts and concentrated
  solutions. {II}. {P}olymer melts and concentrated solutions. \emph{Acta
  Polym.} \textbf{1984}, \emph{35}, 115--119\relax
\mciteBstWouldAddEndPuncttrue
\mciteSetBstMidEndSepPunct{\mcitedefaultmidpunct}
{\mcitedefaultendpunct}{\mcitedefaultseppunct}\relax
\EndOfBibitem
\bibitem[Kr{\"o}ger(2015)]{Kroeger2015}
Kr{\"o}ger,~M. Simple, admissible, and accurate approximants of the inverse
  {L}angevin and {B}rillouin functions, relevant for strong polymer
  deformations and flows. \emph{J. Nonnewton. Fluid Mech.} \textbf{2015},
  \emph{223}, 77--87\relax
\mciteBstWouldAddEndPuncttrue
\mciteSetBstMidEndSepPunct{\mcitedefaultmidpunct}
{\mcitedefaultendpunct}{\mcitedefaultseppunct}\relax
\EndOfBibitem
\bibitem[Grest and Kremer(1986)Grest, and
  Kremer]{GrestKremer1986MDSimPolymHeatBath}
Grest,~G.~S.; Kremer,~K. Molecular {D}ynamics simulation for polymers in the
  presence of a heat bath. \emph{Phys. Rev. A} \textbf{1986}, \emph{33},
  3628--3631\relax
\mciteBstWouldAddEndPuncttrue
\mciteSetBstMidEndSepPunct{\mcitedefaultmidpunct}
{\mcitedefaultendpunct}{\mcitedefaultseppunct}\relax
\EndOfBibitem
\bibitem[Kremer and Grest(1990)Kremer, and
  Grest]{KremerGrest1990DynEntangLinPolyMelt}
Kremer,~K.; Grest,~G.~S. Dynamics of entangled linear polymer melts: {A}
  {M}olecular-{D}ynamics simulation. \emph{J. Chem. Phys.} \textbf{1990},
  \emph{92}, 5057--5086\relax
\mciteBstWouldAddEndPuncttrue
\mciteSetBstMidEndSepPunct{\mcitedefaultmidpunct}
{\mcitedefaultendpunct}{\mcitedefaultseppunct}\relax
\EndOfBibitem
\bibitem[Faller \latin{et~al.}(1999)Faller, Kolb, and
  M{\"u}ller-Plathe]{FallerKolbMullerPlathe1999LocalChainOrder}
Faller,~R.; Kolb,~A.; M{\"u}ller-Plathe,~F. Local chain ordering in amorphous
  polymer melts: influence of chain stiffness. \emph{Phys. Chem. Chem. Phys.}
  \textbf{1999}, \emph{1}, 2071--2076\relax
\mciteBstWouldAddEndPuncttrue
\mciteSetBstMidEndSepPunct{\mcitedefaultmidpunct}
{\mcitedefaultendpunct}{\mcitedefaultseppunct}\relax
\EndOfBibitem
\bibitem[Gr{\o}nbech-Jensen and Farago(2013)Gr{\o}nbech-Jensen, and
  Farago]{GronJensFarago2013GJF}
Gr{\o}nbech-Jensen,~N.; Farago,~O. A simple and effective {V}erlet-type
  algorithm for simulating {L}angevin dynamics. \emph{Mol. Phys.}
  \textbf{2013}, \emph{111}, 983--991\relax
\mciteBstWouldAddEndPuncttrue
\mciteSetBstMidEndSepPunct{\mcitedefaultmidpunct}
{\mcitedefaultendpunct}{\mcitedefaultseppunct}\relax
\EndOfBibitem
\bibitem[Verlet(1967)]{Verlet1967}
Verlet,~L. Computer ''experiments'' on classical fluids. {I}. {T}hermodynamical
  properties of {L}ennard-{J}ones molecules. \emph{Phys. Rev.} \textbf{1967},
  \emph{159}, 98--103\relax
\mciteBstWouldAddEndPuncttrue
\mciteSetBstMidEndSepPunct{\mcitedefaultmidpunct}
{\mcitedefaultendpunct}{\mcitedefaultseppunct}\relax
\EndOfBibitem
\bibitem[Press \latin{et~al.}(1992)Press, Teukolsky, Vetterling, and
  Flannery]{PressTeukolskyVetterlingFlannery1992NumRecInC}
Press,~W.~H.; Teukolsky,~S.~A.; Vetterling,~W.~T.; Flannery,~B.~P.
  \emph{Numerical recipies in {C}. {T}he art of scientific computing}, 2nd ed.;
  Cambridge University Press: New York, 1992\relax
\mciteBstWouldAddEndPuncttrue
\mciteSetBstMidEndSepPunct{\mcitedefaultmidpunct}
{\mcitedefaultendpunct}{\mcitedefaultseppunct}\relax
\EndOfBibitem
\bibitem[Plimpton(1995)]{Plimpton1995LAMMPS}
Plimpton,~S. Fast parallel algorithms for short-range {M}olecular {D}ynamics.
  \emph{J. Comput. Phys.} \textbf{1995}, \emph{117}, 1--19\relax
\mciteBstWouldAddEndPuncttrue
\mciteSetBstMidEndSepPunct{\mcitedefaultmidpunct}
{\mcitedefaultendpunct}{\mcitedefaultseppunct}\relax
\EndOfBibitem
\bibitem[Svaneborg()]{Fetters2006KGMap}
Svaneborg,~C. Kuhn mapping for Kremer-Grest polymer models.
  http://polymer.zqex.dk/\relax
\mciteBstWouldAddEndPuncttrue
\mciteSetBstMidEndSepPunct{\mcitedefaultmidpunct}
{\mcitedefaultendpunct}{\mcitedefaultseppunct}\relax
\EndOfBibitem
\bibitem[Svaneborg and Everaers(2020)Svaneborg, and
  Everaers]{SvaneborgkEveraers2020KGModel}
Svaneborg,~C.; Everaers,~R. Characteristic time and length scales in melts of
  {K}remer-{G}rest bead-spring polymers with wormlike bending stiffness.
  \emph{Macromolecules} \textbf{2020}, \emph{53}, 1917--1941\relax
\mciteBstWouldAddEndPuncttrue
\mciteSetBstMidEndSepPunct{\mcitedefaultmidpunct}
{\mcitedefaultendpunct}{\mcitedefaultseppunct}\relax
\EndOfBibitem
\bibitem[Fetters \latin{et~al.}(2007)Fetters, Lohse, and
  Colby]{FettersLohseColby2007ChainDimEntaSpac}
Fetters,~L.~J.; Lohse,~D.~J.; Colby,~R.~H. In \emph{Physical Properties of
  Polymers Handbook}; Mark,~J.~E., Ed.; Springer Science+Business Media, LLC,
  2007; Chapter 25, pp 447--454\relax
\mciteBstWouldAddEndPuncttrue
\mciteSetBstMidEndSepPunct{\mcitedefaultmidpunct}
{\mcitedefaultendpunct}{\mcitedefaultseppunct}\relax
\EndOfBibitem
\bibitem[Ch{\'a}vez and Saalw{\"a}chter(2011)Ch{\'a}vez, and
  Saalw{\"a}chter]{ChavezSaalwachter2011TimeDomainNMRObservZPolymDyn}
Ch{\'a}vez,~F.~V.; Saalw{\"a}chter,~K. Time-domain {NMR} observation of
  entangled polymer dynamics: analytical theory of signal functions.
  \emph{Macromolecules} \textbf{2011}, \emph{44}, 1560--1569\relax
\mciteBstWouldAddEndPuncttrue
\mciteSetBstMidEndSepPunct{\mcitedefaultmidpunct}
{\mcitedefaultendpunct}{\mcitedefaultseppunct}\relax
\EndOfBibitem
\bibitem[Lin(1987)]{Lin1987}
Lin,~Y.-H. Number of entanglement strands per cubed tube diameter, a
  fundamental aspect of topological universality in polymer viscoelasticity.
  \emph{Macromolecules} \textbf{1987}, \emph{20}, 3080\relax
\mciteBstWouldAddEndPuncttrue
\mciteSetBstMidEndSepPunct{\mcitedefaultmidpunct}
{\mcitedefaultendpunct}{\mcitedefaultseppunct}\relax
\EndOfBibitem
\bibitem[Kavassalis and Noolandi(1987)Kavassalis, and
  Noolandi]{KavassalisNoolandi1987}
Kavassalis,~T.; Noolandi,~J. New view of entanglements in dense polymer
  systems. \emph{Phys. Rev. Lett.} \textbf{1987}, \emph{59}, 2674\relax
\mciteBstWouldAddEndPuncttrue
\mciteSetBstMidEndSepPunct{\mcitedefaultmidpunct}
{\mcitedefaultendpunct}{\mcitedefaultseppunct}\relax
\EndOfBibitem
\bibitem[Rosa and Everaers(2014)Rosa, and
  Everaers]{RosaEveraers2014RingPolymInMelt}
Rosa,~A.; Everaers,~R. Ring polymers in melt state: the physics of crumpling.
  \emph{Phys. Rev. Lett.} \textbf{2014}, \emph{112}, 118302\relax
\mciteBstWouldAddEndPuncttrue
\mciteSetBstMidEndSepPunct{\mcitedefaultmidpunct}
{\mcitedefaultendpunct}{\mcitedefaultseppunct}\relax
\EndOfBibitem
\bibitem[Fetters \latin{et~al.}(1994)Fetters, Lohse, Richter, Witten, and
  Zirkel]{FettersLohseRichterWittenZirkel1994Connection}
Fetters,~L.~J.; Lohse,~D.~J.; Richter,~D.; Witten,~T.~A.; Zirkel,~A. Connection
  between polymer molecular weight, density, chain dimensions, and melt
  viscoelastic properties. \emph{Macromolecules} \textbf{1994}, \emph{27},
  4639\relax
\mciteBstWouldAddEndPuncttrue
\mciteSetBstMidEndSepPunct{\mcitedefaultmidpunct}
{\mcitedefaultendpunct}{\mcitedefaultseppunct}\relax
\EndOfBibitem
\bibitem[Hsu and Kremer(2019)Hsu, and Kremer]{HsuKremer2019ClusterZPoints}
Hsu,~H.-P.; Kremer,~K. Clustering of entanglement points in highly strained
  polymer melts. \emph{Macromolecules} \textbf{2019}, \emph{52},
  6756--6772\relax
\mciteBstWouldAddEndPuncttrue
\mciteSetBstMidEndSepPunct{\mcitedefaultmidpunct}
{\mcitedefaultendpunct}{\mcitedefaultseppunct}\relax
\EndOfBibitem
\bibitem[Jacobson and Stockmayer(1950)Jacobson, and
  Stockmayer]{JacobsonStockmayer1950}
Jacobson,~H.; Stockmayer,~W. Intramolecular reaction in polycondensations. {I}.
  {T}he theory of linear systems. \emph{J. Chem. Phys.} \textbf{1950},
  \emph{18}, 1600--1606\relax
\mciteBstWouldAddEndPuncttrue
\mciteSetBstMidEndSepPunct{\mcitedefaultmidpunct}
{\mcitedefaultendpunct}{\mcitedefaultseppunct}\relax
\EndOfBibitem
\bibitem[Sukumaran \latin{et~al.}(2005)Sukumaran, Grest, Kremer, and
  Everaers]{SukuGrestKremerEveraers2005IdentPPMesh}
Sukumaran,~S.~K.; Grest,~G.~S.; Kremer,~K.; Everaers,~R. Identifying the
  primitive path mesh in entangled polymer liquids. \emph{J. Polym. Sci. B:
  Polym. Phys.} \textbf{2005}, \emph{403}, 917--933\relax
\mciteBstWouldAddEndPuncttrue
\mciteSetBstMidEndSepPunct{\mcitedefaultmidpunct}
{\mcitedefaultendpunct}{\mcitedefaultseppunct}\relax
\EndOfBibitem
\bibitem[Not()]{Note_CriticalBondLength}
Let us consider two pairs of LJ beads in space. In each pair, beads are
  connected by FENE springs. Let us set $\sigma=1$, $\varepsilon=1$, $k=30$,
  $k_B=1$, $T=1$, $R_0=1.5$. The energy barrier, which has to be overpassed for
  the bonds to slide through each other, can be calculated as the difference
  between the energy in the transition state, i.e. when the bonds appear in the
  same plane, crossing perpendicularly, and the energy in the state, when the
  bonds are located infinitely far away from each other. If the bond lengths
  are determined in such the way that they provide the minimum energy to an
  immediate configuration of the system, then the energy threshold to overpass
  in the equilibrium state is equal to $\approx
  75\,k_B\,T$.~\cite{SukuGrestKremerEveraers2005IdentPPMesh} To estimate the
  bond length, which could lead to the bond sliding through each other due to
  thermal fluctuations, let us consider the system of $5\,000\,000$ LJ
  beads~(hence, $\approx 5\,000\,000$ FENE bonds), which dynamics is simulated
  for $10\,000\,000$ time steps. The probability $P$ of that a single bonds
  crossing occurs is equal to the product of the number of bonds, the number of
  simulation time steps and the probability $\exp(-U/k_B\,T)$ for two bonds to
  slide through each other, where $U$ is the energy barrier. Seeking for the
  value of $U$, leading to $P \sim 10^{-3}-10^{-4}$, one finds $U \approx
  30-31\,k_B\,T$. Giving the potential barrier to overpass, let us consider
  again two pairs of LJ beads. Let one of them have fixed length, while other
  have the length determined in agreement with the minimum energy criteria.
  That fixed value is adjusted to the given value of the energy barrier. Then,
  one can estimate it as $\approx 1.4\,\sigma$.\relax
\mciteBstWouldAddEndPunctfalse
\mciteSetBstMidEndSepPunct{\mcitedefaultmidpunct}
{}{\mcitedefaultseppunct}\relax
\EndOfBibitem
\bibitem[Doi and Edwards(1986)Doi, and Edwards]{DoiEdwards1986TheorPolymDyn}
Doi,~M.; Edwards,~S. \emph{The theory of polymer dynamics}; Clarendon Press:
  Oxford, 1986\relax
\mciteBstWouldAddEndPuncttrue
\mciteSetBstMidEndSepPunct{\mcitedefaultmidpunct}
{\mcitedefaultendpunct}{\mcitedefaultseppunct}\relax
\EndOfBibitem
\bibitem[Pak and Flory(1979)Pak, and Flory]{PakFlory1979Stress1DStrPDMS}
Pak,~H.; Flory,~P.~J. Relationship of stress to uniaxial strain in crosslinked
  poly(dimethylsi1oxane) over the full range from large compressions to high
  elongations. \emph{J. Polym. Sci.: Polymer Physics Edition} \textbf{1979},
  \emph{17}, 1845--1854\relax
\mciteBstWouldAddEndPuncttrue
\mciteSetBstMidEndSepPunct{\mcitedefaultmidpunct}
{\mcitedefaultendpunct}{\mcitedefaultseppunct}\relax
\EndOfBibitem
\bibitem[Xu and Mark(1990)Xu, and Mark]{XuMark1990}
Xu,~P.; Mark,~J. Biaxial extension studies using inflation of sheets of
  unimodal model networks. \emph{Rubber Chem. Technol.} \textbf{1990},
  \emph{63}, 276--284\relax
\mciteBstWouldAddEndPuncttrue
\mciteSetBstMidEndSepPunct{\mcitedefaultmidpunct}
{\mcitedefaultendpunct}{\mcitedefaultseppunct}\relax
\EndOfBibitem
\bibitem[Langley(1968)]{Langley1968}
Langley,~N.~R. Elastically effective strand density in polymer networks.
  \emph{Macromolecules} \textbf{1968}, \emph{1}, 348--352\relax
\mciteBstWouldAddEndPuncttrue
\mciteSetBstMidEndSepPunct{\mcitedefaultmidpunct}
{\mcitedefaultendpunct}{\mcitedefaultseppunct}\relax
\EndOfBibitem
\bibitem[Treloar(1973)]{Treloar1973ElastPropRub}
Treloar,~L. The elasticity and related properties of rubbers. \emph{Rep. Prog.
  Phys.} \textbf{1973}, \emph{36}, 755--826\relax
\mciteBstWouldAddEndPuncttrue
\mciteSetBstMidEndSepPunct{\mcitedefaultmidpunct}
{\mcitedefaultendpunct}{\mcitedefaultseppunct}\relax
\EndOfBibitem
\bibitem[Treloar(1974)]{Treloar1974ElastPropRub}
Treloar,~L. The elasticity and related properties of rubbers. \emph{Rubber
  Chem. Technol.} \textbf{1974}, \emph{47}, 625--696\relax
\mciteBstWouldAddEndPuncttrue
\mciteSetBstMidEndSepPunct{\mcitedefaultmidpunct}
{\mcitedefaultendpunct}{\mcitedefaultseppunct}\relax
\EndOfBibitem
\bibitem[Treloar(1974)]{Treloar1974MechRubElast}
Treloar,~L.~R. The mechanics of rubber elasticity. \emph{J. Polym. Sci.:
  Polymer Symposia} \textbf{1974}, \emph{48}, 107--123\relax
\mciteBstWouldAddEndPuncttrue
\mciteSetBstMidEndSepPunct{\mcitedefaultmidpunct}
{\mcitedefaultendpunct}{\mcitedefaultseppunct}\relax
\EndOfBibitem
\bibitem[Gumbrell \latin{et~al.}(1953)Gumbrell, Mullins, and
  Rivlin]{GumbrellMullinsRivlin1953}
Gumbrell,~S.; Mullins,~L.; Rivlin,~R. Departures of the elastic behaviour of
  rubbers in simple extension from the kinetic theory. \emph{Trans. Faraday
  Soc.} \textbf{1953}, \emph{49}, 1495--1505\relax
\mciteBstWouldAddEndPuncttrue
\mciteSetBstMidEndSepPunct{\mcitedefaultmidpunct}
{\mcitedefaultendpunct}{\mcitedefaultseppunct}\relax
\EndOfBibitem
\bibitem[Han \latin{et~al.}(1999)Han, Horka, and McKenna]{HanHorkaMcKenna1999}
Han,~W.~H.; Horka,~F.; McKenna,~G.~B. Mechanical and swelling behaviors of
  rubber: a comparison of some molecular models with experiment. \emph{Math.
  Mech. Solids} \textbf{1999}, \emph{4}, 139--167\relax
\mciteBstWouldAddEndPuncttrue
\mciteSetBstMidEndSepPunct{\mcitedefaultmidpunct}
{\mcitedefaultendpunct}{\mcitedefaultseppunct}\relax
\EndOfBibitem
\bibitem[Ronca and Allegra(1975)Ronca, and Allegra]{RoncaAllegra1975}
Ronca,~G.; Allegra,~G. An approach to rubber elasticity with internal
  constraints. \emph{J. Chem. Phys.} \textbf{1975}, \emph{63}, 4990--4997\relax
\mciteBstWouldAddEndPuncttrue
\mciteSetBstMidEndSepPunct{\mcitedefaultmidpunct}
{\mcitedefaultendpunct}{\mcitedefaultseppunct}\relax
\EndOfBibitem
\bibitem[Boyer and Miller(1987)Boyer, and Miller]{BoyerMiller1987}
Boyer,~R.~F.; Miller,~R.~L. Correlations involving the Mooney-Rivlin
  $\mathrm{C}_2$ constant and the number of chain atoms between physical
  entanglements, $N_c$. \emph{Polymer} \textbf{1987}, \emph{28}, 399--407\relax
\mciteBstWouldAddEndPuncttrue
\mciteSetBstMidEndSepPunct{\mcitedefaultmidpunct}
{\mcitedefaultendpunct}{\mcitedefaultseppunct}\relax
\EndOfBibitem
\bibitem[Svaneborg(2012)]{Everaers2012TopoVSRheo}
Svaneborg,~C. Topological versus rheological entanglement length in
  primitive-path analysis protocols, tube models, and slip-link models.
  \emph{Phys. Rev. E} \textbf{2012}, \emph{86}, 022801\relax
\mciteBstWouldAddEndPuncttrue
\mciteSetBstMidEndSepPunct{\mcitedefaultmidpunct}
{\mcitedefaultendpunct}{\mcitedefaultseppunct}\relax
\EndOfBibitem
\bibitem[Kl{\"u}ppel(1992)]{Kluppel1992}
Kl{\"u}ppel,~M. Trapped engtanglements in polymer networks and their influence
  on the stress-strain behaviour up to large extensions. \emph{Progr. Colloid
  Polym. Sci.} \textbf{1992}, \emph{90}, 137--143\relax
\mciteBstWouldAddEndPuncttrue
\mciteSetBstMidEndSepPunct{\mcitedefaultmidpunct}
{\mcitedefaultendpunct}{\mcitedefaultseppunct}\relax
\EndOfBibitem
\bibitem[Rubinstein and Colby(2003)Rubinstein, and
  Colby]{RubinColby2003PolymerPhysics}
Rubinstein,~M.; Colby,~R.~H. \emph{Polymer physics}; Oxford University Press:
  New York, 2003\relax
\mciteBstWouldAddEndPuncttrue
\mciteSetBstMidEndSepPunct{\mcitedefaultmidpunct}
{\mcitedefaultendpunct}{\mcitedefaultseppunct}\relax
\EndOfBibitem
\end{mcitethebibliography}
